\documentclass[10pt]{iopart} 

\usepackage{iopams}
\usepackage[T1]{fontenc}
\usepackage[utf8]{inputenc}
\usepackage{textcomp}
\usepackage{float}
\usepackage{graphicx}
\usepackage{tabularx}
\usepackage{subfigure}
\usepackage{grffile}
\usepackage{caption}
\usepackage[dvipsnames]{xcolor}
\usepackage{ulem}
\usepackage{soul}
\usepackage{enumerate}
\usepackage{hyperref}
\usepackage{url}
\usepackage{cite}
\usepackage{comment}
\usepackage{todonotes}
\usepackage{multirow}
\usepackage{booktabs}
\usepackage[export]{adjustbox}

\newcommand{\rev}[1]{{{#1}}}

\setulcolor{blue}
\setstcolor{red}

\begin{document}

\title{\rev{Positive} streamer discharge simulations in humid air: uncertainty in input data and sensitivity analysis}

\author{Baohong Guo$^{1,2}$, Hemaditya Malla$^{2}$, Alejandro Malag\'{o}n-Romero$^{2}$, Jannis Teunissen$^{2,*}$}

\address{$^1$ School of Electrical Engineering and Automation, Fuzhou University, Fuzhou, People’s Republic of China}
\address{$^2$ Centrum Wiskunde \& Informatica (CWI), Amsterdam, The Netherlands}

\ead{jannis.teunissen@cwi.nl}

\vspace{10pt}

\begin{indented}
\item[]
\today
\end{indented}

\begin{abstract}
  We study how the choice of input data affects simulations of positive streamers in humid air, focusing on H$_2$O cross sections, photoionization models, and chemistry sets.
  Simulations are performed in air with a mole fraction of 0\%, 3\% or 10\% H$_2$O using an axisymmetric fluid model.
  Five H$_2$O cross section sets are considered, which lead to significant differences in the resulting electron attachment coefficient.
  As a result, the streamer velocity can vary by up to about 50\% with 10\% H$_2$O.
  We compare results with three photoionization models: the Naidis model for humid air, the Aints model for humid air, and the standard Zheleznyak model for dry air.
  With the Naidis and in particular the Aints model, there is a significant reduction in photoionization with higher humidities.
  This results in higher streamer velocities and maximal electric fields, and it can also cause streamer branching in our axisymmetric simulations.
  Three humid air chemistry sets are considered.
  Differences between these sets, particularly in the formation of water clusters around positive ions, cause the streamer velocity to vary by up to about 50\% with 10\% H$_2$O.
  A sensitivity analysis is performed to identify the most important chemical reactions in these chemistries.
\end{abstract}

%
%

\ioptwocol

\section{Introduction}
\label{sec:intro}

Streamers are fast-propagating and filamentary ionization fronts featuring an enhanced electric field at their tips~\cite{nijdam2020a}. 
This field enhancement enables streamers to propagate in sub-breakdown background fields and therefore to serve as precursors to more energetic phenomena such as sparks and lightning leaders~\cite{gallimberti1979, malagon-romero2019}.
Streamers are also fundamental blocks of some transient luminous events (TLEs) that occur in the upper atmosphere, including sprites and blue jets~\cite{gordillo-vazquez2021/atmre}. 
Although streamers are low-temperature plasmas, the strong electric field at the streamer head can accelerate electrons to really high energies and thus, can trigger high-temperature chemical reactions at nearly room temperature~\cite{wang2020a}. 
This property has been harnessed for numerous applications such as plasma medicine~\cite{vonwoedtke2020}, agriculture~\cite{attri2020plasma}, industrial surface treatments~\cite{polonskyi2021}, air cleaning~\cite{kim2004}, plasma-assisted combustion~\cite{starikovskaia2014}, and non-thermal plasma actuators~\cite{zhang2019c}.

The gas composition affects streamer properties such as velocity, radius, breakdown field, and branching characteristics, as well as the plasma species produced by streamers~\cite{nijdam2014a, bouwman2022, li2022a, guo2023f}.
Frequently, water vapor is a component of gas mixtures. 
When studying streamer-like discharges in atmospheric air for plasma medicine and agriculture applications, in supersaturated air inside a thundercloud, or in high-voltage power transmission systems under complex environmental conditions, humidity is an important factor that cannot be neglected.
Below, we briefly discuss some relevant experimental and computational studies that account for air humidity.

In the late 1970s, the pioneering experimental works by Griffiths and Phelps~\cite{phelps1976, griffiths1976a}, Les Renardiers group~\cite{lesrenardiers1977/elektra, lesrenardiers1978/elektra} and Gallimberti~\cite{gallimberti1979}, showed that water vapor hampered streamer propagation in long air gaps.
Therefore, the stability field (i.e., the minimal average electric field required for bridging an air gap) increased with increasing water vapor content~\cite{allen1991, hui2008, mikropoulos2008}.
For example, in~\cite{allen1991} such a stability field was measured at 5.24\,kV/cm at an absolute humidity of 11\,g/m$^3$, and it increased by about 1.3\% for every 1\,g/m$^3$ increase in the humidity.
Furthermore, the field required for a streamer to cause breakdown also increased with increasing air humidity at a rate of about 0.56\% per g/m$^3$~\cite{mikropoulos2008}.
Zhao \textit{et al} systematically investigated the effect of humidity on the evolution of streamer dynamics and discharge instabilities under repetitive pulses~\cite{zhao2023, zhao2024}, finding increased branching in both primary and secondary streamers in humid air.
In addition, air humidity leads to the formation of OH radicals in streamer discharges.
It was observed that an increase in humidity resulted in increased OH density but a decrease in O$_3$ production due to lower atomic oxygen levels~\cite{ono2003a, nakagawa2011, singleton2016}.
Increased humidity was also found to enhance fast gas heating for streamer discharges in humid air~\cite{ono2010a, komuro2014a, komuro2015a}.

There are also numerous computational studies regarding how humidity affects streamer propagation in atmospheric air, using 1.5D (assuming a constant radius of the streamer channel)~\cite{aleksandrov1998, hui2008}, as well as 2D~\cite{komuro2018a, malagon-romero2022, starikovskiy2022, ren2022} and 3D `discharge tree'~\cite{chen2023} models.
These simulations showed that the influence of humidity became noticeable in long streamer discharges.
As air humidity increased, the streamer propagated more slowly with a smaller radius, and the background field required for sustaining streamer propagation was also higher, in agreement with previous experimental findings~\cite{phelps1976, griffiths1976a, lesrenardiers1977/elektra, lesrenardiers1978/elektra, gallimberti1979, allen1991, hui2008, mikropoulos2008}.
This humidity effect on streamer properties was primarily attributed to reduced conductivity in the streamer channel, due to dissociative recombination of electrons with positive cluster ions $\rm H_3O^+(H_2O)_n$ and enhanced three-body electron attachment to O$_2$ molecules~\cite{malagon-romero2022, starikovskiy2022}.
The above effects can be weakened by increasing the gas temperature.
According to~\cite{komuro2018a}, higher gas temperatures reduced the rates of electron recombination and three-body attachment reactions, which in turn increased the conductivity of the streamer channel.
Furthermore, air humidity had a significant impact on the electron swarm parameters needed for fluid streamer simulations, thereby influencing streamer properties, as discussed in~\cite{chen2018a}.

One of the main challenges in modeling streamers (or other discharges) in humid air is selecting appropriate input data.
For example, different electron-neutral cross sections for H$_2$O were used in~\cite{komuro2013, komuro2018a, malagon-romero2022, starikovskiy2022, liu2017}, as well as different cross sections for N$_2$ and O$_2$.
In~\cite{aleksandrov1999}, a simplified reaction set was used for streamer simulations in humid air.
To obtain more complete and reliable kinetic data, a recommended database was constructed for humid air plasma chemistry in~\cite{sieck2000, herron2001}.
Based on these early humid air chemistries, various sets of chemical reactions were further developed in subsequent simulation studies~\cite{komuro2013, komuro2018a, malagon-romero2022, starikovskiy2022, liu2017} under different discharge conditions.
Of particular importance are the cross sections for three-body electron attachment to H$_2$O, as well as the reactions related to the production of cluster ions and the corresponding electron-ion recombination rates~\cite{aleksandrov2022}.
Furthermore, most of the above-mentioned simulations did not account for the effect of H$_2$O on photoionization, despite its significance.

In this paper, we investigate how the choice of input data affects streamer properties such as the streamer velocity, optical diameter, and maximal electric field.
We consider different H$_2$O cross section sets, different photoionization models and different sets of chemical reactions to study their effects on single positive streamers in humid air.

\section{Simulation model}
\label{sec:sim-model}

We perform axisymmetric simulations of positive streamers with the \texttt{Afivo-streamer} code~\cite{teunissen2017}, using a drift-diffusion fluid model with the local field approximation.
The model is briefly described below, for further details see e.g.~\cite{teunissen2017, bagheri2018}.

The simulations are performed in gas mixtures containing N$_2$ and O$_2$ at a 4:1 ratio together with H$_2$O.
We consider three humidities, with a mole fraction of 0\%, 3\% and 10\% H$_2$O, as listed in table~\ref{tab:gas-comp}.
Note that we slightly vary the gas temperature and pressure to obtain humid air with 10\% H$_2$O but keep the gas number density $N_0 = 2.41 \times 10^{25} \, \textrm{m}^{-3}$ the same.

\begin{table}
\centering
\caption{Gas compositions considered in this paper, with percentages relative to the gas number density $N_0 = 2.41 \times 10^{25} \, \textrm{m}^{-3}$.
Case 1 corresponds to dry air.
Cases 2 and 3 correspond to 86\% and 96\% relative humidity, respectively.
}
\label{tab:gas-comp}
\begin{tabular*}{0.48\textwidth}{@{\extracolsep{\fill}}cccccc}
  \br
  Case & H$_2$O & N$_2$ & O$_2$ & P (bar) & T (K) \\
  \mr
  1 & 0.0\% & 80.0\% & 20.0\% & 1.0 & 300 \\
  2 & 3.0\% & 77.6\% & 19.4\% & 1.0 & 300 \\
  3 & 10.0\% & 72.0\% & 18.0\% & 1.067 & 320 \\
  \br
\end{tabular*}
\end{table}


\subsection{Equations}
\label{sec:equations}

In our model, the electron density $n_\mathrm{e}$ evolves in time as
\begin{equation}
\label{eq:dd-electrons}
    \partial_t n_\mathrm{e} = -\nabla \cdot (\mathbf{\Gamma}_\mathrm{drift} + \mathbf{\Gamma}_\mathrm{diff}) + S_\mathrm{e} + S_\mathrm{ph}\,,
\end{equation}
where $\mathbf{\Gamma}_\mathrm{drift} = -\mu_\mathrm{e} \mathbf{E} n_\mathrm{e}$ and $\mathbf{\Gamma}_\mathrm{diff} = - D_\mathrm{e} \nabla n_\mathrm{e}$ are the drift and diffusive fluxes of electrons, $\mu_\mathrm{e}$ the electron mobility, $D_\mathrm{e}$ the diffusion coefficient, and $\mathbf{E}$ the electric field.
Furthermore, $S_\mathrm{e}$ is an electron source term due to reactions (e.g., ionization or attachment) described in section~\ref{sec:chemistry-sets}, and $S_\mathrm{ph}$ is the photo-ionization source term described in section~\ref{sec:photionization}.
Electron transport coefficients are assumed to depend on the local electric field, and they were computed with BOLSIG$-$~\cite{hagelaar2005} using the cross sections described in section~\ref{sec:electr-neutr-cross}.

We consider time scales up to about $200 \, \textrm{ns}$ and found that including ion mobilities made no significant difference in our results.
The motion of ions is therefore neglected, so that ions and neutral densities $n_j$ (numbered by $j=1,\ldots,n$) evolve in time as
\begin{equation}
\label{eq:dd-ions}
    \partial_t n_j = S_j\,,
\end{equation}
where $S_j$ is a source/sink term due to reactions.
\rev{Gas heating is not considered in the model because we consider only a single voltage pulse.
Most heating would occur after the pulse, and furthermore the energy deposited by the simulated positive streamers would increase the gas temperature by at most a few Kelvin, see e.g.~\cite{malla2023}.}

The electric field $\mathbf{E}$ is calculated in the electrostatic approximation as $\mathbf{E} = -\nabla \phi$. The electrostatic potential $\phi$ is obtained by solving the Poisson equation
\begin{equation}
\label{eq:poisson-equation}
    \nabla^2 \phi = -{\rho}/{\epsilon_0}\,,
\end{equation}
using a geometric multigrid method~\cite{teunissen2018, teunissen2023}, where $\rho$ is the space charge density and $\epsilon_0$ is the vacuum permittivity.

\subsection{Computational domain and initial condition}    
\label{sec:domain-init-condition}

\begin{figure}
    \centering
    \includegraphics[width=1\linewidth]{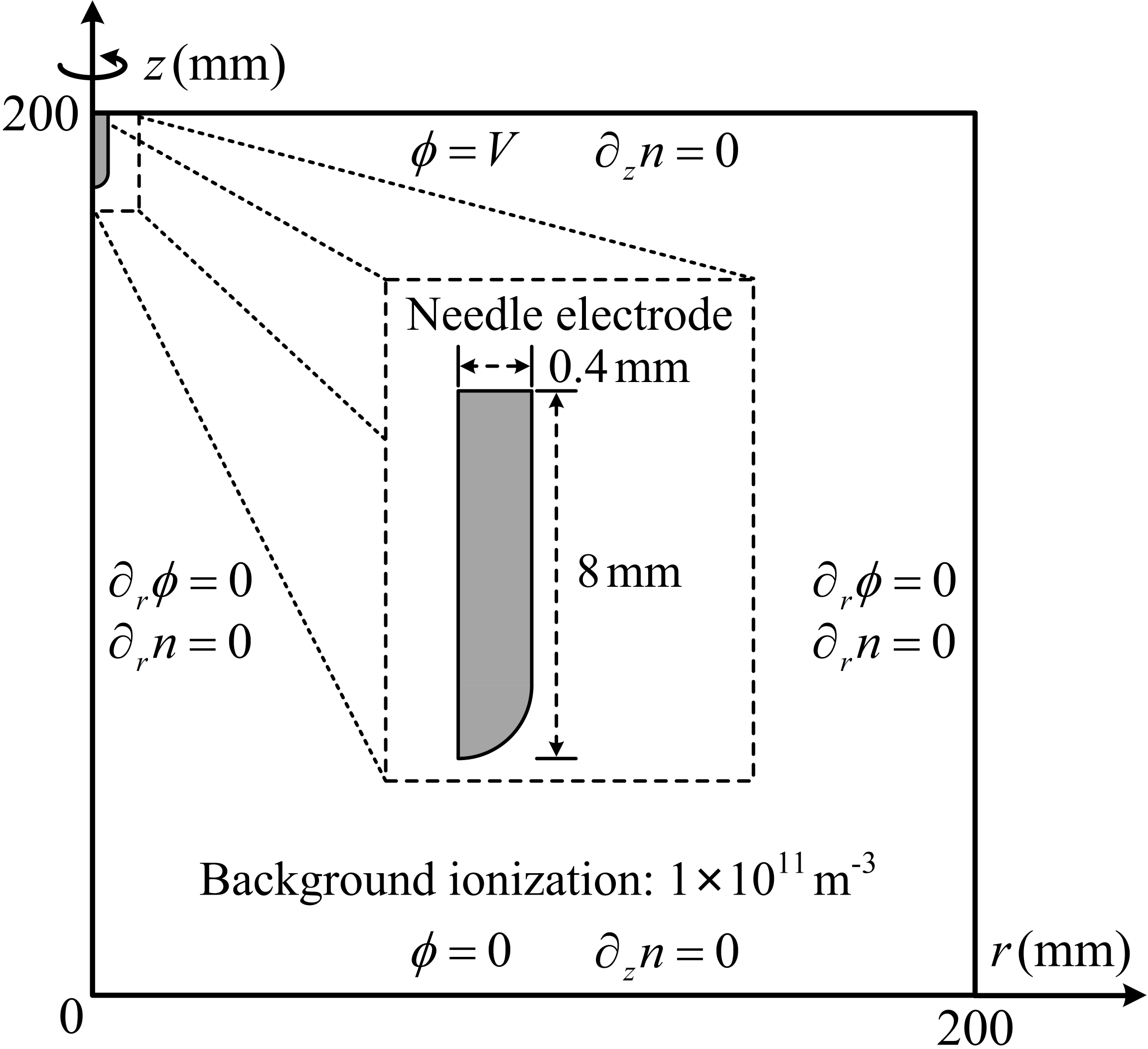}
    \caption{Schematic view of the 2D axisymmetric computational domain, containing a plate-plate geometry with a needle electrode.
    Boundary conditions for the electrostatic potential $\phi$ and the species densities $n_\mathrm{e}$ or $n_j$ are indicated at the domain boundaries.}
    \label{fig:simulation-domain}
\end{figure}

We use a 2D axisymmetric computational domain measuring $200 \, \textrm{mm}$ in the $r$ and $z$ directions, which contains a plate-plate geometry with a needle electrode (8\,mm in length and 0.8\,mm in diameter), as illustrated in figure~\ref{fig:simulation-domain}.
The bottom plate is grounded, whereas a constant high voltage $V$ is applied on the upper plate and the needle electrode.
The axial electric field away from the needle electrode relaxes to the average electric field between the two plate electrodes, which is here defined as the background electric field $E_\mathrm{bg}$:
\begin{equation}
\label{eq:bg-field}
    E_\mathrm{bg} = V / d\,,
\end{equation}
where $d = 200$\,mm is the distance between two plate electrodes.

For the electrostatic potential, a homogeneous Neumann boundary condition is used on the outer radial boundary.
For all species densities, homogeneous Neumann boundary conditions are used on all domain boundaries, including the needle electrode.
Secondary electron emission from electrodes due to ions and photons is not included.

As an initial condition, homogeneous background ionization with a density of $10^{11}\,\mathrm m^{-3}$ for both electrons and N$_2^+$ ions is included.
All other ion densities are initially zero.
The background ionization provides the first electrons for streamer inception near the electrode tip, where there is significant electric field enhancement.
\rev{The simulations are not sensitive to this background ionization density, because photoionization quickly becomes the dominant source of non-local free electrons after inception, see e.g.~\cite{guo2023, li2021a, nijdam2011}.}

For computational efficiency the simulations are performed using adaptive mesh refinement (AMR).
We use the same refinement criteria as in previous works (see e.g.\cite{teunissen2017}), namely to refine when $\alpha(E) \Delta x > 1.0$ and to de-refine when $\alpha(E) \Delta x < 0.125$, where $\alpha(E)$ is the field-dependent ionization coefficient and $\Delta x$ is the grid spacing.
The minimum grid spacing in the simulations was $\Delta x = 1.5 \, \mu\mathrm{m}$.
Around the needle electrode, we enforce $\Delta x < 50 \, \mu\textrm{m}$.

\subsection{Electron-neutral cross sections for N$_2$, O$_2$ and H$_2$O}
\label{sec:electr-neutr-cross}

Electron-neutral collisions are the dominant process in streamer discharges, and energy-dependent probabilities of such collisions are described by cross sections.
In a fluid model these cross sections are used indirectly through a Boltzmann solver to obtain electron transport coefficients (such as the mobility) and rates for electron-neutral reactions.
BOLSIG$-$~\cite{hagelaar2005} with the temporal growth model was used to obtain such data.

For most atoms and molecules, there is still quite some uncertainty in the electron-neutral cross sections.
For example, in~\cite{li2021a} streamer simulations were compared using different cross section sets for dry air (80\% N$_2$, 20\% O$_2$), and it was found that the streamer velocity could vary by about 10\% when compared at the same position.
Although the gas mixtures considered here consist mostly of N$_2$ and O$_2$, we here want to focus on the uncertainty in cross sections for H$_2$O.
All simulations are therefore performed using Phelps' cross sections for N$_2$ and O$_2$~\cite{chemistry_phelps_database, phelps1985, lawton1978}, whereas for H$_2$O we consider the cross section sets described below.

\rev{We used all available H$_2$O cross section sets from \url{lxcat.net} except for the Hayashi database.
It should be noted that many of these cross section sets (including the Hayashi one) lack rotational cross sections, as explained below.}
We probably could have included other data as well (from e.g.,~\cite{DeUrquijo_2014, Yousfi_1996, Budde_2022}).
However, our goal here is simply to illustrate how much variation there will be in streamer simulations using different H$_2$O cross section sets, \rev{rather than conducting a full sensitivity analysis of all H$_2$O cross sections}.
For a more extensive discussion of H$_2$O cross sections and their accuracy, we refer to~\cite{Ness_2012, DeUrquijo_2014, kawaguchi2016, Budde_2022}.

\rev{The cross section sets described below differ in the dissociative electron attachment reactions that they include.
For example, in the Itikawa data there are separate processes for the formation of O$^-$, OH$^-$ and H$^-$, whereas in the Phelps data these processes are grouped into a single process.
However, for the discharges considered in this paper, the three-body attachment reaction
\begin{equation}
    \label{eq:three-body-O2-H2o}
    \rm e + O_2 + H_2O \to O_2^- + H_2O
\end{equation}
is more important.
As this process involves both O$_2$ and H$_2$O it is not included in the H$_2$O cross section sets.
Instead, its rate coefficient is assumed to be a multiple (six or seven in this paper, depending on the chemistry) of the three-body attachment reaction
\begin{equation}
    \label{eq:three-body-O2}
    \rm e + O_2 + O_2 \to O_2^- + O_2.
\end{equation}
}


\subsubsection{Kawaguchi H$_2$O cross sections}
\label{sec:kawaguchi}

The cross section set of Kawaguchi \textit{et al}~\cite{kawaguchi2016} was compiled based on several previous sources (too many to list here), and it contains about 40 cross sections for rotational, vibrational and electronic excitations, ionization and electron attachment.
With Monte Carlo swarm simulations, the authors demonstrate that their set reproduces experimental swarm measurements better than those of~\cite{Yousfi_1996, Itikawa_2005, DeUrquijo_2014}.
What furthermore makes this set attractive to use is that they also provide a set with an isotropic scattering model.
The tabulated cross sections were obtained by contacting the authors directly, but the authors plan to also make them available at \url{lxcat.net}.

\subsubsection{Morgan and Phelps H$_2$O cross sections}
\label{sec:morgan-phelps}

Through \url{lxcat.net}~\cite{pitchford2017}, we obtained the Morgan~\cite{chemistry_morgan_database} and Phelps~\cite{chemistry_phelps_database_H2O} cross section sets for H$_2$O.
For both of these, there is the following warning: \textit{``This cross section set should not be used in Boltzmann calculations because it does not include explicitly rotational cross sections (very important in H$_2$O).
Morgan used an analytical form for these cross sections.''}
Although values for the ``continuous approximation to rotation'' (CAR) are given by Morgan, we were not able to reconstruct a reasonable looking rotational cross section using the formulas in~\cite{Morgan_1990}, in particular due to uncertainty in the units used.
For comparison, we nevertheless include these sets without rotational cross sections.

We remark that in the recently published paper by Budde \textit{et al}~\cite{Budde_2022} the lack of rotational cross sections for H$_2$O is extensively addressed.
The authors provide a complete set containing 147 rotational cross sections, out of a total of 163 cross sections.
The set is tested against swarm measurements using two two-term Boltzmann solvers, and the authors obtain good agreement.
At the time of writing, this set was not yet available on~\url{lxcat.net}, so we did not include it in our comparison.

\subsubsection{Triniti H$_2$O cross sections}
\label{sec:triniti}

We also obtained the Triniti set~\cite{chemistry_triniti_database} through \url{lxcat.net}, which was obtained from the EEDF Boltzmann solver~\cite{eedf_software}.
The cross section file includes a reference to Yousfi \textit{et al} 1987, which we could not find, but this suggests that these cross sections could be related to the ones published later in~\cite{Yousfi_1996}.

\subsubsection{Itikawa H$_2$O cross sections}
\label{sec:itikawa}

The review paper by Itikawa \textit{et al}~\cite{Itikawa_2005, chemistry_itikawa_database} provides an overview of available cross sections for H$_2$O together with recommendations, which are available at \url{lxcat.net}.
An update of these cross sections was recently given in~\cite{song2021}.
We want to briefly mention two caveats when using this data.
First, due to the strong dipole moment of H$_2$O, elastic scattering is highly anisotropic and sharply peaked in the forward direction.
Although a total momentum transfer cross section is provided, the cross sections cannot directly be used in a Boltzmann solver that assumes isotropic scattering, since the set also includes rotational excitations that have to be described anisotropically.
A second caveat is that some cross sections are only given up to a certain energy (which can be as low as $5 \, \textrm{eV}$) at which they are still nonzero.
If such cross sections are used in a Boltzmann solver, it depends on the particular implementation how they will be extrapolated above this energy range.

Although the data of~\cite{Itikawa_2005} should only be used with an appropriate anisotropic scattering model, we include it here in an improper way, just for comparison with the Morgan and Phelps data.
For this comparison, we have removed all rotational cross sections from the Itikawa set.
\rev{However, even when these rotational cross sections are included together with anisotropic scattering, the resulting transport data shows large differences compared to experimental swarm data~\cite{kawaguchi2016}.}
We note that several publications regarding streamer simulations in humid air have previously used the Morgan, Phelps or Itikawa cross sections for H$_2$O from \url{lxcat.net} directly~\cite{komuro2013, komuro2018a, liu2017, malagon-romero2022, starikovskiy2022}, without addressing the issues outlined above.

\subsubsection{Comparison of electron transport coefficients}
\label{sec:transport-coefficients}

\begin{figure*}
    \centering
    \includegraphics[width=1\linewidth]{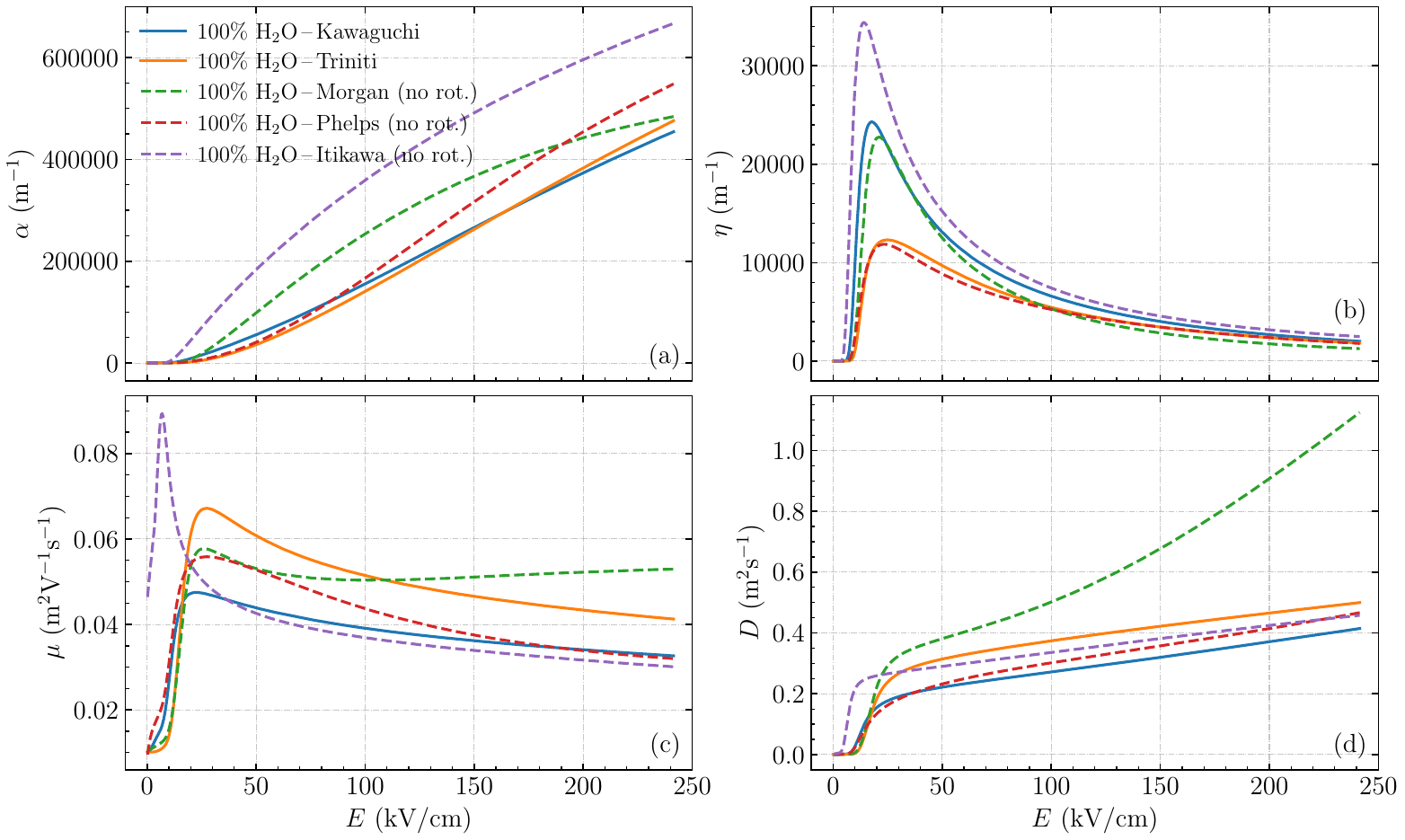}
    \caption{Comparison of electron transport coefficients for pure H$_2$O gas at $1 \, \textrm{bar}$ and $300 \, \textrm{K}$ using different sets of H$_2$O cross sections.
    (a) Ionization coefficient $\alpha$, (b) attachment coefficient $\eta$, (c) mobility $\mu$, and (d) diffusion coefficient $D$.
    Results for H$_2$O cross section sets without rotational cross sections, i.e., the Morgan, Phelps and Itikawa data, are indicated with dashed lines.}
    \label{fig:transport-data-pure-H2O}
\end{figure*}

We have used the cross section sets described above to compute electron transport coefficients using BOLSIG-~\cite{hagelaar2005}.
Figure~\ref{fig:transport-data-pure-H2O} shows results for a hypothetical scenario with pure H$_2$O gas at $1 \, \textrm{bar}$ and $300 \, \textrm{K}$.
There is considerable variation between the five sets for all transport parameters, which is not that surprising since the Morgan, Phelps and Itikawa data are used in an inappropriate way, as discussed above.
The Morgan, Phelps and Itikawa data show higher ionization coefficients compared to the other two sets.
The Morgan data leads to a substantially higher diffusion coefficient and a rather flat mobility.
The Itikawa data yields a higher attachment coefficient and a higher peak mobility at a lower electric field.
Furthermore, the Kawaguchi and Triniti sets agree quite well in terms of the ionization coefficient.
For the attachment coefficient, we find good agreement between Kawaguchi and Morgan and between Triniti and Phelps.

\begin{figure*}
    \centering
    \includegraphics[width=1\linewidth]{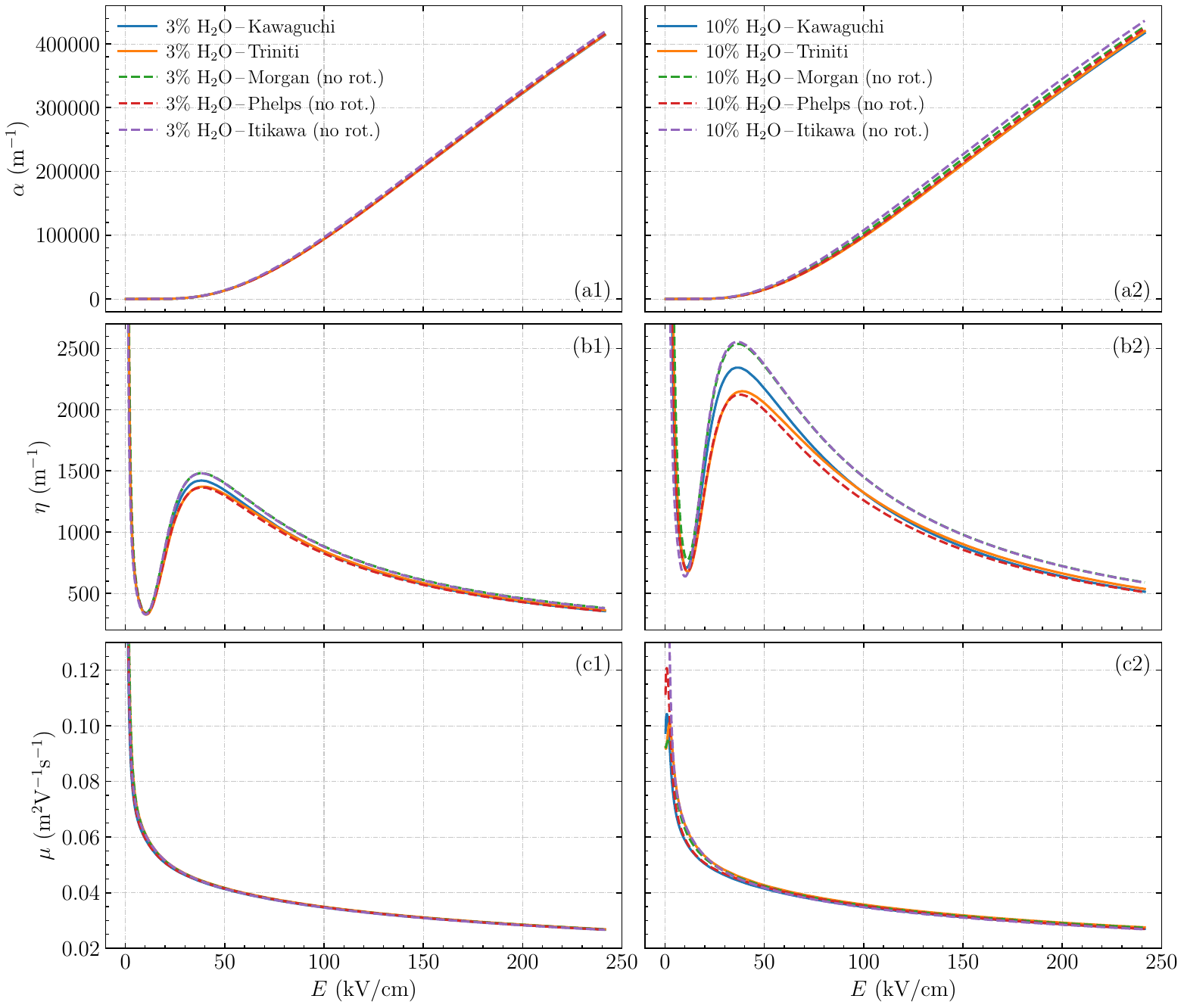}
    \caption{Comparison of electron transport coefficients for air with 3\% (the left row) and 10\% H$_2$O (the right row) at $1 \, \textrm{bar}$ and $300 \, \textrm{K}$ using different sets of H$_2$O cross sections.
    (a) Ionization coefficient $\alpha$, (b) attachment coefficient $\eta$, and (c) mobility $\mu$.
    Results for H$_2$O cross section sets without rotational cross sections, i.e., the Morgan, Phelps and Itikawa data, are indicated with dashed lines.}
    \label{fig:transport-data-0.03-0.10H2O}
\end{figure*}

Figure~\ref{fig:transport-data-0.03-0.10H2O} shows results for air with 3\% and 10\% H$_2$O at $1 \, \textrm{bar}$ and $300 \, \textrm{K}$.
As expected, for both air humidities the differences in the transport coefficients between the five sets are now much smaller, due to the smaller contribution of the H$_2$O cross sections.
The attachment coefficient now has a peak at low electric fields due to three-body attachment reactions given by \rev{equations~(\ref{eq:three-body-O2-H2o}) and (\ref{eq:three-body-O2})},
where for this comparison reaction~(\ref{eq:three-body-O2-H2o}) was assumed to have a rate coefficient six times that of reaction~(\ref{eq:three-body-O2}).
For air with 10\% H$_2$O, the maximum difference in the resulting attachment coefficients is about 20\%, whereas the variation in the ionization coefficient is below 5\%.
For the case of 3\% H$_2$O, the differences in the transport coefficients are even smaller and difficult to see, with the maximum variation in the attachment coefficient being about 8\%.
Note that regardless of the specific H$_2$O cross section set used, a decrease in the mole fraction of H$_2$O from 10\% to 3\% results in a lower attachment coefficient, but hardly affects other transport parameters.

\subsection{Photoionization}
\label{sec:photionization}

Photoionization in humid air resembles that in dry air: excited N$_2$ molecules can emit photons with enough energy to non-locally ionize O$_2$ molecules, see e.g.~\cite{Stephens_2016}.
We consider two approaches for modeling photoionization in humid air, which are described below.
For comparison, we also briefly describe the dry air case.

\subsubsection{Dry air - Zheleznyak's model}
\label{sec:photoi-dry-zheleznyak}

In dry air, the production $I$ (per unit time and volume) and absorption $f(r)$ (per unit distance) of these photons are here approximated using Zheleznyak's model~\cite{zheleznyak1982}.
In this model, $I$ is proportional to the electron impact ionization source term $S_\mathrm{i}$:
\begin{equation}
\label{eq:ph-zhelez-src}
    I = \frac{p_q}{p+p_q} \xi S_\mathrm{i}\,,
\end{equation}
where $p_q= 40\,\mathrm{mbar}$ is the quenching pressure, $p$ the gas pressure, \rev{$p_q/(p+p_q)$ is the quenching factor for which we will use the symbol $f_q$} and $\xi$ is a proportionality factor, which is here set to $\xi = 0.075$ as in e.g.~\cite{bagheri2019, wang2023}.
Furthermore, the absorption function is given by
\begin{equation}
\label{eq:ph-zhelez-absfunc}
    f_\mathrm{air}(r) = \frac{\exp(-k_1 \, p_{\mathrm{O}_2} r) - \exp(-k_2 \, p_{\mathrm{O}_2} r)}{r \ln(k_2 / k_1)}\,,
\end{equation}
where $k_1 = 0.035 \, \mathrm{cm}^{-1} \, \mathrm{Torr}^{-1}$,
$k_2 = 2 \, \mathrm{cm}^{-1} \, \mathrm{Torr}^{-1}$ and $p_{\mathrm{O}_2}$ is
the partial pressure of O$_2$.
We remark that what actually matters is the number density of O$_2$ molecules, but we here follow the usual assumption of the gas being at room temperature.

\subsubsection{Humid air - Naidis model}
\label{sec:photoi-humid-naidis}

Naidis~\cite{naidis2006} proposed a simple modification of Zheleznyak's model, in which water molecules lead to extra absorption:
\begin{equation}
\label{eq:ph-zhelez-absfunc}
    f_\mathrm{Naidis}(r) = \exp(-k_3 \, p_{\mathrm{H}_2\mathrm{O}} \, r) \times f_\mathrm{air}(r)\,,
\end{equation}
where $k_3 = 0.26 \, \mathrm{cm}^{-1} \, \mathrm{Torr}^{-1}$ and $p_{\mathrm{H}_2\mathrm{O}}$ is the partial pressure of H$_2$O.
Note that with this extra term, the integral (from 0 to $\infty$) over the absorption function is less than unity, because the photons absorbed by H$_2$O do not contribute to photoionization.

\subsubsection{Humid air - Aints model}
\label{sec:photoi-humid-aints}

Aints \textit{et al}~\cite{Aints_2008} performed measurements on photoionization in humid air that were not consistent with equation~(\ref{eq:ph-zhelez-absfunc}) for small values of $k_3 \, p_{\mathrm{H}_2\mathrm{O}} \, r$.
They therefore proposed a different absorption function which we here write as
\begin{equation}
\label{eq:ph-aints-absfunc}
    f_\mathrm{Aints}(r) = \frac{\exp(- K_1 r) - \exp(-K_2 r)}{r \ln(K_2 / K_1)}\,,
\end{equation}
where $K_1 = k_1 \, p_{\mathrm{O}_2} + k_4 \, p_{\mathrm{H}_2\mathrm{O}}$ and $K_2 = k_2 \, p_{\mathrm{O}_2} + k_5 \, p_{\mathrm{H}_2\mathrm{O}}$, with $k_4 = 0.13 \, \mathrm{cm}^{-1} \, \mathrm{Torr}^{-1}$ and $k_5 = 0.57 \, \mathrm{cm}^{-1} \, \mathrm{Torr}^{-1}$.
Note that this absorption function has an integral of unity.
The authors instead introduced additional quenching by H$_2$O in equation~(\ref{eq:ph-zhelez-src}): 
\begin{equation}
\label{eq:ph-aints-src}
    I_\mathrm{Aints} = \left(1 + \frac{p-p_{\mathrm{H}_2\mathrm{O}}}{p_q} + \frac{p_{\mathrm{H}_2\mathrm{O}}}{p_{q,\mathrm{H}_2\mathrm{O}}}\right)^{-1} \xi S_{\mathrm{i}}\,,
\end{equation}
which was fitted against their data to obtain $p_{q,\mathrm{H}_2\mathrm{O}} = 0.3 \, \mathrm{Torr}$.
\rev{Here the quenching factor $f_{q,\mathrm{Aints}} = \left(1 + \frac{p-p_{\mathrm{H}_2\mathrm{O}}}{p_q} + \frac{p_{\mathrm{H}_2\mathrm{O}}}{p_{q,\mathrm{H}_2\mathrm{O}}}\right)^{-1}$.}
We use the above expressions, but remark that the value of $p_{q,\mathrm{H}_2\mathrm{O}}$ appears unrealistic, since gas molecules do not collide (or interact) frequently enough to warrant such a low quenching pressure~\cite{Li_2024}.

\subsubsection{\rev{Comparison}}
\label{sec:photoi-comparison}

\begin{figure}
    \centering
    \includegraphics[width=1\linewidth]{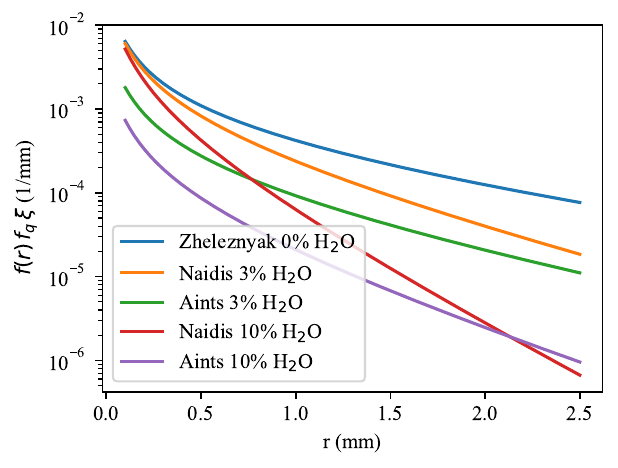}
    \caption{\rev{Comparison of photoionization absorption functions for dry and humid air. 
    Each absorption function has been multiplied by its respective quenching factor $f_q$ and by the proportionality factor $\xi = 0.075$, see equations (\ref{eq:ph-zhelez-src}) and (\ref{eq:ph-aints-src}). 
    The cases shown correspond to the conditions listed in table~\ref{tab:gas-comp}.}}
    \label{fig:photoi-absorption-functions}
\end{figure}

\rev{In figure~\ref{fig:photoi-absorption-functions} we show the absorption functions described above, multiplied by the respective quenching factors $f_q$ and by the proportionality factor $\xi = 0.075$.
These curves indicate how likely photoionization is at some distance $r$ from an ionization event.
For $r \to 0$, the Naidis model agrees with Zheleznyak's model for dry air, whereas there is considerably less photoionization with the Aints model.
For larger distances, the Naidis absorption function decays more quickly than that of Aints, but since the majority of photons are produced at small $r$, there is overall less photoionization with the Aints model.}

\subsubsection{Numerical implementation}
\label{sec:photoi-helmholtz}

We implement photoionization in our fluid model with the Helmholtz approach, as described in~\cite{bourdon2007, luque2007}.
We used an expansion with three Helmholtz terms of the form
\begin{equation}
\label{eq:ph-helmholtz}
    (\nabla^2 - \lambda_i^2) S_\mathrm{ph} = -A_i I\,,
\end{equation}
where $I$ corresponds to equation (\ref{eq:ph-zhelez-src}) or (\ref{eq:ph-aints-src}), and the coefficients $\lambda_i$ and $A_i$ are given in table~\ref{tab:ph-helmholtz-coeff}.
See e.g.\ Appendix A of~\cite{bagheri2018} for more details about this type of expansion.

\begin{table*}
\centering
\caption{Coefficients for the three-term Helmholtz approximations to photoionization, obtained by fitting the respective absorption functions.
All coefficients have been multiplied with the respective partial pressures, so that they are given in units of mm$^{-1}$ and mm$^{-2}$}.
\label{tab:ph-helmholtz-coeff}
\begin{tabular*}{0.85\textwidth}{c@{\extracolsep{\fill}}cllllll}
  \br
  Case & Model & $\lambda_1$ & $\lambda_2$ & $\lambda_3$ (mm$^{-1}$) & $A_1$ & $A_2$ & $A_3$ (mm$^{-2}$) \\
  \mr
  0\% H$_2$O & Zheleznyak & 0.830 & 2.19 & 13.4 & 0.0447 & 1.15 & 110 \\
  \mr
  3\% H$_2$O & Naidis & 1.51 & 3.88 & 32.5 & 0.0863 & 3.89 & 757 \\
  3\% H$_2$O & Aints & 1.21 & 3.59 & 32.6 & 0.0963 & 4.37 & 875 \\
  \mr
  10\% H$_2$O & Naidis & 2.83 & 5.19 & 33.0 & 0.0858 & 3.83 & 697 \\
  10\% H$_2$O & Aints & 1.86 & 4.24 & 33.3 & 0.114 & 5.18 & 1045 \\
  \br
\end{tabular*}
\end{table*}


\subsection{Chemical reactions}    
\label{sec:chemistry-sets}

In addition to electron-neutral cross sections, chemical reactions are another important input data for streamer fluid simulations.
For many reactions, there are significant uncertainties in both the reaction form and the reaction rate coefficient, see e.g.~\cite{turner2015, turner2016}.
Furthermore, previous authors have used different sets of chemical reactions for humid air, and it is often unclear which set is the most suitable and which reactions are most significant.

In this work, we use the three chemistry sets that are described below.
We have made small modifications to these sets so that they are compatible with the electron-neutral cross sections that we use.
We have also included reactions to obtain the emitted light from the second positive system ($\rm N_2(C) \to N_2(B) + {\it h\nu}$), which is responsible for most of the optical emission in air at atmospheric pressure~\cite{pancheshnyi2000}.
Furthermore, we have made corrections to reaction rates when necessary.
All these changes are documented in~\ref{sec:three-chemistry-sets}.

We remark that the choice of chemistry sets was arbitrary, with the sets obtained from recent papers for streamer simulations in dry/humid air.
There are other datasets, e.g., from~\cite{liu2017, sieck2000, herron2001}.
Our purpose here is simply to demonstrate how sensitive streamer simulations are to the use of different datasets.

\subsubsection{Humid air - Malag\'{o}n chemistry set}
\label{sec:humid-chemistry-malagon}

This chemistry set for streamer simulations in humid air is based on the reactions from Malag\'{o}n-Romero \textit{et al}~\cite{malagon-romero2022}, and it is listed in table~\ref{tab:chemistry-set-malagon}.
This set contains 97 chemical reactions, which can be divided into eight different processes: electron impact ionization, electron attachment, electron detachment, negative ion conversion, positive ion conversion, electron-ion recombination, ion-ion recombination, and light emission.
For most reactions of ion-ion recombination, reaction products were not given.

\subsubsection{Humid air - Starikovskiy chemistry set}
\label{sec:humid-chemistry-starikovskiy}

This chemistry set for humid air is based on Starikovskiy \textit{et al}~\cite{starikovskiy2022}, as given in table~\ref{tab:chemistry-set-starikovskiy}.
The set contains only 72 chemical reactions and six processes. 
Compared to the Malag\'{o}n chemistry set described above, it does not include electron detachment and negative ion conversion processes.
Additionally, it includes a single negative ion O$_2^-$ as compared to Malag\'{o}n's which includes seven negative ions.
Finally, the rate coefficient for three-body attachment with H$_2$O (reaction R10) is seven times that of three-body attachment with O$_2$ (reaction R9) rather than six times in the Malag\'{o}n chemistry set. 
Note that for dissociative recombination of electrons with $\rm H_3O^+(H_2O)_n$ ions, the range $n = 1-6$ is used in this set. 
However, in the paper they cite for this reaction~\cite{johnsen1993} $n$ only ranged from 1 to 4.  

\subsubsection{Humid air - Komuro chemistry set}
\label{sec:humid-chemistry-komuro}

This chemistry set for humid air is based on Komuro \textit{et al}~\cite{komuro2013, komuro2018a}, as listed in table~\ref{tab:chemistry-set-komuro}.
Compared to the previous two sets, this set contains many more chemical reactions (191  reactions in total) and eleven types of processes, resulting in some differences.
First, it contains several excited species, summarized in table~\ref{tab:excited-states-komuro}.
Second, it contains electron dissociation reactions (R13--R16), which effectively produce neutral atoms that are needed for neutral species conversion.
Third, it contains neutral species conversion as the reactions were collected for studying the behavior of OH radicals in streamer discharges in~\cite{komuro2013}. 
Furthermore, as stated in~\cite{komuro2018a}, this chemistry set was constructed to investigate the effect of gas temperature rather than humidity.
Therefore, some (important) reactions related to water cluster ions, such as the recombination reactions of electrons with $\rm H_3O^+(H_2O)_n$ ions, $n = 0-2$ were not taken into account.

\subsubsection{Dry air - Guo chemistry set}
\label{sec:dry-chemistry-guo}

This chemistry set from Guo and Teunissen~\cite{guo2023} was constructed for streamer simulations in dry air.
It contains 263 chemical reactions and twelve processes, which were primarily based on the reactions from~\cite{kossyi1992, ono2020}.
This set is not listed in~\ref{sec:three-chemistry-sets} as it is completely the same as the one in~\cite{guo2023}.
For further comparison, we obtained three other chemistry sets for dry air by removing the reactions related to H$_2$O from the above three humid air chemistry sets.

\subsection{Sensitivity tests}
\label{sec:sensitivity-tests}

We use so-called sensitivity tests~\cite{Campolongo_2007} to better understand which chemical reactions are important.
These tests are performed by multiplying individual reaction rate coefficients with a factor $c_i$ for $i = 1, \dots, N_c$, using the following values: $(0.0, 0.25, 0.5, 0.75)$.
With $N_r$ reactions, the total number of additional simulations is thus $N_cN_r$.
We then pick a ``quantity of interest'', here called $Q$, that we compare at some particular time to the case with unmodified rate coeffients $Q_0$.
For reaction $n$ there are then results $Q_{n, i}$ for $i = 1, \dots, N_c$, from which we compute normalized derivatives as
\begin{equation}
\label{eq:sens-test-formula}
    Q'_{n, i} = \frac{1}{Q_0} \frac{Q_{n, i} - Q_0}{\Delta c}\,,
\end{equation}
where $\Delta c = c_i - 1$.
Finally, we determine the mean of the normalized derivatives
\begin{equation}
\label{eq:sensitvity-mu}
    Q'_{n} = \sum_{i = 1}^{N_c} Q'_{n, i}/N_c\,,
\end{equation}
and the sample standard deviation of the $Q'_{n, i}$
\begin{equation}
\label{eq:sensitvity-sigma}
    \sigma = \sqrt{\sum_{i = 1}^{N_c} (Q'_{n, i} - Q'_{n})^2 / (N_c - 1)}\,.
\end{equation}

Note that there are more sophisticated ways to do sensitivity tests in which multiple coefficients are varied at the same time, see e.g.~\cite{turner2016, Berthelot_2017, Terraz_2020}.
However, our purpose here is simply to understand which reactions are important, and not to quantify the uncertainty in results due to uncertainties in the rate coefficients.

\section{Simulation results}
\label{sec:sim-results}

Below we will study how the H$_2$O cross sections, the photoionization model and the chemistry set affect simulations of positive streamers in humid air.
We will compare the following streamer properties:
\begin{itemize}
  \item The maximal electric field $E_\mathrm{max}$ at the streamer head.
  The location of this maximum is defined as the streamer head position.
  \item The velocity $v$, obtained as the time derivative of the streamer head position.
  \item The optical diameter $d$, defined as the full width at half maximum (FWHM) of the time-integrated light emission after a forward Abel transform, see e.g.~\cite{li2021a}.
  This simulated optical diameter can be used for comparison with experimental data.
\end{itemize}

\subsection{Effect of H$_2$O cross sections}
\label{sec:effect-cross-sections}

As described in section~\ref{sec:electr-neutr-cross}, five cross section sets for H$_2$O were considered, namely Kawaguchi, Triniti, Morgan, Phelps, and Itikawa.
We remind the reader that rotational cross sections were not included in the Morgan and Phelps data, and that we left them out for the Itikawa data.
In this section, we investigate the effect of these H$_2$O cross sections on positive streamers in humid air with 3\% and 10\% H$_2$O.
The simulations were performed in a background field of 10\,kV/cm, using the `dry air' Zheleznyak photoionization model and the chemistry set of Malag\'{o}n.

\begin{figure*}
    \centering
    \includegraphics[width=1\linewidth]{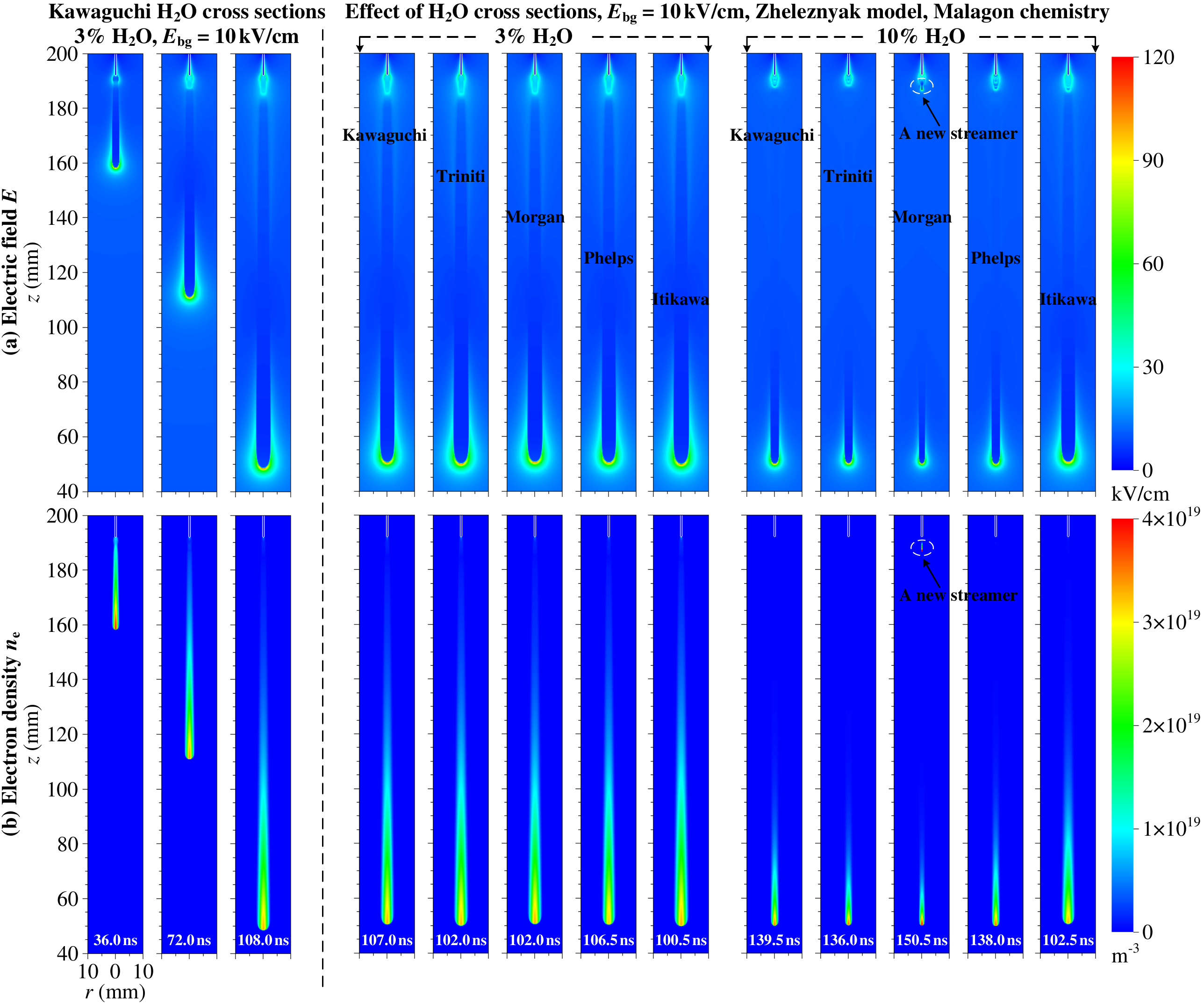}
    \caption{Left: time evolution of (a) the electric field $E$ and (b) electron density $n_\mathrm{e}$ profiles for a positive streamer in a background field of $E_\mathrm{bg} = 10$\,kV/cm in air with 3\% H$_2$O using Kawaguchi H$_2$O cross sections.
    Right: effect of H$_2$O cross sections on positive streamers in $E_\mathrm{bg} = 10$\,kV/cm in air with different humidities (3\% and 10\%), and shown are (a) the electric field $E$ and (b) electron density $n_\mathrm{e}$ profiles for streamers at $z = 50$\,mm.
    All panels are zoomed in into the region where 0 $\leqslant r \leqslant$ 10\,mm and 40 $\leqslant z \leqslant$ 200\,mm.}
    \label{fig:cross-sections-profiles}
\end{figure*}

We first illustrate a typical time evolution in these simulations on the left of figure~\ref{fig:cross-sections-profiles}, which shows the electric field and electron density profiles for the streamer in air with 3\% H$_2$O using the Kawaguchi H$_2$O cross sections.
The streamer starts from the needle electrode and accelerates downwards, with the velocity and the optical diameter increasing approximately linearly with time, from about $0.8\times10^6$\,m/s to $2.3\times10^6$\,m/s and from about 1.2\,mm to 3.7\,mm, respectively.
However, the maximal electric field at the streamer head remains approximately constant at about 110\,kV/cm.
\rev{The ratio between streamer velocity and optical diameter evolves in a similar way as the maximal electric field.}
    
\begin{figure*}
    \centering
    \includegraphics[width=1\linewidth]{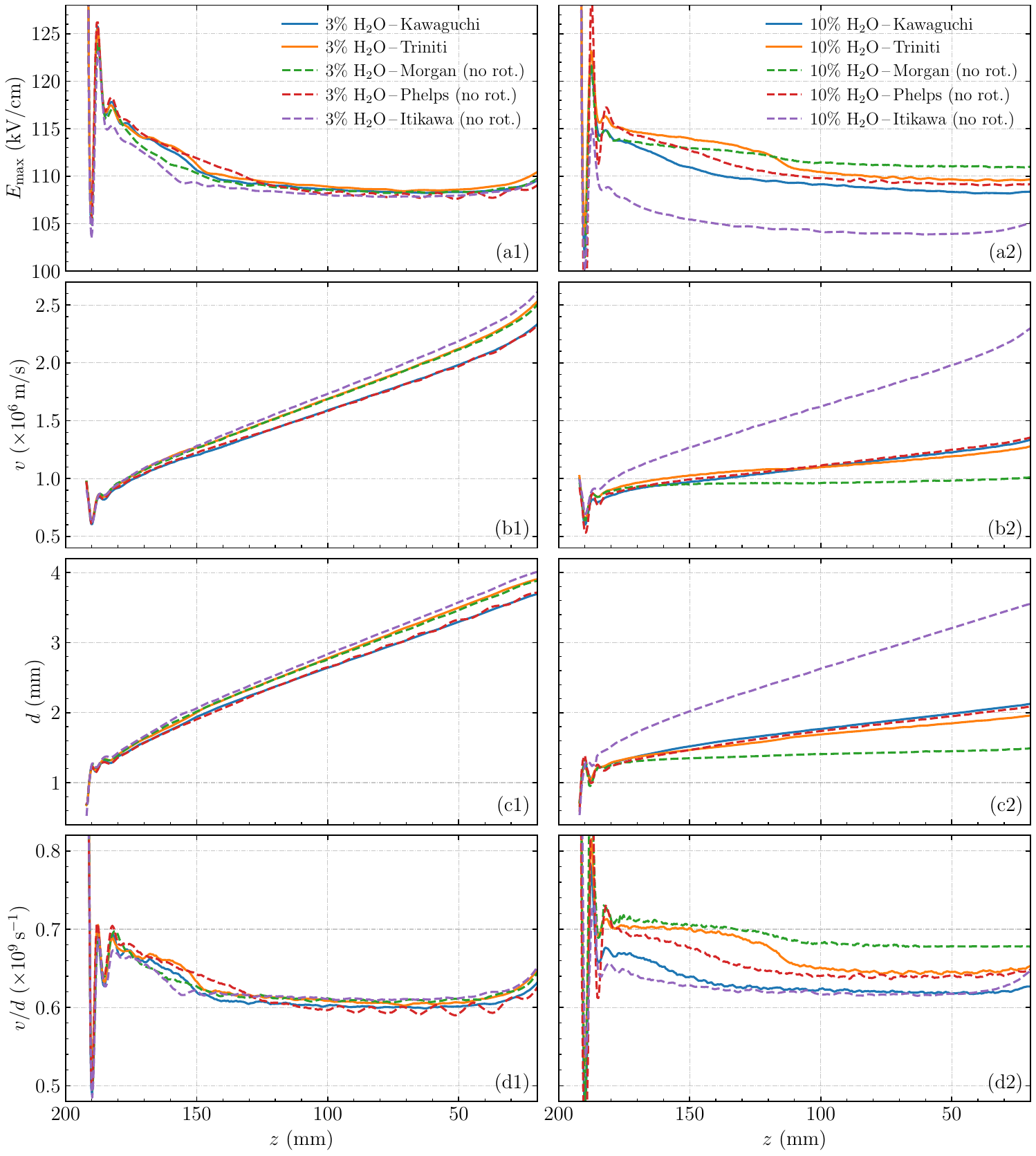}
    \caption{Effect of H$_2$O cross sections on positive streamers in $E_\mathrm{bg} = 10$\,kV/cm in air with 3\% (the left row) and 10\% H$_2$O (the right row).
    Shown are streamer properties of (a) the maximal electric field $E_\mathrm{max}$, (b) streamer velocity $v$, (c) optical diameter $d$, \rev{and (d) ratio between velocity and diameter $v/d$} versus the streamer head position $z$.}
    \label{fig:cross-sections-Emax-v-d-vd}
\end{figure*}

Simulations with different H$_2$O cross sections are compared on the right of figure~\ref{fig:cross-sections-profiles}, 
which shows the electric field and electron density profiles for air with 3\% and 10\% H$_2$O, at the moment the streamers reach $z = 50$\,mm.
In addition, the maximal electric field, streamer velocity, optical diameter, \rev{and ratio between velocity and diameter} are shown versus the streamer head position in figure~\ref{fig:cross-sections-Emax-v-d-vd}.


In air with 3\% H$_2$O, all five different cross sections do not lead to obvious differences in streamer properties, with variations below 10\% when compared at the same streamer length.
This is not that surprising, as differences in the electron transport coefficients are also small with 3\% H$_2$O, as mentioned in section~\ref{sec:transport-coefficients}.
Differences become larger with 10\% H$_2$O in air, and then three groups can be distinguished:
\textcircled{1} the Kawaguchi, Triniti and Phelps cases, which show good agreement; \textcircled{2} the Morgan case; and \textcircled{3} the Itikawa case.
Compared to the cases with 3\% H$_2$O, streamers in group \textcircled{1} become about 40\% slower and thinner, due to a much faster decay of conductivity in the streamer channel.
The Morgan case shows even slower propagation at about $1.0\times10^6$\,m/s and a smaller diameter of about 1.4\,mm, which is attributed to a higher attachment coefficient, as demonstrated in figure~\ref{fig:transport-data-0.03-0.10H2O}.
An interesting phenomenon occurs in the Morgan case where a new positive streamer is generated from the needle electrode behind the previous one.
This phenomenon, recently observed in strongly attaching gases in~\cite{guo2023f} as well, will be further illustrated in~\ref{sec:ionization-reactions}.  
In contrast, the Itikawa case propagates much faster with a larger diameter, similar to those with 3\% H$_2$O, but with a slightly lower $E_\mathrm{max}$. 
Although the attachment coefficient in the Itikawa data is nearly identical to that in the Morgan data, the ionization coefficient is about 5\% higher.
This suggests that ionization reactions play a more significant role than attachment, as discussed in~\ref{sec:ionization-reactions}.
    
In the rest of the paper, we will use the Kawaguchi H$_2$O cross sections, as they can reproduce experimental swarm measurements and because they can be used with an isotropic scattering model, see section~\ref{sec:kawaguchi}.

\subsection{Effect of the photoionization model}
\label{sec:effect-photoionization}

\begin{figure}
    \centering
    \includegraphics[width=1\linewidth]{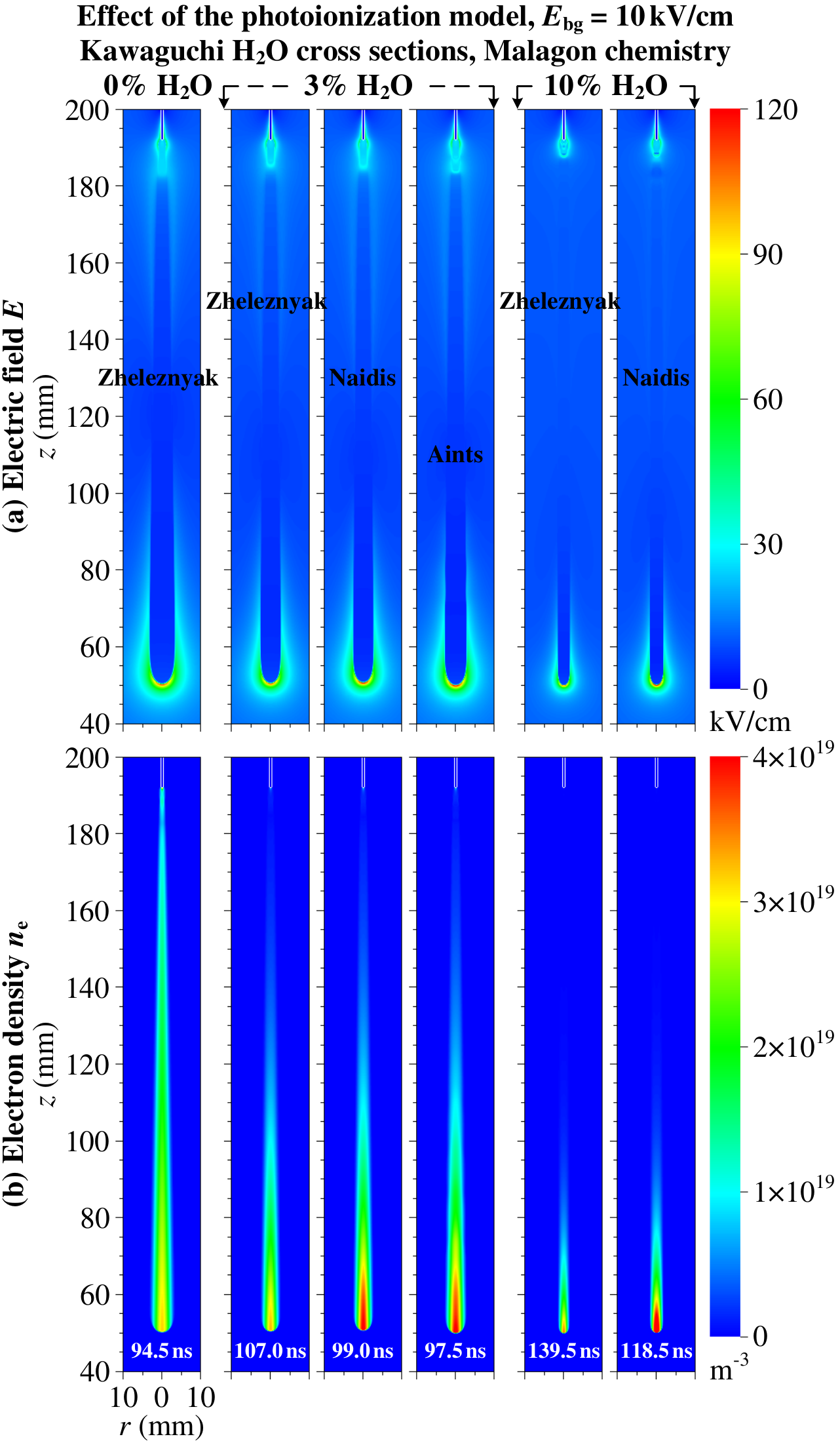}
    \caption{Effect of the photoionization model on positive streamers in $E_\mathrm{bg}=10$\,kV/cm in air with different humidities (0\%, 3\% and 10\%) using Kawaguchi H$_2$O cross sections.
    Shown are (a) the electric field $E$ and (b) electron density $n_\mathrm{e}$ profiles for streamers at $z = 50$\,mm.}
    \label{fig:photoionization-profiles}
\end{figure}

\begin{figure}
    \centering
    \includegraphics[width=1\linewidth]{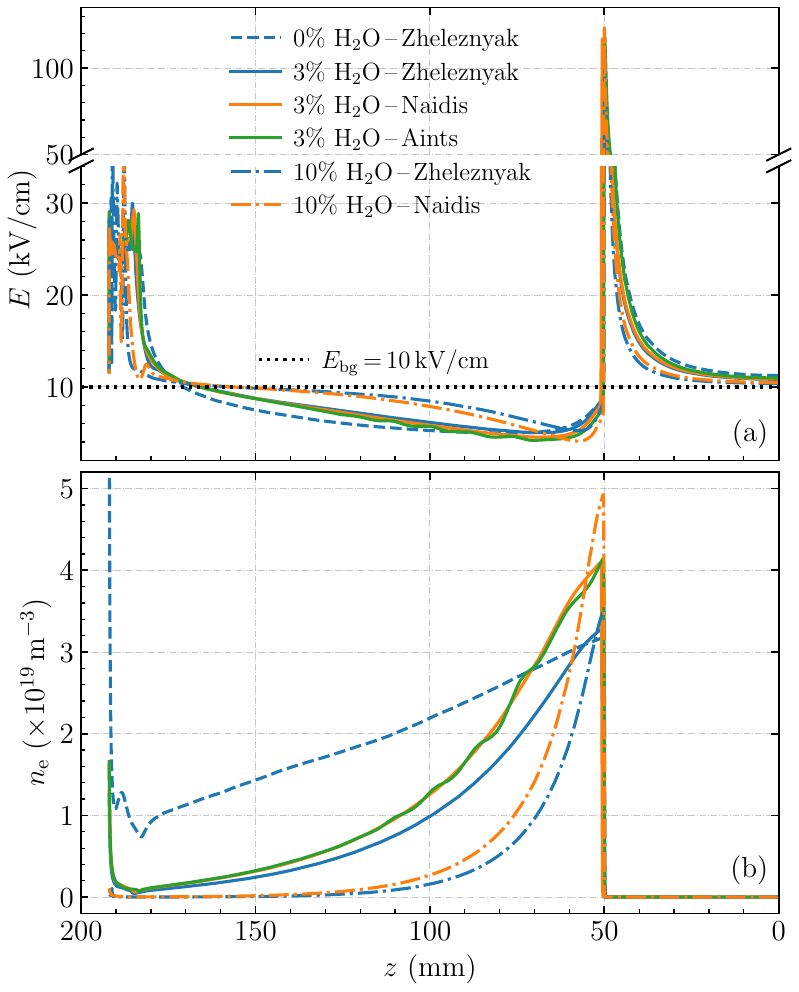}
    \caption{Effect of the photoionization model.
    Shown are (a) the electric field $E$ and (b) electron density $n_\mathrm{e}$ along $z$ axis for streamers at $z = 50$\,mm corresponding to figure~\ref{fig:photoionization-profiles}.}
    \label{fig:photoionization-Ech-ne}
\end{figure}

\begin{figure}
    \centering
    \includegraphics[width=1\linewidth]{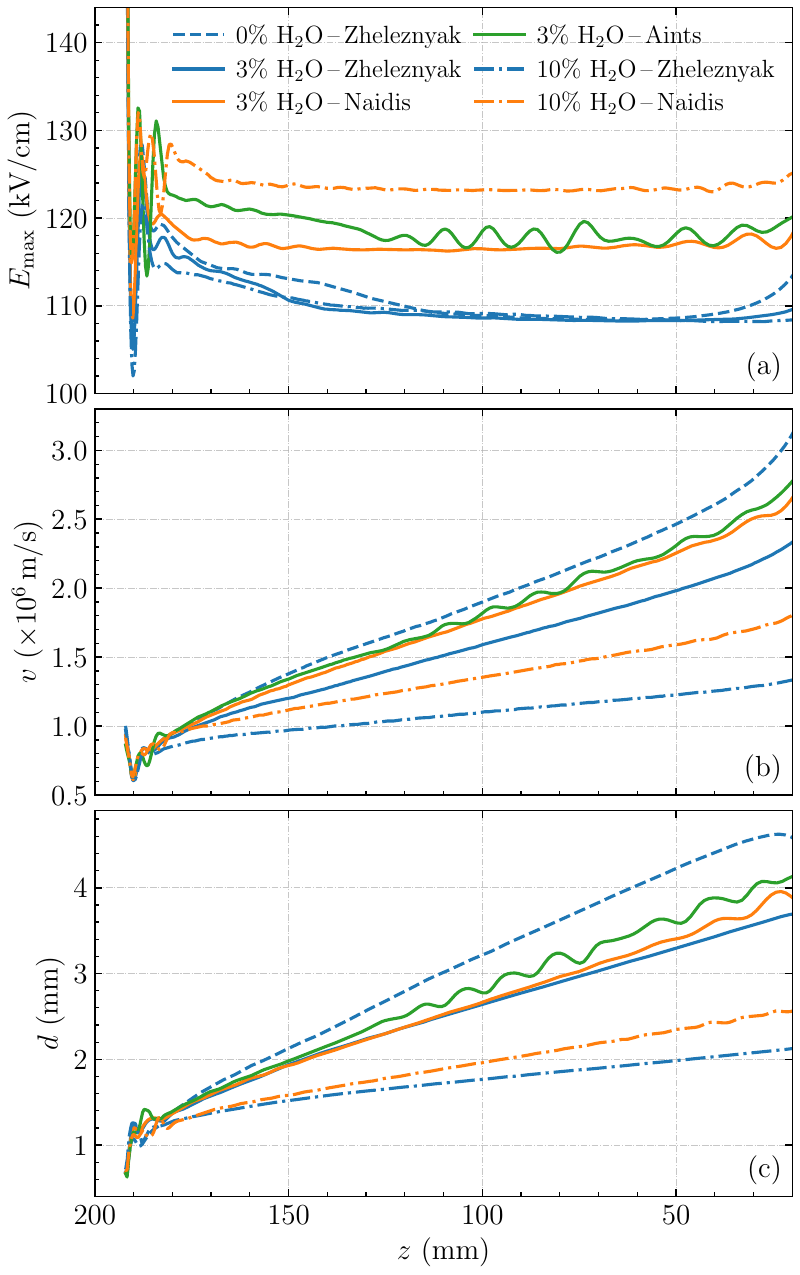}
    \includegraphics[width=1\linewidth]{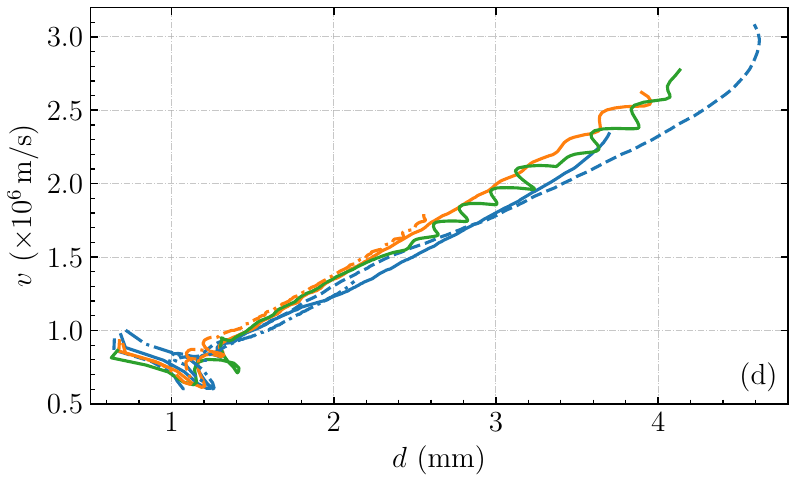}
    \caption{Effect of the photoionization model.
    Shown are streamer properties of (a) the maximal electric field $E_\mathrm{max}$, (b) streamer velocity $v$, and (c) optical diameter $d$ versus the streamer head position $z$. 
    \rev{Panel (d) shows the streamer velocity $v$ versus optical diameter $d$.}}
    \label{fig:photoionization-Emax-v-d}
  \end{figure}

We now compare different photoionization models to understand how important changes in photoionization due to humidity are.
Three approaches are considered: the Naidis and Aints models for humid air discussed in section~\ref{sec:photionization}, and for comparison also the standard Zheleznyak model for dry air described in section~\ref{sec:photoi-dry-zheleznyak}.
We performed simulations with these models using the Malag\'{o}n chemistry and the Kawaguchi H$_2$O cross sections, in a background field of 10\,kV/cm.
Figure~\ref{fig:photoionization-profiles} shows the electric field and electron density profiles when the streamers have reached $z = 50 \, \textrm{mm}$, and corresponding axial profiles are shown in figure~\ref{fig:photoionization-Ech-ne}.
Furthermore, figure~\ref{fig:photoionization-Emax-v-d} shows streamer properties as a function of the streamer head position $z$\rev{, as well as the relation between the streamer velocity $v$ and optical diameter $d$}.

There is a clear effect of photoionization on $E_\mathrm{max}$ and other streamer properties.
If the standard Zheleznyak model is used, $E_\mathrm{max}$ is almost the same for air with 0\%, 3\% and 10\% H$_2$O.
However, other streamer properties such as velocity and diameter vary significantly, with a higher humidity leading to a reduced velocity and diameter.
With the Naidis and Aints models there is less photoionization, which leads to an increase in $E_\mathrm{max}$ of up to about 15\%.
Due to the higher field at the streamer tip, the electron density in the streamer channel is also higher.
Furthermore, the streamer velocity is substantially higher, for example about 30\% higher with the Naidis model and 10\% H$_2$O than with the Zheleznyak model.
The effect on the streamer diameter is smaller, with differences of up to about 15\% between the Naidis and Zheleznyak cases for 10\% H$_2$O.
\rev{Although there is significant variation in diameter and velocity, the relationship between these parameters is not very sensitive to the photoionization model (or the H$_2$O concentration), as shown in figure~\ref{fig:photoionization-Emax-v-d}(d).}

The reduction in photoionization with the Naidis and Aints models can cause the axisymmetric streamers to become unstable and eventually branch, which happens with the Aints model and 10\% H$_2$O.
Since we cannot simulate branching with the axisymmetric fluid model used here, this case was not included in the figures.
Performing realistic simulations of the branching dynamics of streamers in humid air would require fully 3D simulations, as well as a stochastic treatment of photoionization~\cite{wang2023, Marskar_2020, bagheri2019}, which we leave for future work.

That a reduction in photoionization can lead to faster streamers was also found in previous work~\cite{li2021a}.
Perhaps this is most easily understood from the scenario in which there is a very large amount of photoionization.
In such cases, the streamer will have less field enhancement, and thus become less conductive, which means that it will propagate more slowly.

\subsection{Effect of the chemistry set}
\label{sec:effect-chemistry}

\begin{figure*}
    \centering
    \includegraphics[width=1\linewidth]{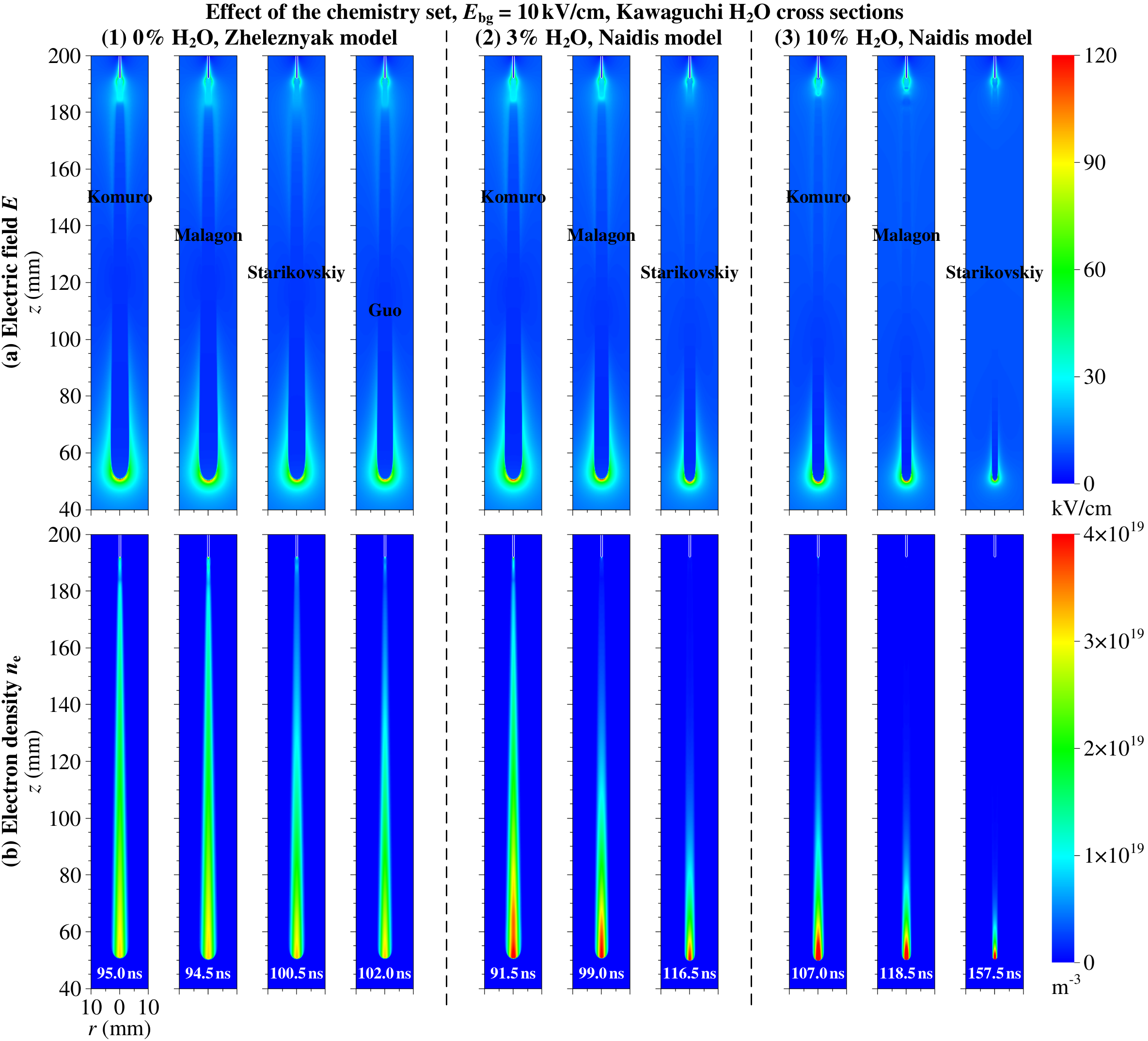}
    \caption{Effect of the chemistry set on positive streamers in $E_\mathrm{bg}=10$\,kV/cm in air with different humidities (0\%, 3\% and 10\%) using Kawaguchi H$_2$O cross sections.
    Shown are (a) the electric field $E$ and (b) electron density $n_\mathrm{e}$ profiles for streamers at $z = 50$\,mm.}
    \label{fig:chemistry-profiles}
\end{figure*}

\begin{figure}
    \centering
    \includegraphics[width=1\linewidth]{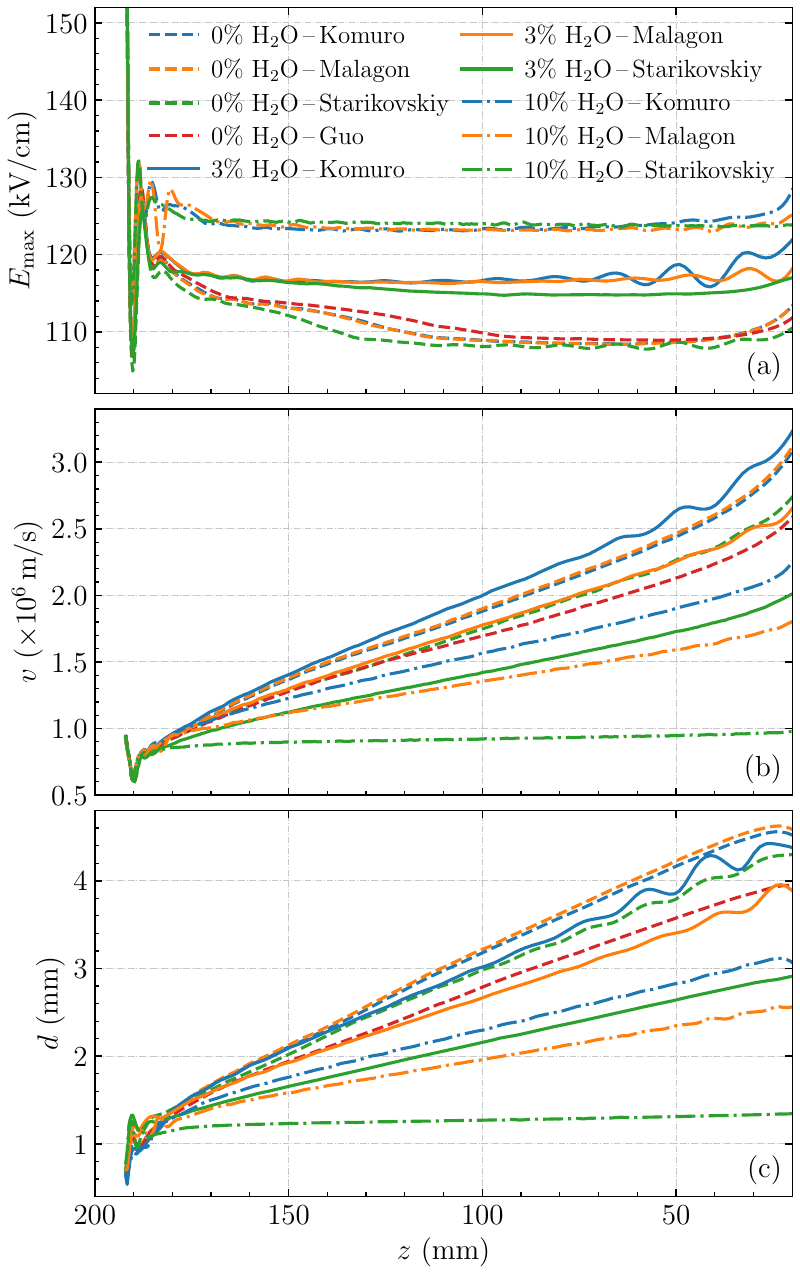}
    \caption{Effect of the chemistry set. 
    Shown are streamer properties of (a) the maximal electric field $E_\mathrm{max}$, (b) streamer velocity $v$, and (c) optical diameter $d$ versus the streamer head position $z$.}
    \label{fig:chemistry-Emax-v-d}
\end{figure}

In this section, we study the effect of chemical reactions on positive streamers in humid air.
We used the chemistry sets described in section~\ref{sec:chemistry-sets} with varying humidity levels to perform a total of ten streamer simulations, for which the conditions are summarized in table~\ref{tab:chemistry-set-details}.
Figure~\ref{fig:chemistry-profiles} shows the electric field and electron density profiles when the streamers have reached $z=50$\,mm, and figure~\ref{fig:chemistry-Emax-v-d} shows streamer properties as a function of the streamer head position $z$.

\begin{table}
\centering
\caption{Simulation parameters for comparison of chemistry sets. 
  All simulations were performed in a background field of $E_\mathrm{bg}=10$\,kV/cm using the Kawaguchi H$_2$O cross sections. The cases with 0\% H$_2$O are performed with the Zhelezhnyak photoionization model.
}
\label{tab:chemistry-set-details}
\begin{tabular*}{0.48\textwidth}{l@{\extracolsep{\fill}}ll}
  \br
  Chemistry & Photoionization model & H$_2$O (\%) \\
  \mr
  Guo & Zhelezhnyak & 0 \\
  Komuro & Zhelezhnyak, Naidis & 0, 3, 10 \\
  Malag\'{o}n & Zhelezhnyak, Naidis & 0, 3, 10 \\
  Starikovskiy & Zhelezhnyak, Naidis & 0, 3, 10 \\
  \br
\end{tabular*}
\end{table}

The chemistry set also has an important effect on streamer properties. 
We first look at the maximal electric field at the streamer head.
For each humidity considered here, there are almost no differences in $E_\mathrm{max}$ with different chemistry sets.
However, an increase in air humidity leads to a higher $E_\mathrm{max}$.

We then focus on the streamer velocity and diameter.
As a first test, we compare results in dry air (0\% H$_2$O) with the four considered chemistry sets, since these chemistries also differ in the included reactions for oxygen and nitrogen species.
Streamers with the Komuro and Malag\'{o}n chemistries agree very well in this case, whereas the Starikovskiy and Guo chemistries lead to slower streamers.
At $z=50$\,mm, the Starikovskiy case is about 8\% slower, whereas the Guo case is about 12\% slower.
In streamer diameter these differences are about 10\% and 14\%, respectively.

Next, we compare results in humid air with 3\% and 10\% H$_2$O using the Komuro, Malag\'{o}n and Starikovskiy chemistries.
The differences in streamer velocity then become larger, with the Komuro case leading to the fastest streamers and the Starikovskiy case leading to the slowest ones.
With 3\% H$_2$O, the Malag\'{o}n case is about 15\% slower than the Komuro case at $z=50$\,mm, and the Starikovskiy case is about 35\% slower.
With 10\% H$_2$O, these differences are about 15\% and 50\%, respectively.
The differences in streamer diameter are approximately the same as those in the velocity, with the slower streamer having a smaller diameter.
Note that the Starikovskiy case with 10\% H$_2$O is approximately a steady positive streamer, whose properties do not vary in time, as discussed in~\cite{li2022a}.

Regardless of the particular chemistry, an increase in the percentage of H$_2$O leads to slower streamers with a smaller diameter, as was also found in previous work~\cite{malagon-romero2022, starikovskiy2022}.
In the results presented here, streamers have propagated for about a hundred ns.
On such short timescales, the main effect of a chemistry on streamer propagation is through reactions that affect the electron density inside the streamer channel, such as electron attachment, detachment and electron-ion recombination reactions.
Below, in section~\ref{sec:sensitivity-analysis}, we will present a sensitivity analysis to identify which reactions in each of the chemistry sets are most important in this respect.

\subsection{Sensitivity analysis of chemical reactions}
\label{sec:sensitivity-analysis}

Using the method described in section~\ref{sec:sensitivity-tests}, we have studied the sensitivity of streamer simulations to individual chemical reactions.
The total number of electrons $N_\mathrm{e}$ at $t = 100 \, \textrm{ns}$ was here used as the ``quantity of interest''.
We exclude electron impact ionization reactions from this analysis; their effect is discussed in~\ref{sec:ionization-reactions}.
All simulations were performed in air containing 3\% H$_2$O, using a background field of 10\,kV/cm, the Aints photoionization model and the Kawaguchi cross sections for H$_2$O.

\begin{figure}
    \centering
    \includegraphics[width=1\linewidth]{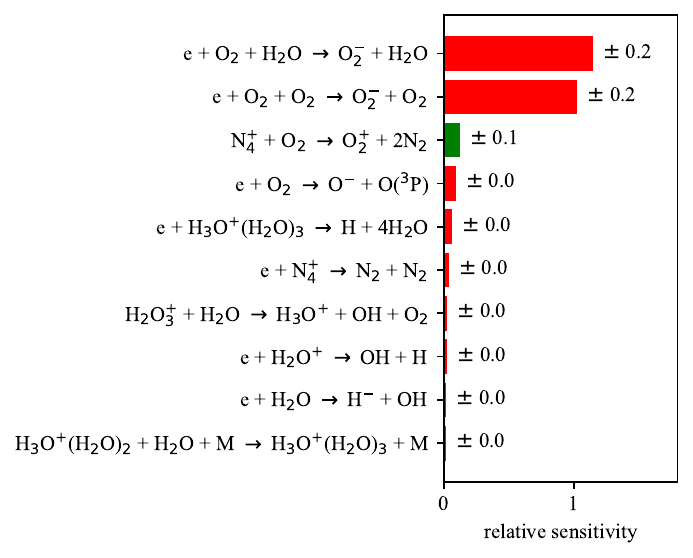}
    \caption{Sensitivity analysis of the Komuro chemistry. 
    The effect of individual reactions on the total number of electrons $N_\mathrm{e}$ at $t = 100 \, \textrm{ns}$ was determined using the method described in section~\ref{sec:sensitivity-tests}. 
    The bars indicate the magnitude of the derivative $Q'_{n}$, see equation~(\ref{eq:sensitvity-mu}), and their color indicates the sign of the derivative $Q'_{n}$ (red: negative, green: positive).
    The numbers next to the bars indicate the standard deviations $\sigma$, see equation~(\ref{eq:sensitvity-sigma}).
    }
    \label{fig:chemistry-barplot-komuro}
\end{figure}

\begin{figure}
    \centering
    \includegraphics[width=1\linewidth]{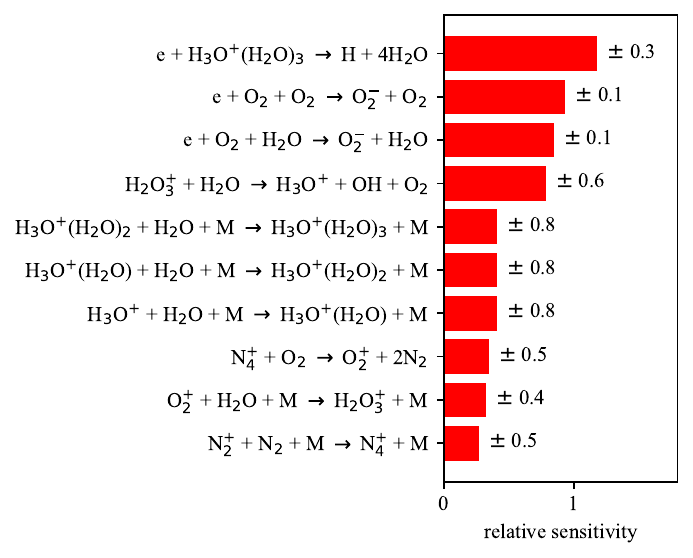}
    \caption{Sensitivity analysis of the Malag\'{o}n chemistry, analogous to figure~\ref{fig:chemistry-barplot-komuro}.}
    \label{fig:chemistry-barplot-malagon}
\end{figure}

\begin{figure}
    \centering
    \includegraphics[width=1\linewidth]{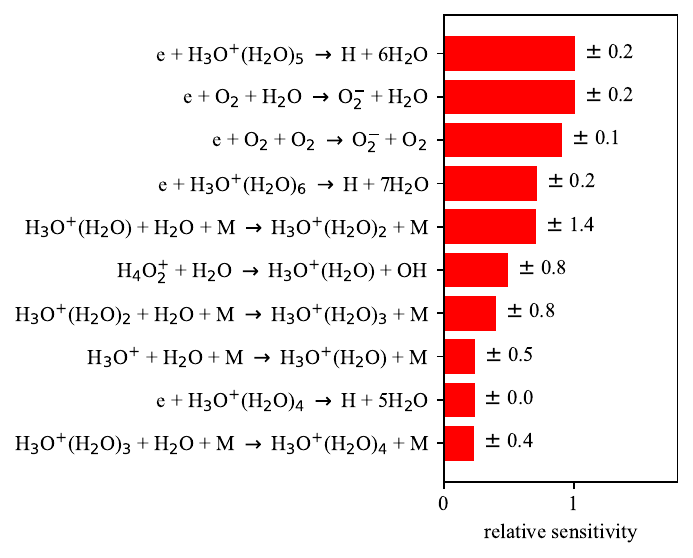}
    \caption{Sensitivity analysis of the Starikovskiy chemistry, analogous to figure~\ref{fig:chemistry-barplot-komuro}.}
    \label{fig:chemistry-barplot-starikovskiy}
\end{figure}

In figures~\ref{fig:chemistry-barplot-komuro}--\ref{fig:chemistry-barplot-starikovskiy}, the ten most important reactions from the Komuro, Malag\'{o}n and Starikovskiy chemistry sets are shown.
`Most important' is here defined as those reactions that lead to largest $Q'_{n}$ (see equation (\ref{eq:sensitvity-mu})), which is an indication of the relative change in the quantity of interest with respect to a change in the reaction rate coefficient.

Almost all of the important reactions decrease $N_\mathrm{e}$, as indicated by the red color in the bar plots.
The two three-body attachment reactions given by equations (\ref{eq:three-body-O2}--\ref{eq:three-body-O2-H2o}) are about equally important for all considered chemistries.
However, there is a major difference in the importance of water clusters around positive ions.
In the Komuro chemistry, the formation of single clusters through $\rm O_2^+ + H_2O + M \to O_2^+(H_2O) + M$ and $\rm O_4^+ + H_2O \to O_2^+(H_2O) + O_2$ is included, but these reactions have no strong effect on the later discharge evolution.
In contrast, water clusters play a very important role in the Malag\'{o}n and Starikovskiy chemistries.
The Malag\'{o}n chemistry includes reactions up to the formation of $\rm H_3O^+(H_2O)_3$, whereas the Starikovskiy chemistry includes reactions up to $\rm H_3O^+(H_2O)_6$.
For both chemistries, the recombination of electrons with such clusters are the reactions with the highest sensitivity.

The standard deviations $\sigma$ of the `sensitivities', as defined by equation (\ref{eq:sensitvity-sigma}), are given by the numbers next to the bars in figures~\ref{fig:chemistry-barplot-komuro}--\ref{fig:chemistry-barplot-starikovskiy}.
Note that there are several reactions related to water clusters that have a large standard deviation.
This is caused by the following phenomenon: if there is a chain of reactions (as there is for water cluster formation), the simulation might not be sensitive to changes in the rate of one reaction in the chain.
But if one reaction in the chain is fully disabled, the whole chain is stopped, which has a large effect on the result.
In our tests, we include such a `disabling' of reactions (namely by multiplying the rate coefficient by zero), which therefore leads to the large $\sigma$ values.

Sensitivity analysis can be a useful tool for reducing the number of reactions in a set.
If we are only interested in the total number of electrons $N_\mathrm{e}$ at $100 \, \textrm{ns}$, our results show that the effect of humidity in the Komuro chemistry can be described by two three-body attachment reactions.
For the Malag\'{o}n and Starikovskiy chemistries, the formation of water clusters around positive ions and the consecutive recombination of electrons with these clusters is also important.
Since electrons predominantly recombine with the largest clusters, it might be possible to simplify the chemistry by leaving out the intermediate states.

\section{\rev{Discussion}}
\label{sec:discussion}

\rev{Our results have illustrated how sensitive positive streamer simulations in humid air are to differences in the H$_2$O cross sections, the photoionization model and the included chemical reactions.
We consider this a first step, with the next step being to compare simulations against experiments to identify the input data that leads to the best agreement.
However, such comparisons are rather challenging, as discussed in depth by Starikovskiy \textit{et al}~\cite{starikovskiy2022}.
We list some of the main challenges below:
\begin{itemize}
    \item Experiments have often been performed under ambient conditions, with limited variation in the H$_2$O mole fraction and the gas pressure.
    \item Humidity reduces the amount of photoionization in front of positive streamers, which increases streamer branching. 
    Branching streamers can only be simulated with a 3D model. 
    This is computationally expensive and the comparison against experiments then has to be performed in a statistical manner.
    \item Experimental results can be sensitive to the voltage waveform, the effect of previous pulses, or the particular electrode geometry. 
    It is challenging to accurately include all these factors in a simulation model.
\end{itemize}
Despite these challenges, Starikovskiy \textit{et al}~\cite{starikovskiy2022} compared their axisymmetric simulations of positive streamers in humid air against several experimental measurements~\cite{phelps1976, Allen_1999, hui2008, mikropoulos2008, Meng_2017}, considering both the streamer stability field (average electric field required for crossing a gap) and the streamer velocity.
However, the spread in experimental results they observed was quite considerable, which makes it difficult to give recommendations for input data based on these measurements.
Below, we briefly discuss how experimental studies on streamer discharges could potentially lead to such recommendations.

\textbf{Cross sections} In humid air under standard conditions the mole fraction of H$_2$O is typically below 3\%.
The effect of H$_2$O on the electron transport data is therefore small, as shown in section~\ref{sec:transport-coefficients}, and it seems unlikely that experiments on streamer discharges could help to identify the most suitable H$_2$O cross sections.
However, if it were experimentally possible to study streamers in steam (i.e., up to 100\% H$_2$O produced by heating water), a comparison between experiments and simulations would probably give quite some information about the validity of the different cross section sets, since differences in electron transport data are then large, as shown in figure~\ref{fig:transport-data-pure-H2O}.

\textbf{Chemistry} For the positive streamers that we have simulated, the main effect of an increase in humidity was an increased electron loss rate.
The two main mechanisms responsible were increased three-body attachment and electron-ion recombination.
The relative importance of three-body processes can be changed by adjusting the pressure, and the rate of electron-ion recombination depends on the degree of ionization.
Measuring the decay of conductivity in a streamer channel under varying conditions (e.g., voltage and pressure) could therefore help assess the validity of different plasma chemistries, similar to the work of~\cite{Spencer_1987}.
Multi-pulse experiments~\cite{Li_2018} could also reveal more about plasma species that are involved in the removal or production of free electrons due to attachment, detachment, recombination or dissociation reactions.

\textbf{Photoionization} Based on the results in~\cite{wang2023}, we expect that the reduction in photoionization in humid air leads to significantly more streamer branching, and it would be interesting to test this hypothesis experimentally.
One challenge is that humidity affects the electron conductivity decay in streamer channels, which will also have an effect on streamer branching.}

\section{\rev{Conclusions}}     
\label{sec:con-and-outlook}

We have studied how strongly simulations of positive streamers in humid air depend on the choice of input data.
We considered different cross section sets for H$_2$O, different approximations for photoionization in humid air, and different chemistry sets.
Simulations were performed in air containing a mole fraction of 0\%, 3\% or 10\% H$_2$O at 300\,K and (approximately) 1\,bar, using an axisymmetric fluid model.
Our focus was on the streamer propagation stage, and time scales of up to about 200\,ns were considered.
The underlying motivation was to understand to what extent uncertainty about input data can be resolved by experimentally studying streamers in humid air.

When five different H$_2$O cross sections were used, the most significant differences were in the resulting electron attachment coefficient.
In air with 3\% H$_2$O, these cross sections led to relatively minor differences in streamer properties, with variations below 10\%.
Whereas for 10\% H$_2$O, the differences in streamer properties became substantially larger, with streamer velocity varying by up to about 50\%.
We did not use all cross section sets properly, \rev{as rotational cross sections were missing for some of the sets}, and because one of the sets is only supposed to be used with an anisotropic scattering model.
\rev{The Kawaguchi set was recommended since it accurately reproduces experimental swarm measurements and can be used with an isotropic scattering model.}

In humid air, there is less photoionization than in dry air.
We compared two photoionization models for humid air, one by Naidis and the other one by Aints, with the standard Zheleznyak model for dry air.
With less photoionization, the streamer had a higher maximal electric field at its tip, leading to a higher electron density in the channel and (in our simulations) also a higher velocity.
The effect of different photoionization models on the streamer velocity was comparable to the effect of using different H$_2$O cross sections.
Furthermore, the reduction in photoionization in some cases led to branching, in particular with the Aints model.

Finally, we performed simulations with three different humid air chemistry sets, and for comparison we also included one dry air chemistry set.
These sets differed significantly in the formation of water clusters around positive ions, which play an important role in electron-ion recombination.
This led to differences of up to about 50\% in streamer velocity for 10\% H$_2$O, whereas the differences in velocity in dry air were only about 10--20\%.
Furthermore, we identified the most important chemical reactions from the three chemistry sets for humid air with a sensitivity analysis.

\rev{This study was intended as a first step, in which we studied how sensitive streamer simulation results in humid air are to differences in the input data.
In future work, (new) experimental data could be used to compare with simulations, allowing for recommendations to be made on the choice of input data.
It would be interesting to experimentally study streamers in humid air with a mole fraction of 10\% H$_2$O or more, which is only possible at temperatures above room temperature.
This could shed more light on the importance of water clusters and their effect on the conductivity of streamer channels.}

\ack
This work was carried out on the Dutch national e-infrastructure with the support of SURF Co-operative.
BG and HM were supported by project 17183 (``Plasma for Plants'') financed by the Dutch Research Council NWO.
BG was also supported by the National Natural Science Foundation of China (52407165) \rev{and the Research Starting Grant of Fuzhou University (XRC-24125)}.
A.M.R. was supported by a Ram\'{o}n Areces Foundation Grant BEVP34A6840.
We thank Dr.~Behnaz Bagheri for fruitful discussions regarding humid air chemistry.

\section*{Data availability statement}


\rev{The data that support the findings of this study are openly available at the following URL/DOI: \url{https://doi.org/10.5281/zenodo.14758636}.}

\appendix

\section{Effect of ionization reactions}
\label{sec:ionization-reactions}

In our analysis, we have focused on the effect of humidity-related input data.
However, the presented simulation results will also be sensitive to the input data used for N$_2$ and O$_2$, of which the most significant effect will probably be changes in the ionization coefficient $\alpha$.
To get an idea of the sensitivity of our simulations to such changes, we have performed simulations in which all ionization reaction rate coefficients were multiplied by a factor $k_\mathrm{ion}$ whereas all other reactions were kept unchanged, as was also done in~\cite{wei2023}.
We modified $k_\mathrm{ion}$ in the range (1.0, 0.75, 0.6, 0.5, 0.4, 0.25), where $k_\mathrm{ion}=1.0$ corresponds to the unchanged $\alpha$.
The simulations were performed in air with 3\% H$_2$O in a background field of 10\,kV/cm, using the Malag\'{o}n chemistry, the Aints photoionization model and the Kawaguchi H$_2$O cross sections.

\begin{figure}
    \centering
    \includegraphics[width=1\linewidth]{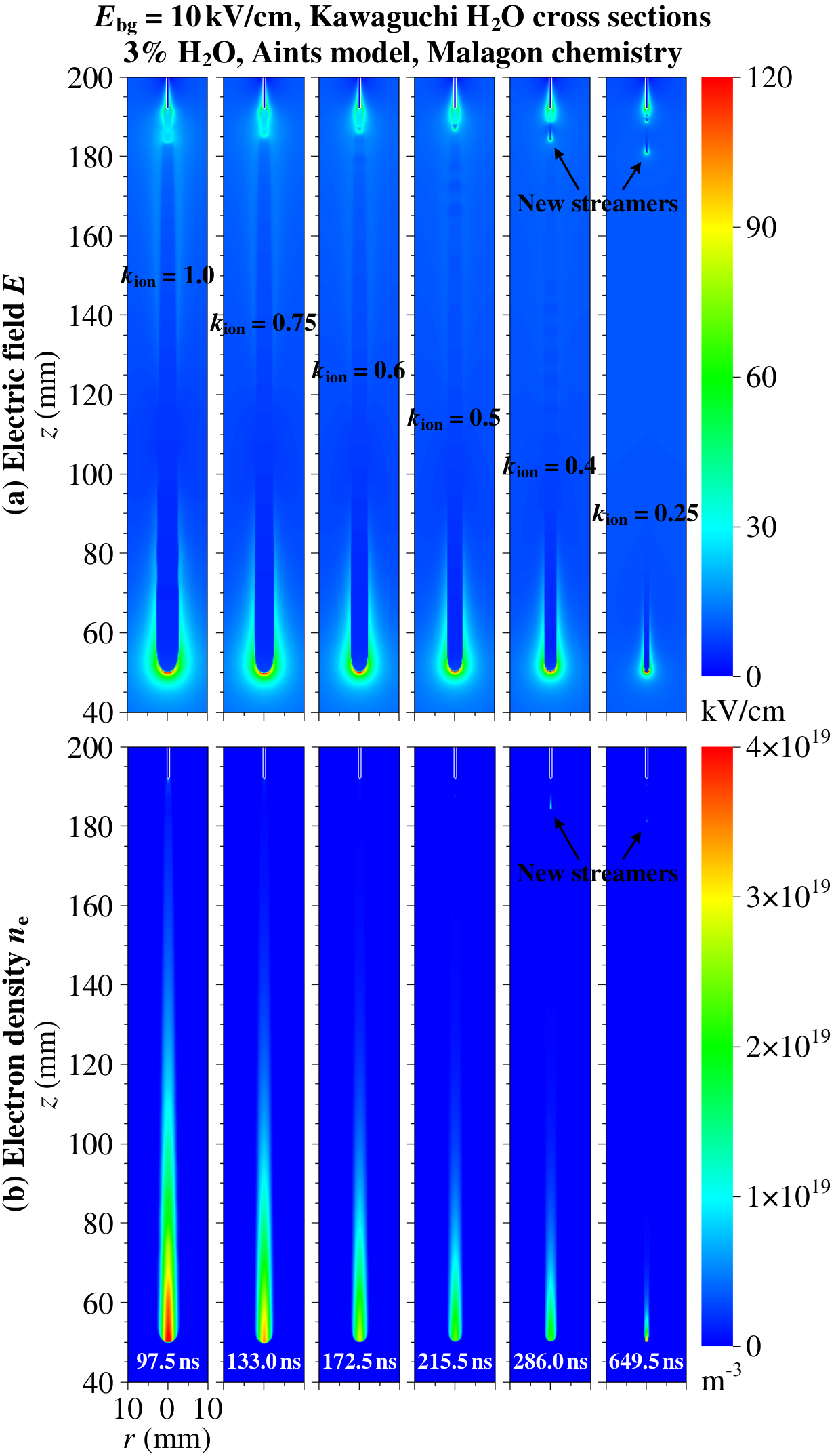}
    \caption{Effect of the ionization coefficient on positive streamers in $E_\mathrm{bg}=10$\,kV/cm in air with 3\% H$_2$O using the Aints photoionization model.
    Shown are (a) the electric field $E$ and (b) electron density $n_\mathrm{e}$ profiles for streamers at $z = 50$\,mm.}
    \label{fig:ionization-rate-profiles}
\end{figure}

\begin{figure}
    \centering
    \includegraphics[width=1\linewidth]{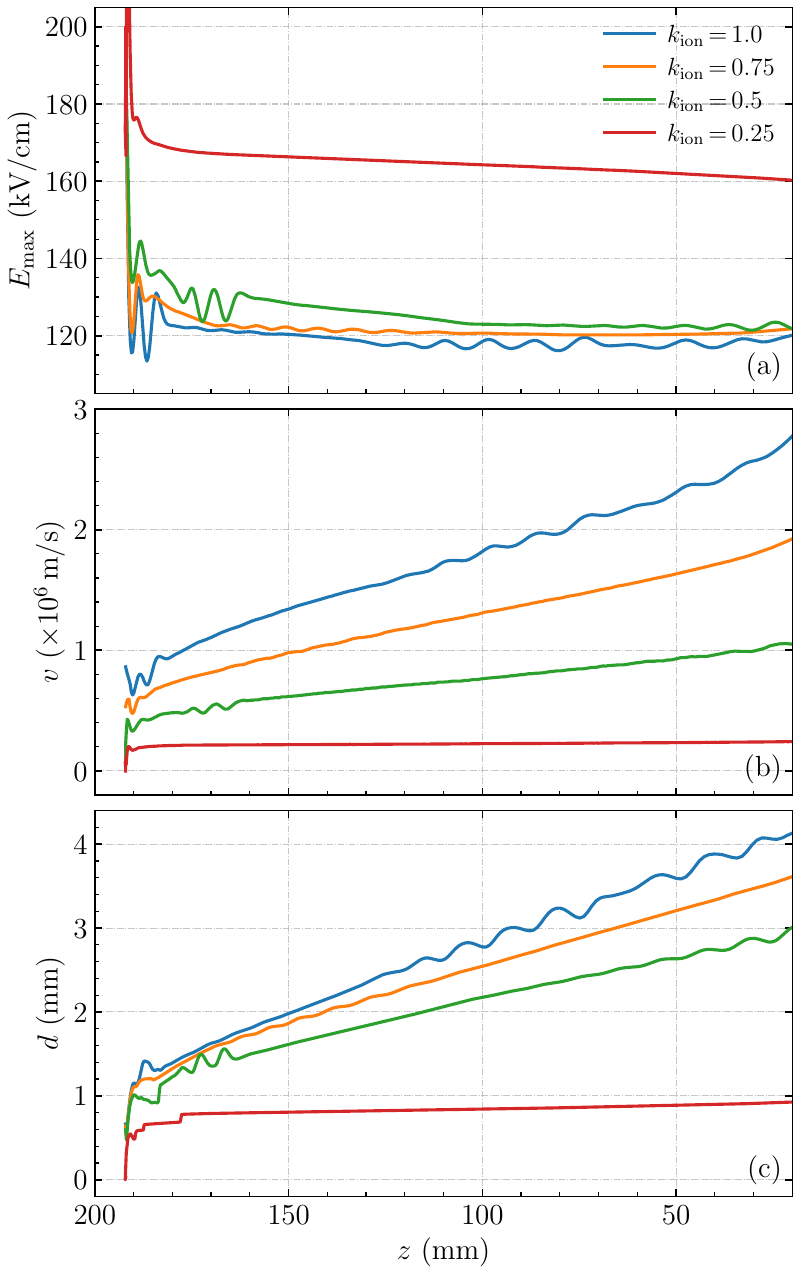}
    \caption{Effect of the ionization coefficient.
    Shown are streamer properties of (a) the maximal electric field $E_\mathrm{max}$, (b) streamer velocity $v$, and (c) optical diameter $d$ versus the streamer head position $z$.}
    \label{fig:ionization-rate-Emax-v-d}
  \end{figure}

Figure~\ref{fig:ionization-rate-profiles} shows the electric field and electron density profiles for different $k_\mathrm{ion}$ values when the streamers have propagated up to $z=50$\,mm, and figure~\ref{fig:ionization-rate-Emax-v-d} shows streamer properties as a function of $z$.
From these results, we conclude that the use of different input data for N$_2$ and O$_2$, resulting in a different ionization coefficient $\alpha$, could have the following effects:
\begin{itemize}
  \item The streamer velocity is rather sensitive to the ionization coefficient $\alpha$, and it seems to be approximately proportional to $\alpha$.
  \item The electron density in the streamer channel is also approximately proportional to $\alpha$ (in figure~\ref{fig:ionization-rate-profiles} this is hard to see due to the clipping of color values).
  \item The streamer diameter and the maximal electric field at the streamer tip are both much less sensitive to $\alpha$.
\end{itemize}

Furthermore, when $k_\mathrm{ion}$ is reduced below 0.5, new streamers can emerge behind the previous ones. 
The reduction in $\alpha$ leads to a decrease in the electron density inside the streamer channel, resembling the effect of increasing the attachment coefficient~\cite{francisco2021a}.
This causes the electric field behind the streamer head to quickly relax back to the background electric field, allowing the formation of a new streamer.

\section{Three chemistry sets for humid air}
\label{sec:three-chemistry-sets}

Tables~\ref{tab:chemistry-set-malagon},~\ref{tab:chemistry-set-starikovskiy} and~\ref{tab:chemistry-set-komuro} show three (Malag\'{o}n, Starikovskiy, and Komuro) chemistry sets for humid air used in the paper.
Comments for each set are appended at the end of each table. 
Table~\ref{tab:excited-states-komuro} lists the excited states of N$_2$, O$_2$ and H$_2$O used in the Komuro chemistry set, with their activation energies.

\newcounter{nombre}
\renewcommand{\thenombre}{\arabic{nombre}}
\setcounter{nombre}{0}
\newcounter{nombresub}
\renewcommand{\thenombresub}{\arabic{nombresub}}
\setcounter{nombresub}{0}
\newcommand{\Rnum}[1][]{\refstepcounter{nombre}#1R\thenombre}
\newcommand{\fnum}[1][]{\refstepcounter{nombresub}#1$f_{\thenombresub}$($E/N$)}

\begin{table*}
\renewcommand{\arraystretch}{1.1}
\centering
\captionsetup{width=1\textwidth}
\caption{The chemistry set for humid air from Malag\'{o}n-Romero \textit{et al}~\cite{malagon-romero2022}.
Reaction rate coefficients are in units of $\mathrm{cm^3\,s^{-1}}$ and $\mathrm{cm^6\,s^{-1}}$ for two-body and three-body reactions, respectively.
The symbol M denotes a neutral molecule (either N$_2$, O$_2$ or H$_2$O).
The reduced electric field $E/N$ is here expressed dimensionless, in units of Td (Townsend).
$k_\mathrm{B}$ is the Boltzmann constant.
$T$(K) and $T_e$(K) are gas and electron temperatures, respectively. 
}
\label{tab:chemistry-set-malagon}
\begin{tabular*}{1\textwidth}{l@{\extracolsep{\fill}}ll}
 \br
 No. & Reaction & Reaction rate coefficient~($\mathrm{cm^3\,s^{-1}}$ or $\mathrm{cm^6\,s^{-1}}$) \\
 \mr
 \multicolumn{3}{l}{(1) Electron impact ionization} \\
 \Rnum & $\rm e + N_2 \to 2e + N_2^+$ & \fnum \\
 \Rnum & $\rm e + O_2 \to 2e + O_2^+$ & \fnum \\ 
 \Rnum & $\rm e + H_2O \to 2e + H_2O^+$ & \fnum \\  
 \Rnum & $\rm e + H_2O \to 2e + H^+ + OH$ & \fnum \\ 
 \Rnum & $\rm e + H_2O \to 2e + OH^+ + H$ & \fnum \\
 \Rnum & $\rm e + H_2O \to 2e + O^+ + H_2$ & \fnum \\ 
 \Rnum & $\rm e + H_2O \to 2e + H_2^+ + O$ & \fnum \\ 
 \Rnum & $\rm e + H_2O \to 3e + O^{2+} + H_2$ & \fnum \\
 \mr
 \multicolumn{3}{l}{(2) Electron attachment} \\
 \Rnum & $\rm e + O_2 + O_2 \to O_2^- + O_2$ & \fnum \\
 \Rnum & $\rm e + O_2 \to O^- + O$ & \fnum \\
 \Rnum & $\rm e + H_2O \to H^- + OH$ & \fnum \\
 \Rnum & $\rm e + H_2O \to OH^- + H$ & \fnum \\
 \Rnum & $\rm e + H_2O \to O^- + H_2$ & \fnum \\
 \Rnum & $\rm e + O_2 + H_2O \to O_2^- + H_2O$ & \fnum \ = $6 \times f_{9}$($E/N$) \\
 \mr
 \multicolumn{3}{l}{(3) Electron detachment} \\
 \Rnum & $\rm O_2^- + M \to e + O_2 + M$ & $1.24\times10^{-11}\exp(-(\frac{179}{8.8+E/N})^2)$ \\
 \Rnum & $\rm O^- + N_2 \to e + N_2O$ & $1.16\times10^{-12}\exp(-(\frac{48.9}{11+E/N})^2)$ \\
 \Rnum & $\rm H^- + O_2 \to e + HO_2$ & $1.2\times10^{-9}$ \\
 \mr
 \multicolumn{3}{l}{(4) Negative ion conversion} \\
 \Rnum & $\rm O^- + O_2 + M \to O_3^- + M$ & $1.1\times10^{-30}\exp(-(\frac{E/N}{65})^2)$ \\ 
 \Rnum & $\rm O^- + O_2 \to O_2^- + O$ & $6.96\times10^{-11}\exp(-(\frac{198}{5.6+E/N})^2)$ \\
 \Rnum & $\rm O^- + H_2O \to OH^- + OH$ & $6.0\times10^{-13}$ \\
 \Rnum & $\rm H^- + O_2 \to O^- + OH$ & $1.0\times10^{-11}$ \\
 \Rnum & $\rm H^- + O_2 \to O_2^- + H$ & $1.0\times10^{-11}$ \\
 \Rnum & $\rm H^- + H_2O \to OH^- + H_2$ & $3.8\times10^{-9}$ \\
 \Rnum & $\rm O_2^- + H_2O + M \to O_2^-(H_2O) + M$ & $2.2\times10^{-28}$ \\
 \Rnum & $\rm O_2^-(H_2O) + H_2O + M \to O_2^-(H_2O)_2 + M$ & $5.0\times10^{-28}$ \\
 \Rnum & $\rm O_2^-(H_2O)_2 + H_2O + M \to O_2^-(H_2O)_3 + M$ & $5.0\times10^{-29}$ \\
 \Rnum & $\rm O_2^-(H_2O) + M \to O_2^- + H_2O + M$ & $5.91\times10^{-9}\exp(-\frac{1.28\times10^{-19}}{k_\mathrm{B}(T+\frac{E/N}{0.18})})$ \\ 
 \Rnum & $\rm O_2^-(H_2O)_2 + M \to O_2^-(H_2O) + H_2O + M$ & $1.34\times10^{-8}\exp(-\frac{5.8\times10^{-20}}{k_\mathrm{B}(T+\frac{E/N}{0.18})})$ \\ 
 \Rnum & $\rm O_2^-(H_2O)_3 + M \to O_2^-(H_2O)_2 + H_2O + M$ & $1.34\times10^{-9}\exp(-\frac{4.49\times10^{-20}}{k_\mathrm{B}(T+\frac{E/N}{0.18})})$ \\ 
 \mr
 \multicolumn{3}{l}{(5) Positive ion conversion} \\
 \Rnum & $\rm N_2^+ + N_2 + M \to N_4^+ + M$ & $5.0\times10^{-29}(\frac{300}{T})^{2}$ \\
 \Rnum & $\rm N_4^+ + O_2 \to O_2^+ + 2N_2$ & $2.5\times10^{-10}$ \\
 \Rnum & $\rm O_2^+ + O_2 + M \to O_4^+ + M$ & $2.4\times10^{-30}(\frac{300}{T})^{3}$ \\
 \Rnum & $\rm O_2^+ + H_2O + M \to O_2^+(H_2O) + M$ & $2.6\times10^{-28}$ \\
 \Rnum & $\rm O_2^+(H_2O) + H_2O \to H_3O^+ + OH + O_2$ & $3.0\times10^{-10}$ \\
 \Rnum & $\rm H_3O^+ + H_2O + M \to H_3O^+(H_2O) + M$ & $3.0\times10^{-27}$ \\
 \Rnum & $\rm H_3O^+(H_2O) + H_2O + M \to H_3O^+(H_2O)_2 + M$ & $3.0\times10^{-27}$ \\
 \Rnum & $\rm H_3O^+(H_2O)_2 + H_2O + M \to H_3O^+(H_2O)_3 + M$ & $3.0\times10^{-27}$ \\ 
 \br
\end{tabular*}
\end{table*}

\addtocounter{table}{-1}

\begin{table*}
\renewcommand{\arraystretch}{1.1}
\centering
\captionsetup{width=1\textwidth}
\caption{(Continued.)}
\begin{tabular*}{1\textwidth}{l@{\extracolsep{\fill}}ll}
 \br
 No. & Reaction & Reaction rate coefficient~($\mathrm{cm^3\,s^{-1}}$ or $\mathrm{cm^6\,s^{-1}}$) \\
 \mr
 \multicolumn{3}{l}{(6) Electron-ion recombination} \\
 \Rnum & $\rm e + O_4^+ \to O_2 + O_2$ & $1.4\times10^{-6}(\frac{300}{T_e})^{0.5}$ \\
 \Rnum & $\rm e + H_3O^+(H_2O)_3 \to H + 4H_2O$ & $6.5\times10^{-6}(\frac{300}{T_e})^{0.5}$ \\
 \mr
 \multicolumn{3}{l}{(7) Ion-ion recombination} \\
 \Rnum & $\rm N_2^+ + O^- \to 2N + O$ & $1.0\times10^{-7}$ \\
 \Rnum & $\rm N_2^+ + O_2^- \to 2N + O_2$ & $1.0\times10^{-7}$ \\
 \Rnum & $\rm N_2^+ + O_3^- \to 2N + O_3$ & $1.0\times10^{-7}$ \\
 \Rnum & $\rm N_2^+ + O_2^-(H_2O) \to neutrals$ & $1.0\times10^{-7}$ \\
 \Rnum & $\rm N_2^+ + O_2^-(H_2O)_2 \to neutrals$ & $1.0\times10^{-7}$ \\
 \Rnum & $\rm N_2^+ + O_2^-(H_2O)_3 \to neutrals$ & $1.0\times10^{-7}$ \\
 \Rnum & $\rm N_4^+ + O^- \to 2N_2 + O$ & $1.0\times10^{-7}$ \\
 \Rnum & $\rm N_4^+ + O_2^- \to 2N_2 + O_2$ & $1.0\times10^{-7}$ \\
 \Rnum & $\rm N_4^+ + O_3^- \to 2N_2 + O_3$ & $1.0\times10^{-7}$ \\
 \Rnum & $\rm N_4^+ + O_2^-(H_2O) \to neutrals$ & $1.0\times10^{-7}$ \\
 \Rnum & $\rm N_4^+ + O_2^-(H_2O)_2 \to neutrals$ & $1.0\times10^{-7}$ \\
 \Rnum & $\rm N_4^+ + O_2^-(H_2O)_3 \to neutrals$ & $1.0\times10^{-7}$ \\
 \Rnum & $\rm O_2^+ + O^- \to 2O + O$ & $1.0\times10^{-7}$ \\
 \Rnum & $\rm O_2^+ + O_2^- \to 2O + O_2$ & $1.0\times10^{-7}$ \\
 \Rnum & $\rm O_2^+ + O_3^- \to 2O + O_3$ & $1.0\times10^{-7}$ \\
 \Rnum & $\rm O_2^+ + O_2^-(H_2O) \to neutrals$ & $1.0\times10^{-7}$ \\
 \Rnum & $\rm O_2^+ + O_2^-(H_2O)_2 \to neutrals$ & $1.0\times10^{-7}$ \\
 \Rnum & $\rm O_2^+ + O_2^-(H_2O)_3 \to neutrals$ & $1.0\times10^{-7}$ \\
 \Rnum & $\rm O_4^+ + O^- \to 2O_2 + O$ & $1.0\times10^{-7}$ \\
 \Rnum & $\rm O_4^+ + O_2^- \to 2O_2 + O_2$ & $1.0\times10^{-7}$ \\
 \Rnum & $\rm O_4^+ + O_3^- \to 2O_2 + O_3$ & $1.0\times10^{-7}$ \\
 \Rnum & $\rm O_4^+ + O_2^-(H_2O) \to neutrals$ & $1.0\times10^{-7}$ \\
 \Rnum & $\rm O_4^+ + O_2^-(H_2O)_2 \to neutrals$ & $1.0\times10^{-7}$ \\
 \Rnum & $\rm O_4^+ + O_2^-(H_2O)_3 \to neutrals$ & $1.0\times10^{-7}$ \\
 \Rnum & $\rm O_2^+(H_2O) + O^- \to neutrals$ & $1.0\times10^{-7}$ \\
 \Rnum & $\rm O_2^+(H_2O) + O_2^- \to neutrals$ & $1.0\times10^{-7}$ \\
 \Rnum & $\rm O_2^+(H_2O) + O_3^- \to neutrals$ & $1.0\times10^{-7}$ \\
 \Rnum & $\rm O_2^+(H_2O) + O_2^-(H_2O) \to neutrals$ & $1.0\times10^{-7}$ \\
 \Rnum & $\rm O_2^+(H_2O) + O_2^-(H_2O)_2 \to neutrals$ & $1.0\times10^{-7}$ \\
 \Rnum & $\rm O_2^+(H_2O) + O_2^-(H_2O)_3 \to neutrals$ & $1.0\times10^{-7}$ \\
 \Rnum & $\rm H_3O^+ + O^- \to neutrals$ & $1.0\times10^{-7}$ \\
 \Rnum & $\rm H_3O^+ + O_2^- \to neutrals$ & $1.0\times10^{-7}$ \\
 \Rnum & $\rm H_3O^+ + O_3^- \to neutrals$ & $1.0\times10^{-7}$ \\
 \Rnum & $\rm H_3O^+ + O_2^-(H_2O) \to neutrals$ & $1.0\times10^{-7}$ \\
 \Rnum & $\rm H_3O^+ + O_2^-(H_2O)_2 \to neutrals$ & $1.0\times10^{-7}$ \\
 \Rnum & $\rm H_3O^+ + O_2^-(H_2O)_3 \to neutrals$ & $1.0\times10^{-7}$ \\
 \Rnum & $\rm H_3O^+(H_2O) + O^- \to neutrals$ & $1.0\times10^{-7}$ \\
 \Rnum & $\rm H_3O^+(H_2O) + O_2^- \to neutrals$ & $1.0\times10^{-7}$ \\
 \Rnum & $\rm H_3O^+(H_2O) + O_3^- \to neutrals$ & $1.0\times10^{-7}$ \\
 \Rnum & $\rm H_3O^+(H_2O) + O_2^-(H_2O) \to neutrals$ & $1.0\times10^{-7}$ \\
 \Rnum & $\rm H_3O^+(H_2O) + O_2^-(H_2O)_2 \to neutrals$ & $1.0\times10^{-7}$ \\
 \Rnum & $\rm H_3O^+(H_2O) + O_2^-(H_2O)_3 \to neutrals$ & $1.0\times10^{-7}$ \\
 \br
\end{tabular*}
\end{table*}

\addtocounter{table}{-1}

\begin{table*}
\renewcommand{\arraystretch}{1.1}
\centering
\captionsetup{width=1\textwidth}
\caption{(Continued.)}
\begin{tabular*}{1\textwidth}{l@{\extracolsep{\fill}}ll}
 \br
 No. & Reaction & Reaction rate coefficient~($\mathrm{cm^3\,s^{-1}}$ or $\mathrm{cm^6\,s^{-1}}$) \\
 \mr
 \Rnum & $\rm H_3O^+(H_2O)_2 + O^- \to neutrals$ & $1.0\times10^{-7}$ \\
 \Rnum & $\rm H_3O^+(H_2O)_2 + O_2^- \to neutrals$ & $1.0\times10^{-7}$ \\
 \Rnum & $\rm H_3O^+(H_2O)_2 + O_3^- \to neutrals$ & $1.0\times10^{-7}$ \\
 \Rnum & $\rm H_3O^+(H_2O)_2 + O_2^-(H_2O) \to neutrals$ & $1.0\times10^{-7}$ \\
 \Rnum & $\rm H_3O^+(H_2O)_2 + O_2^-(H_2O)_2 \to neutrals$ & $1.0\times10^{-7}$ \\
 \Rnum & $\rm H_3O^+(H_2O)_2 + O_2^-(H_2O)_3 \to neutrals$ & $1.0\times10^{-7}$ \\
 \Rnum & $\rm H_3O^+(H_2O)_3 + O^- \to neutrals$ & $1.0\times10^{-7}$ \\
 \Rnum & $\rm H_3O^+(H_2O)_3 + O_2^- \to neutrals$ & $1.0\times10^{-7}$ \\
 \Rnum & $\rm H_3O^+(H_2O)_3 + O_3^- \to neutrals$ & $1.0\times10^{-7}$ \\
 \Rnum & $\rm H_3O^+(H_2O)_3 + O_2^-(H_2O) \to neutrals$ & $1.0\times10^{-7}$ \\
 \Rnum & $\rm H_3O^+(H_2O)_3 + O_2^-(H_2O)_2 \to neutrals$ & $1.0\times10^{-7}$ \\
 \Rnum & $\rm H_3O^+(H_2O)_3 + O_2^-(H_2O)_3 \to neutrals$ & $1.0\times10^{-7}$ \\ 
 \mr
 \multicolumn{3}{l}{(8) Light emission} \\
 \Rnum & $\rm e + N_2 \to e + N_2(C)$ & \fnum \\
 \Rnum & $\rm N_2(C) + N_2 \to N_2 + N_2$ & $1.3\times10^{-11}$ \\
 \Rnum & $\rm N_2(C) + O_2 \to N_2 + O_2$ & $3.0\times10^{-10}$ \\
 \Rnum & $\rm N_2(C) \to N_2(B) + {\it h\nu}$ & $1/(42\,\mathrm{ns})$ \\
 \br
\end{tabular*}
\begin{tabular*}{1\textwidth}{@{\extracolsep{\fill}}l}
1. In~\cite{malagon-romero2022}, a total ionization reaction like R3 was used for H$_2$O. Here more ionization reaction channels (R3--R8) \\
are included when using Kawaguchi H$_2$O cross sections, whereas for other H$_2$O cross sections only R3 is used. \\
2. In~\cite{malagon-romero2022}, rate coefficients for R30 and R31 were $5.0\times10^{-29}(\frac{300}{T})^{3}$ and $2.5\times10^{-10}(\frac{300}{T})^{3}$, respectively. We check \\
their cited paper~\cite{aleksandrov1999} and have corrected them to $5.0\times10^{-29}(\frac{300}{T})^{2}$ and $2.5\times10^{-10}$, respectively. \\
3. R94--R97 are included for light emission, which were taken from~\cite{pancheshnyi2005}. \\
\end{tabular*}
\end{table*}

\setcounter{nombre}{0}
\setcounter{nombresub}{0}

\begin{table*}
\renewcommand{\arraystretch}{1.1}
\centering
\captionsetup{width=1\textwidth}
\caption{The chemistry set for humid air from Starikovskiy \textit{et al}~\cite{starikovskiy2022}.
}
\label{tab:chemistry-set-starikovskiy}
\begin{tabular*}{1\textwidth}{l@{\extracolsep{\fill}}ll}
 \br
 No. & Reaction & Reaction rate coefficient~($\mathrm{cm^3\,s^{-1}}$ or $\mathrm{cm^6\,s^{-1}}$) \\ 
 \mr
 \multicolumn{3}{l}{(1) Electron impact ionization} \\
 \Rnum & $\rm e + N_2 \to 2e + N_2^+$ & \fnum \\
 \Rnum & $\rm e + O_2 \to 2e + O_2^+$ & \fnum \\ 
 \Rnum & $\rm e + H_2O \to 2e + H_2O^+$ & \fnum \\ 
 \Rnum & $\rm e + H_2O \to 2e + H^+ + OH$ & \fnum \\ 
 \Rnum & $\rm e + H_2O \to 2e + OH^+ + H$ & \fnum \\
 \Rnum & $\rm e + H_2O \to 2e + O^+ + H_2$ & \fnum \\ 
 \Rnum & $\rm e + H_2O \to 2e + H_2^+ + O$ & \fnum \\ 
 \Rnum & $\rm e + H_2O \to 3e + O^{2+} + H_2$ & \fnum \\ 
 \mr
 \multicolumn{3}{l}{(2) Electron attachment} \\
 \Rnum & $\rm e + O_2 + O_2 \to O_2^- + O_2$ & \fnum \\
 \Rnum & $\rm e + O_2 + H_2O \to O_2^- + H_2O$ & \fnum \ = $7 \times f_{9}$($E/N$) \\
 \mr
 \multicolumn{3}{l}{(3) Positive ion conversion} \\
 \Rnum & $\rm N_2^+ + N_2 + M \to N_4^+ + M$ & $5.0\times10^{-29}$ \\
 \Rnum & $\rm N_2^+ + O_2 \to O_2^+ + N_2$ & $6.0\times10^{-11}$ \\
 \Rnum & $\rm N_2^+ + H_2O \to H_2O^+ + N_2$ & $2.3\times10^{-9}(\frac{300}{T})^{0.5}$ \\
 \Rnum & $\rm N_2^+ + H_2O \to N_2H^+ + OH$ & $5.0\times10^{-10}(\frac{300}{T})^{0.5}$ \\
 \Rnum & $\rm N_4^+ + O_2 \to O_2^+ + 2N_2$ & $2.5\times10^{-10}$ \\
 \Rnum & $\rm N_4^+ + H_2O \to H_2O^+ + 2N_2$ & $2.4\times10^{-9}(\frac{300}{T})^{0.5}$ \\
 \Rnum & $\rm O_2^+ + N_2 + M \to N_2O_2^+ + M$ & $9.0\times10^{-31}$ \\
 \Rnum & $\rm O_2^+ + O_2 + M \to O_4^+ + M$ & $2.4\times10^{-30}$ \\
 \Rnum & $\rm O_2^+ + H_2O + M \to O_2^+(H_2O) + M$ & $2.6\times10^{-28}(\frac{300}{T})^{4}$ \\
 \Rnum & $\rm O_4^+ + H_2O \to O_2^+(H_2O) + O_2$ & $1.7\times10^{-9}$ \\
 \Rnum & $\rm N_2H^+ + H_2O \to H_3O^+ + N_2$ & $2.6\times10^{-9}(\frac{300}{T})^{0.5}$ \\
 \Rnum & $\rm N_2O_2^+ + N_2 \to O_2^+ + 2N_2$ & $4.3\times10^{-10}$ \\
 \Rnum & $\rm N_2O_2^+ + O_2 \to O_4^+ + N_2$ & $1.0\times10^{-9}$ \\
 \Rnum & $\rm H_2O^+ + O_2 \to O_2^+ + H_2O$ & $4.1\times10^{-10}$ \\
 \Rnum & $\rm H_2O^+ + H_2O \to H_3O^+ + OH$ & $2.1\times10^{-9}(\frac{300}{T})^{0.5}$ \\
 \Rnum & $\rm O_2^+(H_2O) + H_2O \to H_3O^+ + OH + O_2$ & $3.0\times10^{-10}$ \\
 \Rnum & $\rm O_2^+(H_2O) + H_2O \to H_3O^+(OH) + O_2$ & $1.9\times10^{-9}$ \\ 
 \Rnum & $\rm H_3O^+(OH) + H_2O \to H_3O^+(H_2O) + OH$ & $3.0\times10^{-9}$ \\ 
 \Rnum & $\rm H_3O^+ + H_2O + M \to H_3O^+(H_2O) + M$ & $3.2\times10^{-27}(\frac{300}{T})^{4}$ \\
 \Rnum & $\rm H_3O^+(H_2O) + H_2O + M \to H_3O^+(H_2O)_2 + M$ & $7.4\times10^{-27}(\frac{300}{T})^{7.5}$ \\
 \Rnum & $\rm H_3O^+(H_2O)_2 + H_2O + M \to H_3O^+(H_2O)_3 + M$ & $2.5\times10^{-27}(\frac{300}{T})^{8.1}$ \\
 \Rnum & $\rm H_3O^+(H_2O)_3 + H_2O + M \to H_3O^+(H_2O)_4 + M$ & $3.3\times10^{-28}(\frac{300}{T})^{14}$ \\
 \Rnum & $\rm H_3O^+(H_2O)_4 + H_2O + M \to H_3O^+(H_2O)_5 + M$ & $4.0\times10^{-29}(\frac{300}{T})^{15.3}$ \\
 \Rnum & $\rm H_3O^+(H_2O)_5 + H_2O + M \to H_3O^+(H_2O)_6 + M$ & $4.5\times10^{-30}(\frac{300}{T})^{16}$ \\
 \Rnum & $\rm H_3O^+(H_2O) + M \to H_3O^+ + H_2O + M$ & $4.3\times10^{-2}(\frac{300}{T})^{4}\exp(\frac{-16430}{T})$ \\
 \Rnum & $\rm H_3O^+(H_2O)_2 + M \to H_3O^+(H_2O) + H_2O + M$ & $2.1\times10^{-2}(\frac{300}{T})^{7.5}\exp(\frac{-10030}{T})$ \\
 \Rnum & $\rm H_3O^+(H_2O)_3 + M \to H_3O^+(H_2O)_2 + H_2O + M$ & $1.7\times10^{-2}(\frac{300}{T})^{8.1}\exp(\frac{-8320}{T})$ \\
 \Rnum & $\rm H_3O^+(H_2O)_4 + M \to H_3O^+(H_2O)_3 + H_2O + M$ & $1.3\times10^{-4}(\frac{300}{T})^{14}\exp(\frac{-5750}{T})$ \\
 \Rnum & $\rm H_3O^+(H_2O)_5 + M \to H_3O^+(H_2O)_4 + H_2O + M$ & $3.3\times10^{-5}(\frac{300}{T})^{15.3}\exp(\frac{-5000}{T})$ \\
 \Rnum & $\rm H_3O^+(H_2O)_6 + M \to H_3O^+(H_2O)_5 + H_2O + M$ & $4.6\times10^{-6}(\frac{300}{T})^{16}\exp(\frac{-5000}{T})$ \\
 \br
\end{tabular*}
\end{table*}

\addtocounter{table}{-1}

\begin{table*}
\renewcommand{\arraystretch}{1.1}
\centering
\captionsetup{width=1\textwidth}
\caption{(Continued.)}
\begin{tabular*}{1\textwidth}{l@{\extracolsep{\fill}}ll}
 \br
 No. & Reaction & Reaction rate coefficient~($\mathrm{cm^3\,s^{-1}}$ or $\mathrm{cm^6\,s^{-1}}$) \\ 
 \mr
 \multicolumn{3}{l}{(4) Electron-ion recombination} \\
 \Rnum & $\rm e + N_2^+ \to N + N$ & $2.8\times10^{-7}(\frac{300}{T_e})^{0.5}$ \\
 \Rnum & $\rm e + N_4^+ \to N_2 + N_2$ & $2.6\times10^{-6}(\frac{300}{T_e})^{0.41}$ \\
 \Rnum & $\rm e + O_2^+ \to O + O$ & $2.0\times10^{-7}(\frac{300}{T_e})$ \\
 \Rnum & $\rm e + O_4^+ \to O_2 + O_2$ & $4.2\times10^{-6}(\frac{300}{T_e})^{0.48}$ \\
 \Rnum & $\rm e + N_2H^+ \to H + N_2$ & $2.4\times10^{-7}$ \\
 \Rnum & $\rm e + N_2O_2^+ \to N_2 + O_2$ & $1.3\times10^{-6}(\frac{300}{T_e})^{0.5}$ \\ 
 \Rnum & $\rm e + H_2O^+ \to neutrals$ & $4.3\times10^{-7}(\frac{300}{T_e})^{0.7}$ \\
 \Rnum & $\rm e + H_3O^+ \to neutrals$ & $8.0\times10^{-7}(\frac{300}{T_e})^{0.8}$ \\
 \Rnum & $\rm e + H_3O^+(OH) \to neutrals$ & $2.0\times10^{-6}(\frac{300}{T_e})^{0.5}$ \\
 \Rnum & $\rm e + H_3O^+(H_2O) \to H + 2H_2O$ & $2.5\times10^{-6}(\frac{300}{T_e})^{0.5}$ \\ 
 \Rnum & $\rm e + H_3O^+(H_2O)_2 \to H + 3H_2O$ & $4.5\times10^{-6}(\frac{300}{T_e})^{0.5}$ \\ 
 \Rnum & $\rm e + H_3O^+(H_2O)_3 \to H + 4H_2O$ & $6.5\times10^{-6}(\frac{300}{T_e})^{0.5}$ \\ 
 \Rnum & $\rm e + H_3O^+(H_2O)_4 \to H + 5H_2O$ & $8.5\times10^{-6}(\frac{300}{T_e})^{0.5}$ \\ 
 \Rnum & $\rm e + H_3O^+(H_2O)_5 \to H + 6H_2O$ & $10.5\times10^{-6}(\frac{300}{T_e})^{0.5}$ \\ 
 \Rnum & $\rm e + H_3O^+(H_2O)_6 \to H + 7H_2O$ & $12.5\times10^{-6}(\frac{300}{T_e})^{0.5}$ \\ 
 \Rnum & $\rm e + H_3O^+(H_2O) + H_2O \to H + 3H_2O$ & $2.7\times10^{-23}(\frac{300}{T_e})^{2}$ \\ 
 \Rnum & $\rm e + H_3O^+(H_2O)_2 + H_2O \to H + 4H_2O$ & $2.7\times10^{-23}(\frac{300}{T_e})^{2}$ \\
 \Rnum & $\rm e + H_3O^+(H_2O)_3 + H_2O \to H + 5H_2O$ & $2.7\times10^{-23}(\frac{300}{T_e})^{2}$ \\
 \Rnum & $\rm e + H_3O^+(H_2O)_4 + H_2O \to H + 6H_2O$ & $2.7\times10^{-23}(\frac{300}{T_e})^{2}$ \\
 \Rnum & $\rm e + H_3O^+(H_2O)_5 + H_2O \to H + 7H_2O$ & $2.7\times10^{-23}(\frac{300}{T_e})^{2}$ \\
 \Rnum & $\rm e + H_3O^+(H_2O)_6 + H_2O \to H + 8H_2O$ & $2.7\times10^{-23}(\frac{300}{T_e})^{2}$ \\
 \Rnum & $\rm e + O_2^+(H_2O) \to O_2 + H_2O$ & $3.0\times10^{-7}(\frac{300}{T_e})^{0.5}$ \\
 \Rnum & $\rm 2e + N_2^+ \to e + N_2$ & $1.0\times10^{-19}(\frac{300}{T_e})^{4.5}$ \\
 \Rnum & $\rm 2e + O_2^+ \to e + O_2$ & $1.0\times10^{-19}(\frac{300}{T_e})^{4.5}$ \\
 \Rnum & $\rm 2e + H_2O^+ \to e + H_2O$ & $1.0\times10^{-19}(\frac{300}{T_e})^{4.5}$ \\
 \mr
 \multicolumn{3}{l}{(5) Ion-ion recombination} \\ 
 \Rnum & $\rm O_2^- + N_2^+ + M \to O_2 + N_2 + M$ & $2.0\times10^{-25}(\frac{300}{T})^{2.5}$ \\
 \Rnum & $\rm O_2^- + O_2^+ + M \to O_2 + O_2 + M$ & $2.0\times10^{-25}(\frac{300}{T})^{2.5}$ \\
 \Rnum & $\rm O_2^- + H_2O^+ + M \to O_2 + H_2O + M$ & $2.0\times10^{-25}(\frac{300}{T})^{2.5}$ \\
 \mr
 \multicolumn{3}{l}{(6) Light emission} \\
 \Rnum & $\rm e + N_2 \to e + N_2(C)$ & \fnum \\
 \Rnum & $\rm N_2(C) + N_2 \to N_2 + N_2$ & $1.3\times10^{-11}$ \\
 \Rnum & $\rm N_2(C) + O_2 \to N_2 + O_2$ & $3.0\times10^{-10}$ \\
 \Rnum & $\rm N_2(C) \to N_2(B) + {\it h\nu}$ & $1/(42\,\mathrm{ns})$ \\
 \br
\end{tabular*}
\begin{tabular*}{1\textwidth}{@{\extracolsep{\fill}}l}
1. In~\cite{starikovskiy2022}, a total ionization reaction like R3 was used for H$_2$O. Here more ionization reaction channels (R3--R8) \\
are included when using Kawaguchi H$_2$O cross sections. \\
2. In~\cite{starikovskiy2022}, a rate coefficient function was used for three-body attachment reactions R9 and R10. Here their rate \\
coefficients are obtained using BOLSIG$-$. \\
3. In~\cite{starikovskiy2022}, rate coefficients for R29--R40 were pressure- and temperature-dependent. Here pressure-dependent \\
is not taken into account since the pressure in our simulations is (approximately) 1\,bar. \\  
4. In~\cite{starikovskiy2022}, the third body for R66--R68 contained any neutral species. Here we only include the gas molecule M. \\
5. R69--R72 are included for light emission, which were taken from~\cite{pancheshnyi2005}. \\
\end{tabular*}
\end{table*}

\setcounter{nombre}{0}
\setcounter{nombresub}{0}

\begin{table*}
\renewcommand{\arraystretch}{1.1}
\centering
\captionsetup{width=1\textwidth}
\caption{The chemistry set for humid air from Komuro \textit{et al}~\cite{komuro2013, komuro2018a}.
The reactions involving charged ion particles were taken from table 1 of~\cite{komuro2018a} with some additions (R27--R29, R33--R37, R42) from tables 1 and 2 of~\cite{komuro2013}.
Other reactions were taken from tables 1 and 3 of~\cite{komuro2013}.
}
\label{tab:chemistry-set-komuro}
\begin{tabular*}{1\textwidth}{l@{\extracolsep{\fill}}ll}
 \br
 No. & Reaction & Reaction rate coefficient~($\mathrm{cm^3\,s^{-1}}$ or $\mathrm{cm^6\,s^{-1}}$) \\
 \mr
 \multicolumn{3}{l}{(1) Vibrational excitation} \\ 
 \Rnum & $\rm e + N_2 \to e + N_2(\it{v})$ & \fnum \\
 \Rnum & $\rm e + O_2 \to e + O_2(\it{v})$ & \fnum \\
 \Rnum & $\rm e + H_2O \to e + H_2O(\it{v})$ & \fnum \\
 \mr
 \multicolumn{3}{l}{(2) Electron excitation} \\ 
 \Rnum & $\rm e + N_2 \to e + N_2(A_1)$ & \fnum \\
 \Rnum & $\rm e + N_2 \to e + N_2(A_2)$ & \fnum \\
 \Rnum & $\rm e + N_2 \to e + N_2(B)$ & \fnum \\
 \Rnum & $\rm e + N_2 \to e + N_2(a)$ & \fnum \\
 \Rnum & $\rm e + N_2 \to e + N_2(C)$ & \fnum \\
 \Rnum & $\rm e + N_2 \to e + N_2(E)$ & \fnum \\
 \Rnum & $\rm e + O_2 \to e + O_2(a)$ & \fnum \\
 \Rnum & $\rm e + O_2 \to e + O_2(b)$ & \fnum \\
 \Rnum & $\rm e + O_2 \to e + O_2(A)$ & \fnum \\
 \mr
 \multicolumn{3}{l}{(3) Electron dissociation} \\
 \Rnum & $\rm e + N_2 \to e + N(^4S) + N(^2D)$ & \fnum \\
 \Rnum & $\rm e + O_2 \to e + O(^3P) + O(^3P)$ & \fnum \\
 \Rnum & $\rm e + O_2 \to e + O(^3P) + O(^1D)$ & \fnum \\
 \Rnum & $\rm e + O_2 \to e + O(^3P) + O(^1S)$ & \fnum \\ 
 \mr
 \multicolumn{3}{l}{(4) Electron impact ionization} \\
 \Rnum & $\rm e + N_2 \to 2e + N_2^+$ & \fnum \\
 \Rnum & $\rm e + O_2 \to 2e + O_2^+$ & \fnum \\ 
 \Rnum & $\rm e + H_2O \to 2e + H_2O^+$ & \fnum \\ 
 \Rnum & $\rm e + H_2O \to 2e + H^+ + OH$ & \fnum \\ 
 \Rnum & $\rm e + H_2O \to 2e + OH^+ + H$ & \fnum \\
 \Rnum & $\rm e + H_2O \to 2e + O^+ + H_2$ & \fnum \\ 
 \Rnum & $\rm e + H_2O \to 2e + H_2^+ + O$ & \fnum \\ 
 \Rnum & $\rm e + H_2O \to 3e + O^{2+} + H_2$ & \fnum \\
 \mr
 \multicolumn{3}{l}{(5) Electron attachment} \\
 \Rnum & $\rm e + O_2 + O_2 \to O_2^- + O_2$ & \fnum \\
 \Rnum & $\rm e + O_2 \to O^- + O(^3P)$ & \fnum \\
 \Rnum & $\rm e + H_2O \to H^- + OH$ & \fnum \\
 \Rnum & $\rm e + H_2O \to OH^- + H$ & \fnum \\
 \Rnum & $\rm e + H_2O \to O^- + H_2$ & \fnum \\
 \Rnum & $\rm e + O_2 + H_2O \to O_2^- + H_2O$ & \fnum \ = $7 \times f_{25}$($E/N$) \\
 \mr
 \multicolumn{3}{l}{(6) Electron detachment} \\
 \Rnum & $\rm O_2^- + N_2 \to e + N_2 + O_2$ & $1.90\times10^{-12}(\frac{T}{300})^{0.5}\exp(\frac{-4990}{T})$ \\
 \Rnum & $\rm O_2^- + O_2 \to e + O_2 + O_2$ & $2.70\times10^{-10}(\frac{T}{300})^{0.5}\exp(\frac{-5590}{T})$ \\
 \Rnum & $\rm O_2^- + O_2(a) \to e + 2O_2$ & $2.00\times10^{-10}$ \\
 \Rnum & $\rm O_2^- + O(^3P) \to e + O_3$ & $3.00\times10^{-10}$ \\
 \Rnum & $\rm O_2^- + H \to e + HO_2$ & $1.20\times10^{-9}$ \\
 \Rnum & $\rm O^- + O_2(a) \to e + O_3$ & $3.00\times10^{-10}$ \\
 \br
\end{tabular*}
\end{table*}

\addtocounter{table}{-1}

\begin{table*}
\renewcommand{\arraystretch}{1.1}
\centering
\captionsetup{width=1\textwidth}
\caption{(Continued.)}
\begin{tabular*}{1\textwidth}{l@{\extracolsep{\fill}}ll}
 \br
 No. & Reaction & Reaction rate coefficient~($\mathrm{cm^3\,s^{-1}}$ or $\mathrm{cm^6\,s^{-1}}$) \\
 \mr
 \Rnum & $\rm O^- + O(^3P) \to e + O_2$ & $2.00\times10^{-10}$ \\
 \Rnum & $\rm H^- + O_2 \to e + HO_2$ & $1.20\times10^{-9}$ \\
 \Rnum & $\rm H^- + H \to e + H_2$ & $2.00\times10^{-9}$ \\
 \Rnum & $\rm OH^- + O(^3P) \to e + HO_2$ & $2.00\times10^{-10}$ \\
 \Rnum & $\rm OH^- + H \to e + H_2O$ & $1.40\times10^{-9}$ \\
 \mr
 \multicolumn{3}{l}{(7) Negative ion conversion} \\
 \Rnum & $\rm O_2^- + H \to OH^- + O(^3P)$ & $1.50\times10^{-9}$ \\
 \Rnum & $\rm H^- + H_2O \to OH^- + H_2$ & $1.00\times10^{-9}$ \\
 \mr
 \multicolumn{3}{l}{(8) Positive ion conversion} \\
 \Rnum & $\rm N_2^+ + N_2 + M \to N_4^+ + M$ & $5.20\times10^{-29}(\frac{300}{T})^{2.2}$ \\
 \Rnum & $\rm O_2^+ + N_2 + N_2 \to N_2O_2^+ + N_2$ & $9.00\times10^{-31}(\frac{300}{T})^{2}$ \\ 
 \Rnum & $\rm O_2^+ + O_2 + M \to O_4^+ + M$ & $2.40\times10^{-30}(\frac{300}{T})^{3.2}$ \\
 \Rnum & $\rm O_2^+ + H_2O + M \to O_2^+(H_2O) + M$ & $2.60\times10^{-28}(\frac{300}{T})^{4}$ \\
 \Rnum & $\rm N_4^+ + N_2 \to N_2^+ + 2N_2$ & $10^{-14.6+0.0036(T-300)}$ \\
 \Rnum & $\rm N_4^+ + O_2 \to O_2^+ + 2N_2$ & $2.50\times10^{-10}$ \\
 \Rnum & $\rm O_4^+ + N_2 \to N_2O_2^+ + O_2$ & $4.60\times10^{-12}(\frac{T}{300})^{2.5}\exp(\frac{-2650}{T})$ \\
 \Rnum & $\rm O_4^+ + O_2 \to O_2^+ + 2O_2$ & $3.30\times10^{-6}(\frac{300}{T})^{4}\exp(\frac{-5030}{T})$ \\
 \Rnum & $\rm O_4^+ + O_2(a) \to O_2^+ + 2O_2$ & $1.00\times10^{-10}$ \\
 \Rnum & $\rm O_4^+ + O_2(b) \to O_2^+ + 2O_2$ & $1.00\times10^{-10}$ \\
 \Rnum & $\rm O_4^+ + O(^3P) \to O_2^+ + O_3$ & $3.00\times10^{-10}$ \\
 \Rnum & $\rm O_4^+ + O(^1D) \to O_2^+ + O_3$ & $3.00\times10^{-10}$ \\
 \Rnum & $\rm O_4^+ + O(^1S) \to O_2^+ + O_3$ & $3.00\times10^{-10}$ \\
 \Rnum & $\rm O_4^+ + H_2O \to O_2^+(H_2O) + O_2$ & $1.70\times10^{-9}$ \\
 \Rnum & $\rm O_2^+(H_2O) + H_2O \to H_3O^+ + OH + O_2$ & $1.30\times10^{-9}$ \\
 \Rnum & $\rm N_2O_2^+ + N_2 \to O_2^+ + 2N_2$ & $1.10\times10^{-6}(\frac{300}{T})^{5.3}\exp(\frac{-2357}{T})$ \\ 
 \Rnum & $\rm N_2O_2^+ + O_2 \to O_4^+ + N_2$ & $1.00\times10^{-9}$ \\
 \Rnum & $\rm H_3O^+ + H_2O + M \to H_3O^+(H_2O) + M$ & $3.40\times10^{-27}(\frac{300}{T})^{4}$ \\
 \Rnum & $\rm H_3O^+(H_2O) + H_2O + M \to H_3O^+(H_2O)_2 + M$ & $2.30\times10^{-27}(\frac{300}{T})^{4}$ \\
 \Rnum & $\rm H_3O^+(H_2O)_2 + H_2O + M \to H_3O^+(H_2O)_3 + M$ & $2.40\times10^{-27}(\frac{300}{T})^{4}$ \\
 \mr
 \multicolumn{3}{l}{(9) Electron-ion recombination} \\
 \Rnum & $\rm e + N_2^+ \to N(^4S) + N(^4S)$ & $1.80\times10^{-7}(\frac{300}{T_e})^{0.39}$ \\
 \Rnum & $\rm e + N_2^+ \to N_2$ & $4.00\times10^{-12}$ \\
 \Rnum & $\rm e + N_4^+ \to N_2 + N_2$ & $2.00\times10^{-6}(\frac{300}{T_e})^{0.5}$ \\
 \Rnum & $\rm e + O_2^+ \to O(^3P) + O(^3P)$ & $2.00\times10^{-7}(\frac{300}{T_e})^{0.7}$ \\
 \Rnum & $\rm e + O_2^+ \to O_2$ & $4.00\times10^{-12}$ \\
 \Rnum & $\rm e + O_4^+ \to O_2 + O_2$ & $1.40\times10^{-6}(\frac{300}{T_e})^{0.5}$ \\
 \Rnum & $\rm e + N_2O_2^+ \to N_2 + O_2$ & $1.30\times10^{-6}(\frac{300}{T_e})^{0.5}$ \\
 \Rnum & $\rm e + H_2O^+ \to OH + H$ & $3.80\times10^{-7}$ \\
 \Rnum & $\rm e + H_2O^+ \to H_2 + O(^3P)$ & $1.40\times10^{-7}$ \\
 \Rnum & $\rm e + H_2O^+ \to 2H + O(^3P)$ & $1.70\times10^{-7}$ \\
 \Rnum & $\rm e + H_3O^+(H_2O)_3 \to H + 4H_2O$ & $5.50\times10^{-7}(\frac{300}{T_e})^{0.78}$ \\ 
 \mr
 \multicolumn{3}{l}{(10) Ion-ion recombination} \\ 
 \Rnum & $\rm N_2^+ + O^- \to N_2 + O(^3P)$ & $4.00\times10^{-7}$ \\
 \Rnum & $\rm N_2^+ + O_2^- \to N_2 + O_2$ & $1.60\times10^{-7}$ \\
 \br
\end{tabular*}
\end{table*}

\addtocounter{table}{-1}

\begin{table*}
\renewcommand{\arraystretch}{1.1}
\centering
\captionsetup{width=1\textwidth}
\caption{(Continued.)}
\begin{tabular*}{1\textwidth}{l@{\extracolsep{\fill}}ll}
 \br
 No. & Reaction & Reaction rate coefficient~($\mathrm{cm^3\,s^{-1}}$ or $\mathrm{cm^6\,s^{-1}}$) \\ 
 \mr
 \Rnum & $\rm O_2^+ + O^- \to O_2 + O(^3P)$ & $9.60\times10^{-8}$ \\
 \Rnum & $\rm O_2^+ + O_2^- \to O_2 + O_2$ & $4.20\times10^{-7}$ \\
 \Rnum & $\rm O_2^+ + O_2^- + O_2 \to 2O(^3P) + 2O_2$ & $2.00\times10^{-25}(\frac{300}{T})^{2.5}$ \\
 \Rnum & $\rm O_4^+ + O_2^- + O_2 \to 2O(^3P) + 3O_2$ & $2.00\times10^{-25}(\frac{300}{T})^{2.5}$ \\
 \Rnum & $\rm H_2O^+ + O^- \to H_2O + O(^3P)$ & $4.00\times10^{-7}$ \\
 \Rnum & $\rm H_2O^+ + O_2^- \to H_2O + O_2$ & $4.00\times10^{-7}$ \\
 \mr
 \multicolumn{3}{l}{(11) Neutral species conversion} \\
 \Rnum & $\rm N_2(A_1) + O_2 \to N_2 + O_2(b)$ & $7.50\times10^{-13}$ \\
 \Rnum & $\rm N_2(A_1) + O_2 \to N_2 + 2O(^3P)$ & $1.70\times10^{-12}$ \\
 \Rnum & $\rm N_2(A_1) + O_2 \to N_2O + O(^3P)$ & $7.80\times10^{-12}$ \\
 \Rnum & $\rm N_2(A_1) + N_2(A_1) \to N_2 + N_2(B)$ & $7.70\times10^{-11}$ \\
 \Rnum & $\rm N_2(A_1) + N_2(A_1) \to N_2 + N_2(C)$ & $1.60\times10^{-10}$ \\
 \Rnum & $\rm N_2(A_1) + N_2(A_1) \to N_2 + N_2(E)$ & $1.00\times10^{-11}$ \\
 \Rnum & $\rm N_2(A_1) + O(^3P) \to N_2 + O(^3P)$ & $2.00\times10^{-11}$ \\
 \Rnum & $\rm N_2(A_1) + O(^3P) \to N_2 + O(^1S)$ & $3.00\times10^{-11}$ \\
 \Rnum & $\rm N_2(A_1) + O(^3P) \to NO + N(^2D)$ & $7.00\times10^{-12}$ \\
 \Rnum & $\rm N_2(A_1) + H \to N_2 + H$ & $2.10\times10^{-10}$ \\
 \Rnum & $\rm N_2(A_1) + OH \to N_2 + OH$ & $1.00\times10^{-10}$ \\
 \Rnum & $\rm N_2(A_1) + H_2O \to N_2 + H + OH$ & $5.00\times10^{-14}$ \\
 \Rnum & $\rm N_2(A_1) + NO \to N_2 + NO(A)$ & $6.90\times10^{-11}$ \\
 \Rnum & $\rm N_2(A_2) + N_2 \to N_2 + N_2(A_1)$ & $1.00\times10^{-11}$ \\
 \Rnum & $\rm N_2(A_2) + O(^3P) \to N_2 + O(^3P)$ & $2.00\times10^{-11}$ \\
 \Rnum & $\rm N_2(A_2) + O(^3P) \to NO + N(^4S)$ & $7.00\times10^{-12}$ \\
 \Rnum & $\rm N_2(A_2) + H \to N_2 + H$ & $2.10\times10^{-10}$ \\
 \Rnum & $\rm N_2(A_2) + OH \to N_2 + OH$ & $1.00\times10^{-10}$ \\
 \Rnum & $\rm N_2(A_2) + H_2O \to N_2 + H + OH$ & $5.00\times10^{-14}$ \\
 \Rnum & $\rm N_2(A_2) + NO \to N_2 + NO(A)$ & $6.90\times10^{-11}$ \\
 \Rnum & $\rm N_2(B) + O_2 \to N_2 + 2O(^3P)$ & $3.00\times10^{-10}$ \\
 \Rnum & $\rm N_2(B) + N_2 \to N_2 + N_2(A_1)$ & $1.00\times10^{-11}$ \\
 \Rnum & $\rm N_2(B) \to N_2(A_1) + {\it h\nu}$ & $1.50\times10^{5}$ (s$^{-1}$) \\
 \Rnum & $\rm N_2(a) + O_2 \to N_2 + O(^3P) + O(^1D)$ & $2.80\times10^{-11}$ \\
 \Rnum & $\rm N_2(a) + N_2 \to N_2 + N_2$ & $2.00\times10^{-13}$ \\
 \Rnum & $\rm N_2(a) + N_2 \to N_2 + N_2(B)$ & $2.00\times10^{-13}$ \\
 \Rnum & $\rm N_2(a) + H_2O \to N_2 + H + OH$ & $3.00\times10^{-10}$ \\
 \Rnum & $\rm N_2(a) + H_2 \to N_2 + 2H$ & $2.60\times10^{-10}$ \\
 \Rnum & $\rm N_2(a) + NO \to N_2 + N(^4S) + O(^3P)$ & $3.60\times10^{-10}$ \\
 \Rnum & $\rm N_2(C) + O_2 \to N_2 + 2O(^3P)$ & $2.50\times10^{-10}$ \\
 \Rnum & $\rm N_2(C) + N_2 \to N_2 + N_2(B) $ & $1.00\times10^{-11}$ \\
 \Rnum & $\rm N_2(C) + N_2 \to N_2 + N_2(a)$ & $1.00\times10^{-11}$ \\
 \Rnum & $\rm N_2(C) \to N_2(B) + {\it h\nu}$ & $2.80\times10^{7}$ (s$^{-1}$) \\
 \Rnum & $\rm N_2(E) + N_2 \to N_2 + N_2(C)$ & $1.00\times10^{-10}$ \\
 \Rnum & $\rm NO(A) \to NO + {\it h\nu}$ & $5.10\times10^{6}$ (s$^{-1}$) \\
 \Rnum & $\rm N(^4S) + O_2 \to NO + O(^3P)$ & $4.47\times10^{-12}(\frac{T}{300})\exp(\frac{-3270.2}{T})$ \\
 \Rnum & $\rm N(^4S) + O(^3P) + N_2 \to N_2 + NO$ & $6.89\times10^{-33}\exp(\frac{134.7}{T})$ \\
 \Rnum & $\rm N(^4S) + N(^4S) + N_2 \to N_2 + N_2$ & $8.27\times10^{-34}\exp(\frac{500}{T})$ \\
 \Rnum & $\rm N(^4S) + OH \to NO + H$ & $3.80\times10^{-11}\exp(\frac{85.39}{T})$ \\
 \br
\end{tabular*}
\end{table*}

\addtocounter{table}{-1}

\begin{table*}
\renewcommand{\arraystretch}{1.1}
\centering
\captionsetup{width=1\textwidth}
\caption{(Continued.)}
\begin{tabular*}{1\textwidth}{l@{\extracolsep{\fill}}ll}
 \br
 No. & Reaction & Reaction rate coefficient~($\mathrm{cm^3\,s^{-1}}$ or $\mathrm{cm^6\,s^{-1}}$) \\ 
 \mr
 \Rnum & $\rm N(^4S) + HO_2 \to NO + OH$ & $2.19\times10^{-11}$ \\
 \Rnum & $\rm N(^4S) + NO \to N_2 + O(^3P)$ & $3.51\times10^{-11}\exp(\frac{-49.84}{T})$ \\
 \Rnum & $\rm N(^4S) + NO_2 \to NO + NO$ & $2.30\times10^{-12}$ \\
 \Rnum & $\rm N(^4S) + NO_2 \to N_2O + O(^3P)$ & $5.80\times10^{-12}\exp(\frac{-220}{T})$ \\
 \Rnum & $\rm N(^2D) + O_2 \to NO + O(^3P)$ & $1.50\times10^{-12}(\frac{T}{300})^{0.5}$ \\
 \Rnum & $\rm N(^2D) + O_2 \to NO + O(^1D)$ & $9.70\times10^{-12}\exp(\frac{185}{T})$ \\
 \Rnum & $\rm N(^2D) + N_2 \to N_2 + N(^4S)$ & $1.70\times10^{-14}$ \\
 \Rnum & $\rm N(^2D) + O(^3P) \to N(^4S) + O(^3P)$ & $3.30\times10^{-12}\exp(\frac{260}{T})$ \\
 \Rnum & $\rm N(^2D) + H_2O \to NH + OH$ & $4.00\times10^{-11}$ \\
 \Rnum & $\rm N(^2D) + NO \to N_2 + O(^3P)$ & $1.80\times10^{-10}$ \\
 \Rnum & $\rm O_2(a) + O_2 \to O_2 + O_2$ & $2.20\times10^{-18}(\frac{T}{300})^{0.8}$ \\
 \Rnum & $\rm O_2(a) + N_2 \to N_2 + O_2$ & $1.40\times10^{-19}$ \\
 \Rnum & $\rm O_2(a) + O(^3P) \to O_2 + O(^3P)$ & $7.00\times10^{-16}$ \\
 \Rnum & $\rm O_2(a) + N(^4S) \to NO + O(^3P)$ & $2.00\times10^{-14}$ \\
 \Rnum & $\rm O_2(a) + NO \to O_2 + NO$ & $2.50\times10^{-11}$ \\
 \Rnum & $\rm O_2(a) + NO \to NO_2 + O(^3P)$ & $3.49\times10^{-17}$ \\
 \Rnum & $\rm O_2(b) + O_2 \to O_2 + O_2(a)$ & $4.10\times10^{-17}$ \\
 \Rnum & $\rm O_2(b) + N_2 \to N_2 + O_2(a)$ & $2.10\times10^{-15}$ \\
 \Rnum & $\rm O_2(b) + O(^3P) \to O_2 + O(^3P)$ & $8.00\times10^{-14}$ \\
 \Rnum & $\rm O_2(b) + H_2O \to O_2 + H_2O$ & $4.60\times10^{-12}$ \\
 \Rnum & $\rm O_2(b) + NO \to O_2(a) + NO$ & $4.00\times10^{-14}$ \\
 \Rnum & $\rm O_2(b) + O_3 \to 2O_2(a) + O(^3P)$ & $1.80\times10^{-11}$ \\
 \Rnum & $\rm O_2(A) + O_2 \to O_2(b) + O_2(b)$ & $2.90\times10^{-13}$ \\
 \Rnum & $\rm O_2(A) + N_2 \to N_2 + O_2(b)$ & $3.00\times10^{-13}$ \\
 \Rnum & $\rm O_2(A) + O(^3P) \to O_2(b) + O(^1D)$ & $9.00\times10^{-12}$ \\
 \Rnum & $\rm O(^3P) + O_2 + O_2 \to O_2 + O_3$ & $6.01\times10^{-34}(\frac{300}{T})^{2.6}$ \\
 \Rnum & $\rm O(^3P) + O_2 + N_2 \to N_2 + O_3$ & $5.51\times10^{-34}(\frac{300}{T})^{2.6}$ \\
 \Rnum & $\rm O(^3P) + O(^3P) + O_2 \to O_2 + O_2$ & $3.81\times10^{-33}(\frac{300}{T})^{0.63}$ \\
 \Rnum & $\rm O(^3P) + O(^3P) + N_2 \to N_2 + O_2$ & $9.46\times10^{-34}\exp(\frac{484.7}{T})$ \\
 \Rnum & $\rm O(^3P) + OH \to O_2 + H$ & $2.40\times10^{-11}\exp(\frac{110}{T})$ \\
 \Rnum & $\rm O(^3P) + HO_2 \to O_2 + OH$ & $2.70\times10^{-11}\exp(\frac{224}{T})$ \\
 \Rnum & $\rm O(^3P) + NO + N_2 \to N_2 + NO_2$ & $1.03\times10^{-30}(\frac{300}{T})^{2.87}\exp(\frac{780.5}{T})$ \\
 \Rnum & $\rm O(^3P) + NO_2 \to O_2 + NO$ & $5.50\times10^{-12}\exp(\frac{-187.9}{T})$ \\
 \Rnum & $\rm O(^3P) + O_3 \to O_2 + O_2$ & $8.00\times10^{-12}\exp(\frac{-2060}{T})$ \\
 \Rnum & $\rm O(^1D) + O_2 \to O_2 + O(^3P)$ & $3.12\times10^{-11}\exp(\frac{70}{T})$ \\
 \Rnum & $\rm O(^1D) + N_2 \to N_2 + O(^3P)$ & $2.10\times10^{-11}\exp(\frac{115}{T})$ \\
 \Rnum & $\rm O(^1D) + H_2O \to O_2 + H_2$ & $3.57\times10^{-10}$ \\
 \Rnum & $\rm O(^1D) + H_2O \to OH + OH$ & $2.20\times10^{-10}$ \\
 \Rnum & $\rm O(^1D) + H_2 \to H + OH$ & $1.10\times10^{-10}$ \\
 \Rnum & $\rm O(^1D) + H_2O_2 \to O_2 + H_2O$ & $5.20\times10^{-10}$ \\
 \Rnum & $\rm O(^1D) + O_3 \to O_2 + 2O(^3P)$ & $1.20\times10^{-10}$ \\
 \Rnum & $\rm O(^1S) + H_2O \to O_2 + H_2$ & $5.00\times10^{-10}$ \\
 \Rnum & $\rm O(^1S) + H_2O \to H_2O + O(^3P)$ & $3.00\times10^{-10}$ \\
 \Rnum & $\rm O(^1S) + H_2O \to OH + OH$ & $5.00\times10^{-10}$ \\
 \Rnum & $\rm O_3 + H \to O_2 + OH$ & $1.40\times10^{-10}\exp(\frac{-480}{T})$ \\
 \Rnum & $\rm O_3 + OH \to O_2 + HO_2$ & $1.70\times10^{-12}\exp(\frac{-940}{T})$ \\
 \br
\end{tabular*}
\end{table*}

\addtocounter{table}{-1}

\begin{table*}
\renewcommand{\arraystretch}{1.1}
\centering
\captionsetup{width=1\textwidth}
\caption{(Continued.)}
\begin{tabular*}{1\textwidth}{l@{\extracolsep{\fill}}ll}
 \br
 No. & Reaction & Reaction rate coefficient~($\mathrm{cm^3\,s^{-1}}$ or $\mathrm{cm^6\,s^{-1}}$) \\ 
 \mr
 \Rnum & $\rm O_3 + NO \to O_2 + NO_2$ & $3.16\times10^{-12}\exp(\frac{1563}{T})$ \\
 \Rnum & $\rm O_3 + O_3 \to O_2 + O_3 + O(^3P)$ & $7.16\times10^{-10}\exp(\frac{-11200}{T})$ \\
 \Rnum & $\rm H + O_2 + O_2 \to O_2 + HO_2$ & $5.94\times10^{-32}(\frac{300}{T})$ \\
 \Rnum & $\rm H + O_2 + N_2 \to N_2 + HO_2$ & $5.94\times10^{-32}(\frac{300}{T})$ \\
 \Rnum & $\rm H + OH + N_2 \to N_2 + H_2O$ & $6.87\times10^{-31}(\frac{300}{T})^{2}$ \\
 \Rnum & $\rm H + OH + H_2O \to H_2O + H_2O$ & $4.38\times10^{-31}(\frac{300}{T})^{2}$ \\
 \Rnum & $\rm H + HO_2 \to O_2 + H_2$ & $1.75\times10^{-10}\exp(\frac{-1030}{T})$ \\
 \Rnum & $\rm H + HO_2 \to H_2O + O(^3P)$ & $5.00\times10^{-11}\exp(\frac{-866}{T})$ \\
 \Rnum & $\rm H + HO_2 \to OH + OH$ & $7.40\times10^{-10}\exp(\frac{-700}{T})$ \\
 \Rnum & $\rm OH + OH \to H_2O + O(^3P)$ & $6.20\times10^{-14}(\frac{T}{300})^{2.6}\exp(\frac{945}{T})$ \\
 \Rnum & $\rm OH + OH \to H_2O_2$ & $2.60\times10^{-11}$ \\
 \Rnum & $\rm OH + OH + O_2 \to O_2 + H_2O_2$ & $6.05\times10^{-31}(\frac{300}{T})^{3}$ \\
 \Rnum & $\rm OH + OH + N_2 \to N_2 + H_2O_2$ & $6.90\times10^{-31}(\frac{300}{T})^{0.8}$ \\
 \Rnum & $\rm OH + OH + H_2O \to H_2O + H_2O_2$ & $1.54\times10^{-31}(\frac{300}{T})^{2}\exp(\frac{183.6}{T})$ \\
 \Rnum & $\rm OH + HO_2 \to O_2 + H_2O$ & $4.80\times10^{-11}\exp(\frac{250}{T})$ \\
 \Rnum & $\rm OH + NO + O_2 \to O_2 + HNO_2$ & $7.40\times10^{-31}(\frac{300}{T})^{2.4}$ \\
 \Rnum & $\rm OH + NO + N_2 \to N_2 + HNO_2$ & $7.40\times10^{-31}(\frac{300}{T})^{2.4}$ \\
 \Rnum & $\rm OH + NO_2 + O_2 \to O_2 + HNO_3$ & $2.20\times10^{-30}(\frac{300}{T})^{2.9}$ \\
 \Rnum & $\rm OH + NO_2 + N_2 \to N_2 + HNO_3$ & $2.60\times10^{-30}(\frac{300}{T})^{2.9}$ \\
 \Rnum & $\rm HO_2 + NO \to OH + NO_2$ & $3.60\times10^{-12}\exp(\frac{269.4}{T})$ \\
 \Rnum & $\rm HO_2 + NO_2 \to O_2 + HNO_2$ & $1.20\times10^{-13}$ \\
 \Rnum & $\rm HO_2 + HO_2 \to O_2 + H_2O_2$ & $2.20\times10^{-19}\exp(\frac{600.2}{T})$ \\
 \Rnum & $\rm HO_2 + HO_2 + O_2 \to 2O_2 + H_2O_2$ & $1.90\times10^{-33}\exp(\frac{980}{T})$ \\
 \Rnum & $\rm HO_2 + HO_2 + N_2 \to O_2 + N_2 + H_2O_2$ & $1.90\times10^{-33}\exp(\frac{980}{T})$ \\
 \br
\end{tabular*}
\begin{tabular*}{1\textwidth}{@{\extracolsep{\fill}}l}
1. In~\cite{komuro2013}, H$_2$O($v_1$), H$_2$O($v_2$) and H$_2$O($v_3$) were included.
However, in Kawaguchi H$_2$O cross sections the \\
vibrational excited state was only separated into H$_2$O($v_1$) and H$_2$O($v_2$), and here we merge them into H$_2$O($v$). \\
2. Here we do not include electron dissociation reactions for H$_2$O (i.e., R24--R25 in table 1 of~\cite{komuro2013}), as Kawaguchi \\
H$_2$O cross sections did not specify them. \\
3. In~\cite{komuro2018a}, a total ionization reaction like R19 was used for H$_2$O. Here more ionization reaction channels (R19--R24) \\
are included when using Kawaguchi H$_2$O cross sections. \\
4. Here N(S) is replaced by N($^4$S), O(P) and O$_2$(P) by O($^3$P), O(D) and O($^3$D) by O($^1$D), and O$^-$($^2$P) by O$^-$. \\
5. Here we do not include R88 ($\rm N_2^+(B) \to N_2^+ + {\it h\nu}$) in table 3 of~\cite{komuro2013}. \\
6. Rate coefficients for R81--R82 have been corrected from $4.0\times10^{-6}$ to $4.0\times10^{-7}$ by checking their cited paper~\cite{eichwald1997}. \\
7. In~\cite{komuro2013}, the rate coefficient for R120 was $1.38\times10^{-33}\exp(\frac{-502.9}{T})$, and we have revised it to $8.27\times10^{-34}\exp(\frac{500}{T})$ \\ according to~\cite{kossyi1992}. \\
\end{tabular*}
\end{table*}

\begin{table}
\renewcommand{\arraystretch}{1.1}
\centering
\caption{Excited states of N$_2$, O$_2$ and H$_2$O with activation energies and the corresponding effective states used in table~\ref{tab:chemistry-set-komuro}. 
The table is partially based on table 4 of~\cite{komuro2013}.}
\label{tab:excited-states-komuro}
\begin{tabular*}{0.48\textwidth}{l@{\extracolsep{\fill}}cc}
 \br
 \multirow{2}{*}{Excited state} & Activation & \multirow{2}{*}{Effective state} \\ & energy $\epsilon_e$~(eV) \\ 
 \mr
 N$_2$($X$, $v=1$) & 0.29 & N$_2(v)$ \\
 N$_2$($X$, $v=2$) & 0.59 & N$_2(v)$ \\
 N$_2$($X$, $v=3$) & 0.88 & N$_2(v)$ \\
 N$_2$($X$, $v=4$) & 1.17 & N$_2(v)$ \\
 N$_2$($X$, $v=5$) & 1.47 & N$_2(v)$ \\
 N$_2$($X$, $v=6$) & 1.76 & N$_2(v)$ \\
 N$_2$($X$, $v=7$) & 2.06 & N$_2(v)$ \\
 N$_2$($X$, $v=8$) & 2.35 & N$_2(v)$ \\
 N$_2$($A^3\Sigma_u^+$, $v = 0...4$) & 6.17 & N$_2$(A$_1$) \\
 N$_2$($A^3\Sigma_u^+$, $v = 5...9$) & 7.00 & N$_2$(A$_2$) \\
 N$_2$($B^3\Pi_g$) & 7.35 & N$_2$(B) \\
 N$_2$($W^3\Delta_u$) & 7.36 & N$_2$(B) \\
 N$_2$($A^3\Sigma_u^+$, $v>10$) & 7.80 & N$_2$(B) \\
 N$_2$($B'^3\Sigma_u^-$) & 8.16 & N$_2$(B) \\
 N$_2$($a'^1\Sigma_u^-$) & 8.40 & N$_2$(a) \\
 N$_2$($a^1\Pi_g$) & 8.55 & N$_2$(a) \\
 N$_2$($w^1\Delta_u$) & 8.89 & N$_2$(a) \\
 N$_2$($C^3\Pi_u$) & 11.03 & N$_2$(C) \\
 N$_2$($E^3\Sigma_g^+$) & 11.87 & N$_2$(E) \\
 N$_2$($a''^1\Sigma_g^+$) & 12.25 & N$_2$(E) \\
 O$_2$($X$, $v=1$) & 0.19 & O$_2(v)$ \\
 O$_2$($X$, $v=2$) & 0.38 & O$_2(v)$ \\
 O$_2$($X$, $v=3$) & 0.57 & O$_2(v)$ \\
 O$_2$($X$, $v=4$) & 0.75 & O$_2(v)$ \\
 O$_2$($a^1\Delta_g$) & 0.977 & O$_2$(a) \\
 O$_2$($b^1\Sigma_g^+$) & 1.627 & O$_2$(b) \\
 O$_2$($c^1\Sigma_u^-$) & 4.05 & O$_2$(A) \\ 
 O$_2$($A'^3\Delta_u$) & 4.26 & O$_2$(A) \\
 O$_2$($A^3\Sigma_u^+$) & 4.34 & O$_2$(A) \\ 
 H$_2$O($X$, $v=1$) & 0.198 & H$_2$O($v$) \\
 H$_2$O($X$, $v=2$) & 0.466 & H$_2$O($v$) \\
 \br
\end{tabular*}
\end{table}

\clearpage
\section*{References}
\normalem
\bibliography{references}

\begin{thebibliography}{10}

\bibitem{nijdam2020a}
Sander Nijdam, Jannis Teunissen, and Ute Ebert.
\newblock The physics of streamer discharge phenomena.
\newblock {\em Plasma Sources Science and Technology}, 29(10):103001, November
  2020.

\bibitem{gallimberti1979}
I.~Gallimberti.
\newblock The mechanism of the long spark formation.
\newblock {\em Le Journal de Physique Colloques}, 40(C7):C7--193--C7--250, July
  1979.

\bibitem{malagon-romero2019}
A.~{Malag{\'o}n-Romero} and A.~Luque.
\newblock Spontaneous {{Emergence}} of {{Space Stems Ahead}} of {{Negative
  Leaders}} in {{Lightning}} and {{Long Sparks}}.
\newblock {\em Geophysical Research Letters}, 46(7):4029--4038, 2019.

\bibitem{gordillo-vazquez2021/atmre}
F.~J. {Gordillo-V{\'a}zquez} and F.~J. {P{\'e}rez-Invern{\'o}n}.
\newblock {A review of the impact of transient luminous events on the
  atmospheric chemistry: Past, present, and future}.
\newblock {\em Atmospheric Research}, 252:105432, April 2021.

\bibitem{wang2020a}
Douyan Wang and Takao Namihira.
\newblock Nanosecond pulsed streamer discharges: {{II}}. {{Physics}}, discharge
  characterization and plasma processing.
\newblock {\em Plasma Sources Science and Technology}, 29(2):023001, February
  2020.

\bibitem{vonwoedtke2020}
Thomas {von Woedtke}, Steffen Emmert, Hans-Robert Metelmann, Stefan Rupf, and
  Klaus-Dieter Weltmann.
\newblock Perspectives on cold atmospheric plasma ({{CAP}}) applications in
  medicine.
\newblock {\em Physics of Plasmas}, 27(7):070601, July 2020.

\bibitem{attri2020plasma}
Pankaj Attri, Kenji Ishikawa, Takamasa Okumura, Kazunori Koga, and Masaharu
  Shiratani.
\newblock Plasma agriculture from laboratory to farm: A review.
\newblock {\em Processes}, 8(8):1002, 2020.

\bibitem{polonskyi2021}
O.~Polonskyi, T.~Hartig, J.~R. Uzarski, and M.~J. Gordon.
\newblock Precise localization of {{DBD}} plasma streamers using
  topographically patterned insulators for maskless structural and chemical
  modification of surfaces.
\newblock {\em Applied Physics Letters}, 119(21):211601, November 2021.

\bibitem{kim2004}
Hyun-Ha Kim.
\newblock Nonthermal plasma processing for air-pollution control: A historical
  review, current issues, and future prospects.
\newblock {\em Plasma Processes and Polymers}, 1(2):91--110, September 2004.

\bibitem{starikovskaia2014}
S~M Starikovskaia.
\newblock Plasma-assisted ignition and combustion: Nanosecond discharges and
  development of kinetic mechanisms.
\newblock {\em Journal of Physics D: Applied Physics}, 47(35):353001, September
  2014.

\bibitem{zhang2019c}
Cheng Zhang, Bangdou Huang, Zhenbing Luo, Xueke Che, Ping Yan, and Tao Shao.
\newblock Atmospheric-pressure pulsed plasma actuators for flow control: Shock
  wave and vortex characteristics.
\newblock {\em Plasma Sources Science and Technology}, 28(6):064001, May 2019.

\bibitem{nijdam2014a}
S~Nijdam, E~Takahashi, A~H Markosyan, and U~Ebert.
\newblock Investigation of positive streamers by double-pulse experiments,
  effects of repetition rate and gas mixture.
\newblock {\em Plasma Sources Science and Technology}, 23(2):025008, March
  2014.

\bibitem{bouwman2022}
Dennis Bouwman, Jannis Teunissen, and Ute Ebert.
\newblock {{3D}} particle simulations of positive air\textendash methane
  streamers for combustion.
\newblock {\em Plasma Sources Science and Technology}, 31(4):045023, April
  2022.

\bibitem{li2022a}
Xiaoran Li, Baohong Guo, Anbang Sun, Ute Ebert, and Jannis Teunissen.
\newblock A computational study of steady and stagnating positive streamers in
  {{N}}{\textsubscript{2}}\textendash {{O}}{\textsubscript{2}} mixtures.
\newblock {\em Plasma Sources Science and Technology}, 31(6):065011, June 2022.

\bibitem{guo2023f}
Baohong Guo, Ute Ebert, and Jannis Teunissen.
\newblock {{3D}} particle-in-cell simulations of negative and positive
  streamers in
  {{C}}{\textsubscript{4}}{{F}}{\textsubscript{7}}{{N}}--{{CO}}{\textsubscript{2}}
  mixtures.
\newblock {\em Plasma Sources Science and Technology}, 32(11):115001, November
  2023.

\bibitem{phelps1976}
C.~T. Phelps and R.~F. Griffiths.
\newblock Dependence of positive corona streamer propagation on air pressure
  and water vapor content.
\newblock {\em Journal of Applied Physics}, 47(7):2929--2934, July 1976.

\bibitem{griffiths1976a}
R.~F. Griffiths and C.~T. Phelps.
\newblock The effects of air pressure and water vapour content on the
  propagation of positive corona streamers, and their implications to lightning
  initiation: {{CORONA STREAMERS}}.
\newblock {\em Quarterly Journal of the Royal Meteorological Society},
  102(432):419--426, April 1976.

\bibitem{lesrenardiers1977/elektra}
Les~Renardi\`{e}res group.
\newblock Positive discharges in long air gaps at les renardi\`{e}res, 1975
  results and conclusions.
\newblock {\em Elektra}, 53:31--153, 1978.

\bibitem{lesrenardiers1978/elektra}
Les~Renardi\`{e}res group.
\newblock Negative discharges in long air gaps at les renardi\`{e}res.
\newblock {\em Elektra}, 74:67--216, 1978.

\bibitem{allen1991}
N.L. Allen and M.~Boutlendj.
\newblock Study of the electric fields required for streamer propagation in
  humid air.
\newblock {\em IEE Proceedings A Science, Measurement and Technology},
  138(1):37, 1991.

\bibitem{hui2008}
Jianfeng Hui, Zhicheng Guan, Liming {wang}, and Qiuwei {li}.
\newblock Variation of the {{Dynamics}} of {{Positive Streamer}} with
  {{Pressure}} and {{Humidity}} in {{Air}}.
\newblock {\em IEEE Transactions on Dielectrics and Electrical Insulation},
  15(2):382--389, April 2008.

\bibitem{mikropoulos2008}
P.~Mikropoulos, C.~Stassinopoulos, and B.~Sarigiannidou.
\newblock Positive {{Streamer Propagation}} and {{Breakdown}} in {{Air}}: The
  {{Influence}} of {{Humidity}}.
\newblock {\em IEEE Transactions on Dielectrics and Electrical Insulation},
  15(2):416--425, April 2008.

\bibitem{zhao2023}
Zheng Zhao, Qiuyu Gao, Xinlei Zheng, Haowei Zhang, Haotian Zheng, Anbang Sun,
  and Jiangtao Li.
\newblock Evolutions of streamer dynamics and discharge instabilities under
  repetitive pulses in humid air.
\newblock {\em Plasma Sources Science and Technology}, 32(12):125011, December
  2023.

\bibitem{zhao2024}
Zheng Zhao, Qiuyu Gao, Haowei Zhang, Haotian Zheng, Xinlei Zheng, Zihan Sun,
  Anbang Sun, and Jiangtao Li.
\newblock Effects of {{DC}} bias on evolutions of repetitively pulsed streamer
  discharge in humid air.
\newblock {\em Journal of Physics D: Applied Physics}, 57(25):255206, June
  2024.

\bibitem{ono2003a}
Ryo Ono and Tetsuji Oda.
\newblock Dynamics of ozone and {{OH}} radicals generated by pulsed corona
  discharge in humid-air flow reactor measured by laser spectroscopy.
\newblock {\em Journal of Applied Physics}, 93(10):5876--5882, May 2003.

\bibitem{nakagawa2011}
Yusuke Nakagawa, Ryo Ono, and Tetsuji Oda.
\newblock Density and temperature measurement of {{OH}} radicals in
  atmospheric-pressure pulsed corona discharge in humid air.
\newblock {\em Journal of Applied Physics}, 110(7):073304, October 2011.

\bibitem{singleton2016}
Daniel Singleton, Campbell Carter, Scott~J. Pendleton, Christopher Brophy, Jose
  Sinibaldi, John~W. Luginsland, Michael Brown, Emanuel Stockman, and Martin~A.
  Gundersen.
\newblock The effect of humidity on hydroxyl and ozone production by nanosecond
  discharges.
\newblock {\em Combustion and Flame}, 167:164--171, May 2016.

\bibitem{ono2010a}
Ryo Ono, Yoshiyuki Teramoto, and Tetsuji Oda.
\newblock Effect of humidity on gas temperature in the afterglow of pulsed
  positive corona discharge.
\newblock {\em Plasma Sources Science and Technology}, 19(1):015009, January
  2010.

\bibitem{komuro2014a}
Atsushi Komuro and Ryo Ono.
\newblock Two-dimensional simulation of fast gas heating in an atmospheric
  pressure streamer discharge and humidity effects.
\newblock {\em Journal of Physics D: Applied Physics}, 47(15):155202, April
  2014.

\bibitem{komuro2015a}
Atsushi Komuro, Kazunori Takahashi, and Akira Ando.
\newblock Vibration-to-translation energy transfer in atmospheric-pressure
  streamer discharge in dry and humid air.
\newblock {\em Plasma Sources Science and Technology}, 24(5):055020, September
  2015.

\bibitem{aleksandrov1998}
N.~L. Aleksandrov, {\'E}.~M. Bazelyan, and D.~A. Novitskii.
\newblock Influence of moisture on the properties of long streamers in air.
\newblock {\em Technical Physics Letters}, 24(5):367--368, May 1998.

\bibitem{komuro2018a}
Atsushi Komuro, Shuto Matsuyuki, and Akira Ando.
\newblock Simulation of pulsed positive streamer discharges in air at high
  temperatures.
\newblock {\em Plasma Sources Science and Technology}, 27(10):105001, October
  2018.

\bibitem{malagon-romero2022}
Alejandro {Malag{\'o}n-Romero} and Alejandro Luque.
\newblock Streamer propagation in humid air.
\newblock {\em Plasma Sources Science and Technology}, 31(10):105010, October
  2022.

\bibitem{starikovskiy2022}
A~Yu Starikovskiy, E~M Bazelyan, and N~L Aleksandrov.
\newblock The influence of humidity on positive streamer propagation in long
  air gap.
\newblock {\em Plasma Sources Science and Technology}, 31(11):114009, November
  2022.

\bibitem{ren2022}
Xiaodong Ren, Xingliang Jiang, Guolin Yang, Yafei Huang, Jianguo Wu, and
  Zhongyi Yang.
\newblock Effect of {{Environmental Parameters}} on {{Streamer Discharge}} in
  {{Short Air Gap}} between {{Rod}} and {{Plate}}.
\newblock {\em Energies}, 15(3):817, January 2022.

\bibitem{chen2023}
She Chen, Tianwei Wang, Zezhong Wang, Ziyu Yan, Lipeng Zhong, Qiuqin Sun, and
  Feng Wang.
\newblock Experimental {{Study}} and {{3D Modeling}} of {{Positive Streamer
  Discharges}} under the {{Effect}} of {{Humidity}}.
\newblock {\em IEEE Transactions on Dielectrics and Electrical Insulation},
  pages 1--1, 2023.

\bibitem{chen2018a}
Xiaoyue Chen, Wangling He, Xinyu Du, Xiaoqing Yuan, Lei Lan, Xishan Wen, and
  Baoquan Wan.
\newblock Electron swarm parameters and {{Townsend}} coefficients of
  atmospheric corona discharge plasmas by considering humidity.
\newblock {\em Physics of Plasmas}, 25(6):063525, June 2018.

\bibitem{komuro2013}
Atsushi Komuro, Ryo Ono, and Tetsuji Oda.
\newblock Behaviour of {{OH}} radicals in an atmospheric-pressure streamer
  discharge studied by two-dimensional numerical simulation.
\newblock {\em Journal of Physics D: Applied Physics}, 46(17):175206, May 2013.

\bibitem{liu2017}
Lipeng Liu and Marley Becerra.
\newblock Gas heating dynamics during leader inception in long air gaps at
  atmospheric pressure.
\newblock {\em Journal of Physics D: Applied Physics}, 50(34):345202, August
  2017.

\bibitem{aleksandrov1999}
N~L Aleksandrov and E~M Bazelyan.
\newblock Ionization processes in spark discharge plasmas.
\newblock {\em Plasma Sources Science and Technology}, 8(2):285--294, May 1999.

\bibitem{sieck2000}
L.~Wayne Sieck, John~T. Heron, and David~S. Green.
\newblock Chemical {{Kinetics Database}} and {{Predictive Schemes}} for {{Humid
  Air Plasma Chemistry}}. {{Part I}}: {{Positive Ion-Molecule Reactions}}.
\newblock {\em Plasma Chemistry and Plasma Processing}, 20(2):235--258, 2000.

\bibitem{herron2001}
John~T. Herron and David~S. Green.
\newblock Chemical {{Kinetics Database}} and {{Predictive Schemes}} for
  {{Nonthermal Humid Air Plasma Chemistry}}. {{Part II}}. {{Neutral Species
  Reactions}}.
\newblock {\em Plasma Chemistry and Plasma Processing}, 21(3):459--481, 2001.

\bibitem{aleksandrov2022}
N~L Aleksandrov, E~M Bazelyan, A~A Ponomarev, and A~Yu Starikovsky.
\newblock Kinetics of charged species in non-equilibrium plasma in water vapor-
  and hydrocarbon-containing gaseous mixtures.
\newblock {\em Journal of Physics D: Applied Physics}, 55(38):383002, September
  2022.

\bibitem{teunissen2017}
Jannis Teunissen and Ute Ebert.
\newblock Simulating streamer discharges in {{3D}} with the parallel adaptive
  {{Afivo}} framework.
\newblock {\em Journal of Physics D: Applied Physics}, 50(47):474001, November
  2017.

\bibitem{bagheri2018}
B~Bagheri, J~Teunissen, U~Ebert, M~M Becker, S~Chen, O~Ducasse, O~Eichwald,
  D~Loffhagen, A~Luque, D~Mihailova, J~M Plewa, J~{van Dijk}, and M~Yousfi.
\newblock Comparison of six simulation codes for positive streamers in air.
\newblock {\em Plasma Sources Science and Technology}, 27(9):095002, September
  2018.

\bibitem{hagelaar2005}
G~J~M Hagelaar and L~C Pitchford.
\newblock Solving the {{Boltzmann}} equation to obtain electron transport
  coefficients and rate coefficients for fluid models.
\newblock {\em Plasma Sources Science and Technology}, 14(4):722--733, November
  2005.

\bibitem{malla2023}
H~Malla, A~Martinez, U~Ebert, and J~Teunissen.
\newblock Double-pulse streamer simulations for varying interpulse times in
  air.
\newblock {\em Plasma Sources Science and Technology}, 32(9):095006, September
  2023.

\bibitem{teunissen2018}
Jannis Teunissen and Ute Ebert.
\newblock Afivo: {{A}} framework for quadtree/octree {{AMR}} with shared-memory
  parallelization and geometric multigrid methods.
\newblock {\em Computer Physics Communications}, 233:156--166, December 2018.

\bibitem{teunissen2023}
Jannis Teunissen and Francesca Schiavello.
\newblock Geometric multigrid method for solving {{Poisson}}'s equation on
  octree grids with irregular boundaries.
\newblock {\em Computer Physics Communications}, 286:108665, May 2023.

\bibitem{guo2023}
Baohong Guo and Jannis Teunissen.
\newblock A computational study on the energy efficiency of species production
  by single-pulse streamers in air.
\newblock {\em Plasma Sources Science and Technology}, 32(2):025001, February
  2023.

\bibitem{li2021a}
Xiaoran Li, Siebe Dijcks, Sander Nijdam, Anbang Sun, Ute Ebert, and Jannis
  Teunissen.
\newblock Comparing simulations and experiments of positive streamers in air:
  Steps toward model validation.
\newblock {\em Plasma Sources Science and Technology}, 30(9):095002, September
  2021.

\bibitem{nijdam2011}
S~Nijdam, G~Wormeester, E~M {van Veldhuizen}, and U~Ebert.
\newblock Probing background ionization: Positive streamers with varying pulse
  repetition rate and with a radioactive admixture.
\newblock {\em Journal of Physics D: Applied Physics}, 44(45):455201, November
  2011.

\bibitem{chemistry_phelps_database}
Phelps database ({N}\textsubscript{2}, {O}\textsubscript{2})
  \url{www.lxcat.net} (retrieved 13 October 2023).

\bibitem{phelps1985}
A.~V. Phelps and L.~C. Pitchford.
\newblock Anisotropic scattering of electrons by {{N}}{\textsubscript{{2}}} and
  its effect on electron transport.
\newblock {\em Physical Review A}, 31(5):2932--2949, May 1985.

\bibitem{lawton1978}
S.~A. Lawton and A.~V. Phelps.
\newblock Excitation of the {b}\textsuperscript{1}{$\Sigma$}$_g^+$ state of
  {O}\textsubscript{2} by low energy electrons.
\newblock {\em The Journal of Chemical Physics}, 69(3):1055, August 1978.

\bibitem{DeUrquijo_2014}
J.~De~Urquijo, E.~Basurto, A.~M. Ju{\'a}rez, K.~F. Ness, R.~E. Robson, M.~J.
  Brunger, and R.~D. White.
\newblock Electron drift velocities in {{He}} and water mixtures:
  {{Measurements}} and an assessment of the water vapour cross-section sets.
\newblock {\em The Journal of Chemical Physics}, 141(1):014308, July 2014.

\bibitem{Yousfi_1996}
M.~Yousfi and M.~D. Benabdessadok.
\newblock Boltzmann equation analysis of electron-molecule collision cross
  sections in water vapor and ammonia.
\newblock {\em Journal of Applied Physics}, 80(12):6619--6630, December 1996.

\bibitem{Budde_2022}
Maik Budde, Tiago Cunha~Dias, Luca Vialetto, Nuno Pinh{\~a}o, Vasco Guerra, and
  Tiago Silva.
\newblock Electron-neutral collision cross sections for {{H}}
  {\textsubscript{2}} {{O}}: {{I}}. {{Complete}} and consistent set.
\newblock {\em Journal of Physics D: Applied Physics}, 55(44):445205, November
  2022.

\bibitem{Ness_2012}
K.~F. Ness, R.~E. Robson, M.~J. Brunger, and R.~D. White.
\newblock Transport coefficients and cross sections for electrons in water
  vapour: {{Comparison}} of cross section sets using an improved {{Boltzmann}}
  equation solution.
\newblock {\em The Journal of Chemical Physics}, 136(2):024318, January 2012.

\bibitem{kawaguchi2016}
Satoru Kawaguchi, Kazuhiro Takahashi, Kohki Satoh, and Hidenori Itoh.
\newblock Electron transport analysis in water vapor.
\newblock {\em Japanese Journal of Applied Physics}, 55(7S2):07LD03, July 2016.

\bibitem{Itikawa_2005}
Yukikazu Itikawa and Nigel Mason.
\newblock Cross {{Sections}} for {{Electron Collisions}} with {{Water
  Molecules}}.
\newblock {\em Journal of Physical and Chemical Reference Data}, 34(1):1--22,
  March 2005.

\bibitem{pitchford2017}
Leanne~C. Pitchford, Luis~L. Alves, Klaus Bartschat, Stephen~F. Biagi,
  Marie-Claude Bordage, Igor Bray, Chris~E. Brion, Michael~J. Brunger, Laurence
  Campbell, Alise Chachereau, Bhaskar Chaudhury, Loucas~G. Christophorou, Emile
  Carbone, Nikolay~A. Dyatko, Christian~M. Franck, Dmitry~V. Fursa, Reetesh~K.
  Gangwar, Vasco Guerra, Pascal Haefliger, Gerjan J.~M. Hagelaar, Andreas
  Hoesl, Yukikazu Itikawa, Igor~V. Kochetov, Robert~P. McEachran, W.~Lowell
  Morgan, Anatoly~P. Napartovich, Vincent Puech, Mohamed Rabie, Lalita Sharma,
  Rajesh Srivastava, Allan~D. Stauffer, Jonathan Tennyson, Jaime {de Urquijo},
  Jan {van Dijk}, Larry~A. Viehland, Mark~C. Zammit, Oleg Zatsarinny, and
  Sergey Pancheshnyi.
\newblock {{LXCat}}: An {{Open-Access}}, {{Web-Based Platform}} for {{Data
  Needed}} for {{Modeling Low Temperature Plasmas}}.
\newblock {\em Plasma Processes and Polymers}, 14(1-2):1600098, January 2017.

\bibitem{chemistry_morgan_database}
Morgan database ({H}\textsubscript{2}{O}) \url{www.lxcat.net} (retrieved 13
  October 2023).

\bibitem{chemistry_phelps_database_H2O}
Phelps database ({H}\textsubscript{2}{O}) \url{www.lxcat.net} (retrieved 13
  October 2023).

\bibitem{Morgan_1990}
W.L. Morgan and B.M. Penetrante.
\newblock {{ELENDIF}}: {{A}} time-dependent {{Boltzmann}} solver for partially
  ionized plasmas.
\newblock {\em Computer Physics Communications}, 58(1-2):127--152, February
  1990.

\bibitem{chemistry_triniti_database}
Triniti database ({H}\textsubscript{2}{O}) \url{www.lxcat.net} (retrieved 13
  October 2023).

\bibitem{eedf_software}
N.A. Dyatko, I.V. Kochetov, A.P. Napartovich, and A.G. Sukharev.
\newblock {EEDF}: the software package for calculations of the electron energy
  distribution function in gas mixtures.
\newblock \url{http://www.lxcat.net/software/EEDF/}, 2011.
\newblock Accessed: 2024-01-30.

\bibitem{chemistry_itikawa_database}
Itikawa database ({H}\textsubscript{2}{O}) \url{www.lxcat.net} (retrieved 13
  October 2023).

\bibitem{song2021}
Mi-Young Song, Hyuck Cho, Grzegorz~P. Karwasz, Viatcheslav Kokoouline,
  Yoshiharu Nakamura, Jonathan Tennyson, Alexandre Faure, Nigel~J. Mason, and
  Yukikazu Itikawa.
\newblock Cross {{Sections}} for {{Electron Collisions}} with {{H2O}}.
\newblock {\em Journal of Physical and Chemical Reference Data}, 50(2):023103,
  June 2021.

\bibitem{Stephens_2016}
J~Stephens, A~Fierro, S~Beeson, G~Laity, D~Trienekens, R~P Joshi, J~Dickens,
  and A~Neuber.
\newblock Photoionization capable, extreme and vacuum ultraviolet emission in
  developing low temperature plasmas in air.
\newblock {\em Plasma Sources Science and Technology}, 25(2):025024, April
  2016.

\bibitem{zheleznyak1982}
M~B {Zheleznyak}, A~Kh {Mnatsakanyan}, and S~V {Sizykh}.
\newblock Photo-ionization of nitrogen and oxygen mixtures by radiation from a
  gas-discharge.
\newblock {\em High Temperature}, 20(3):357--362, November 1982.

\bibitem{bagheri2019}
B~Bagheri and J~Teunissen.
\newblock The effect of the stochasticity of photoionization on {{3D}} streamer
  simulations.
\newblock {\em Plasma Sources Science and Technology}, 28(4):045013, April
  2019.

\bibitem{wang2023}
Zhen Wang, Siebe Dijcks, Yihao Guo, Martijn Van Der~Leegte, Anbang Sun, Ute
  Ebert, Sander Nijdam, and Jannis Teunissen.
\newblock Quantitative modeling of streamer discharge branching in air.
\newblock {\em Plasma Sources Science and Technology}, 32(8):085007, August
  2023.

\bibitem{naidis2006}
G~V Naidis.
\newblock On photoionization produced by discharges in air.
\newblock {\em Plasma Sources Science and Technology}, 15(2):253--255, May
  2006.

\bibitem{Aints_2008}
M{\"a}rt Aints, Ants Haljaste, Toomas Plank, and Leho Roots.
\newblock Absorption of {{Photo}}-{{Ionizing Radiation}} of {{Corona
  Discharges}} in {{Air}}.
\newblock {\em Plasma Processes and Polymers}, 5(7):672--680, September 2008.

\bibitem{Li_2024}
Xiaoran Li, Siebe Dijcks, Anbang Sun, Sander Nijdam, and Jannis Teunissen.
\newblock Investigation of positive streamers in {{CO}}{\textsubscript{2}}:
  Experiments and {{3D}} particle-in-cell simulations.
\newblock {\em Plasma Sources Science and Technology}, 33(9):095009, September
  2024.

\bibitem{bourdon2007}
A~Bourdon, V~P Pasko, N~Y Liu, S~C{\'e}lestin, P~S{\'e}gur, and E~Marode.
\newblock Efficient models for photoionization produced by non-thermal gas
  discharges in air based on radiative transfer and the {{Helmholtz}}
  equations.
\newblock {\em Plasma Sources Science and Technology}, 16(3):656--678, August
  2007.

\bibitem{luque2007}
Alejandro Luque, Ute Ebert, Carolynne Montijn, and Willem Hundsdorfer.
\newblock Photoionization in negative streamers: {{Fast}} computations and two
  propagation modes.
\newblock {\em Applied Physics Letters}, 90(8):081501, February 2007.

\bibitem{turner2015}
Miles~M Turner.
\newblock Uncertainty and error in complex plasma chemistry models.
\newblock {\em Plasma Sources Science and Technology}, 24(3):035027, June 2015.

\bibitem{turner2016}
Miles~M Turner.
\newblock Uncertainty and sensitivity analysis in complex plasma chemistry
  models.
\newblock {\em Plasma Sources Science and Technology}, 25(1):015003, February
  2016.

\bibitem{pancheshnyi2000}
S.~V. Pancheshnyi, S.~V. Sobakin, S.~M. Starikovskaya, and A.~{\relax Yu}.
  Starikovskii.
\newblock Discharge dynamics and the production of active particles in a
  cathode-directed streamer.
\newblock {\em Plasma Physics Reports}, 26(12):1054--1065, December 2000.

\bibitem{johnsen1993}
R.~Johnsen.
\newblock Electron-temperature dependence of the recombination of
  {{H3O}}+({{H2O}}) {\emph{n}} ions with electrons.
\newblock {\em The Journal of Chemical Physics}, 98(7):5390--5395, April 1993.

\bibitem{kossyi1992}
I~A Kossyi, A~Yu Kostinsky, A~A Matveyev, and V~P Silakov.
\newblock Kinetic scheme of the non-equilibrium discharge in nitrogen-oxygen
  mixtures.
\newblock {\em Plasma Sources Science and Technology}, 1(3):207--220, August
  1992.

\bibitem{ono2020}
Ryo Ono and Atsushi Komuro.
\newblock Generation of the single-filament pulsed positive streamer discharge
  in atmospheric-pressure air and its comparison with two-dimensional
  simulation.
\newblock {\em Journal of Physics D: Applied Physics}, 53(3):035202, January
  2020.

\bibitem{Campolongo_2007}
Francesca Campolongo, Jessica Cariboni, and Andrea Saltelli.
\newblock An effective screening design for sensitivity analysis of large
  models.
\newblock {\em Environmental Modelling \& Software}, 22(10):1509--1518, October
  2007.

\bibitem{Berthelot_2017}
Antonin Berthelot and Annemie Bogaerts.
\newblock Modeling of {{CO}} {\textsubscript{2}} plasma: Effect of
  uncertainties in the plasma chemistry.
\newblock {\em Plasma Sources Science and Technology}, 26(11):115002, October
  2017.

\bibitem{Terraz_2020}
Loann Terraz, Tiago Silva, Antonio {Tejero-del-Caz}, Lu{\'i}s Lemos~Alves, and
  Vasco Guerra.
\newblock Sensitivity {{Analysis}} in {{Plasma Chemistry}}: {{Application}} to
  {{Oxygen Cold Plasmas}} and the {{LoKI Simulation Tool}}.
\newblock {\em The Journal of Physical Chemistry A}, 124(22):4354--4366, June
  2020.

\bibitem{Marskar_2020}
Robert Marskar.
\newblock {{3D}} fluid modeling of positive streamer discharges in air with
  stochastic photoionization.
\newblock {\em Plasma Sources Science and Technology}, 29(5):055007, May 2020.

\bibitem{Allen_1999}
N~L Allen and P~N Mikropoulos.
\newblock Dynamics of streamer propagation in air.
\newblock {\em Journal of Physics D: Applied Physics}, 32(8):913--919, April
  1999.

\bibitem{Meng_2017}
Xiaobo Meng, Hongwei Mei, Liming Wang, Zhicheng Guan, and Jun Zhou.
\newblock Characteristics of streamer propagation along the insulation surface:
  {{Influence}} of air pressure and humidity.
\newblock {\em IEEE Transactions on Dielectrics and Electrical Insulation},
  24(1):391--400, February 2017.

\bibitem{Spencer_1987}
M~N Spencer, J~S Dickinson, and D~J Eckstrom.
\newblock Afterglow conductivity measurements of air and
  {{N}}{\textsubscript{2}} following intense electron-beam excitation.
\newblock {\em Journal of Physics D: Applied Physics}, 20(7):923--932, July
  1987.

\bibitem{Li_2018}
Y~Li, E~M {van Veldhuizen}, G~J Zhang, U~Ebert, and S~Nijdam.
\newblock Positive double-pulse streamers: How pulse-to-pulse delay influences
  initiation and propagation of subsequent discharges.
\newblock {\em Plasma Sources Science and Technology}, 27(12):125003, December
  2018.

\bibitem{wei2023}
Zhenyu Wei, Atsushi Komuro, and Ryo Ono.
\newblock The influence of individual evaluation of electron-impact ionization,
  attachment, and photoionization rates on positive streamer propagation.
\newblock {\em Plasma Processes and Polymers}, page e2300113, October 2023.

\bibitem{francisco2021a}
Hani Francisco, Behnaz Bagheri, and Ute Ebert.
\newblock Electrically isolated propagating streamer heads formed by strong
  electron attachment.
\newblock {\em Plasma Sources Science and Technology}, 30(2):025006, February
  2021.

\bibitem{pancheshnyi2005}
S.~Pancheshnyi, M.~Nudnova, and A.~Starikovskii.
\newblock Development of a cathode-directed streamer discharge in air at
  different pressures: {{Experiment}} and comparison with direct numerical
  simulation.
\newblock {\em Physical Review E}, 71(1):016407, January 2005.

\bibitem{eichwald1997}
O.~Eichwald, M.~Yousfi, A.~Hennad, and M.~D. Benabdessadok.
\newblock Coupling of chemical kinetics, gas dynamics, and charged particle
  kinetics models for the analysis of {{NO}} reduction from flue gases.
\newblock {\em Journal of Applied Physics}, 82(10):4781--4794, November 1997.

\end{thebibliography}

\end{document}